\documentclass [12pt]{article}
\usepackage{amsmath,amssymb,cite}

\usepackage{tabularx}
\usepackage[english]{babel}
\usepackage[dvips]{graphicx}
\usepackage{indentfirst}
\usepackage{epsfig}
\numberwithin{equation}{section}

\begin{document}

\title{Matrix Models, Gauge Theory and Emergent Geometry}

\author{ Rodrigo Delgadillo-Blando$^{a,b}$, Denjoe O'Connor$^{a}$,  Badis Ydri$^{c}$\\
{}\\
$^{a}$School of Theoretical Physics, DIAS,\\ 
10 Burlington Road, Dublin 4, Ireland.\\
{}\\
$^{b}$Departamento de F\'{\i}sica, CINVESTAV-IPN,\\ 
Apdo. 14-740, 07000, M\'exico {\it D.F.} M\'exico.\\ 
{}\\
$^{c}$Institut f\"{u}r Physik, Humboldt-Universit\"{a}t  zu Berlin,\\ 
D-12489 Berlin, Germany.}
\maketitle

\begin{abstract}
We present, theoretical predictions and Monte Carlo simulations, for a
simple three matrix model that exhibits an exotic phase transition.
The nature of the transition is very different if approached from the
high or low temperature side. The high temperature phase is described
by three self interacting random matrices with no background
spacetime geometry. As the system cools there is a phase transition in
which a classical two-sphere condenses to form the background
geometry.  The transition has an entropy jump or latent heat, yet the
specific heat diverges as the transition is approached from low
temperatures. We find no divergence or evidence of critical
fluctuations when the transition is approached from the high
temperature phase.  At sufficiently low temperatures the system is
described by small fluctuations, on a background classical two-sphere, of a
$U(1)$ gauge field coupled to a massive scalar field.  The critical
temperature is pushed upwards as the scalar field mass is increased.
Once the geometrical phase is well established the specific heat takes
the value $1$ with the gauge and scalar fields each contributing
$1/2$.
\end{abstract}
\begin{center}
{\it THE AUTHOR B.Y DEDICATES THIS WORK TO THE MEMORY OF HIS DAUGHTER\nonumber\\ NOUR (24-NOV 06,28-JUN
07)}
\end{center}
\newpage
\tableofcontents
\newpage
\section{Introduction} 
 
All the fundamental laws of physics are now understood in geometrical
terms.  The classical geometry that plays a fundamental role in our
formulation of these laws has been vastly extended in noncommutative
geometry \cite{connes}. However, we still have very little insight
into the origins of spacetime geometry itself.  This situation has
been undergoing a significant evolution in recent years and it now
seems possible to understand geometry as an emergent concept.  The
notion of geometry as an emergent concept is not new, see for example
\cite{Bombelli:1987aa} for an inspiring discussion and
\cite{Seiberg:2006wf,Ambjorn:2006hu} for some recent ideas. We examine
such a phenomenon in the context of a simple three matrix model
\cite{Azuma:2004zq,CastroVillarreal:2004vh,O'Connor:2006wv}.

Matrix models with a background noncommutative geometry have received
attention as an alternative setting for the regularization of field
theories
\cite{Ydri:2001pv,O'Connor:2003aj,Balachandran:2005ew,Grosse:1996mz}
and as the configurations of $D0$ branes in string theory
\cite{Myers:1999ps,Alekseev:2000fd}. In the model studied here, the
situation is quite different. It has no background geometry in the
high temperature phase and the geometry itself emerges as the system
cools, much as a Bose condensate or superfluid emerges as a collective
phenomenon at low temperatures.  The simplicity of the model we study
here allows for a detailed examination of such an exotic
transition. We suspect the characteristic features of the transition
may be generic to this novel phenomenon.

In this article we study both theoretically and numerically a three
matrix model with global $SO(3)$ symmetry whose energy functional or
Euclidean action functional (see eq. (\ref{Action-simpleform})) is a
single trace of a quartic polynomial in the matrices $D_a$.  The model
contains three parameters, the inverse temperature
$\beta=\tilde{\alpha}^4=g^{-2}$, and parameters $m$ and $\mu$ which
provide coefficients for the quartic and quadratic terms in the
potential.

We find that as the parameters are varied the model has a phase
transition with two clearly distinct phases, one {\it geometrical} the
other a {\it matrix phase}.  Small fluctuations in the {\it
  geometrical phase} are those of a Yang-Mills and a scalar field
around a ground state corresponding to a round two-sphere.  In the
{\it matrix phase} there is no background spacetime geometry and the
fluctuations are those of the matrix entries around zero. In this
article we focus on the subset of parameter space where in the large
matrix limit the gauge group is Abelian.

For finite but large $N$, at low temperature, the model exhibits
fluctuations around a fuzzy sphere \cite{HoppeMadore}.  In the
infinite $N$ limit the macroscopic geometry becomes classical.  As the
temperature is increased it undergoes a transition with latent heat
so the entropy jumps, yet the model has critical fluctuations and a divergent
specific heat. As this critical coupling is approached the fuzzy
sphere radius expands to a critical radius and the sphere evaporates.
The neighbourhood of the critical point exhibits all the standard
symptoms of a continuous 2nd order transition, such as large scale
fluctuations, critical slowing down (of the Monte Carlo routine) and
is characterized by a specific heat exponent which we argue is
$\alpha=1/2$, a value consistent with our numerical simulations.  In
the high temperature (strong coupling) phase the model is a matrix model
closely related to zero dimensional Yang-Mills theory. As the transition
is approached from within this phase we find no evidence of critical
fluctuations and no divergence in the specific heat. 

In the {\it geometrical sphere} phase the gauge coupling constant is
$g^2=\beta^{-1}$ and $m$ parameterizes the mass of the scalar field.
For small $m^2$ we find a transition with discontinuous internal
energy $U=<S>/\beta$ (so the entropy jumps across the transition
\cite{emergent-geometry-prl}) while the specific heat is divergent as
the transition is approached from the low temperature fuzzy sphere
phase but finite when approached from the high temperature {\it matrix
  phase}.  The fuzzy sphere emerges as the low temperature ground
state which as one expects is the low entropy phase and the transition
is characterized by both a latent heat and divergent fluctuations.

To our knowledge it is the first clear example of a transition where
the spacetime geometry is emergent.  This transition itself is
extremely unusual. We know of no other physical situation that has a
transition with these features.  Standard transitions are very
dependent on the dimension of the background spacetime and when this
is itself in transition an asymmetry of the approach to criticality is
not so surprising.

By studying the eigenvalues of operators in the theory we establish
that, in the matrix phase, the matrices $D_a$ are characterised by
continuous eigenvalue distributions which undergo a transition to a
point spectrum characteristic of the fuzzy sphere phase as the
temperature is lowered. The point spectrum is consistent with
$D_a=L_a/R$ where $L_a$ are $su(2)$ angular momentum generators in the
irreducible representation given by the matrix size and $R$ is the
radius of the fuzzy sphere. The full model received an initial study
in \cite{O'Connor:2006wv} while a simpler version, invariant under
translations of $D_a$, arises naturally as the configuration of $D0$
branes in the large $k$ limit of a boundary Wess-Zumino-Novikov-Witten
model \cite{Alekseev:2000fd} and has been studied numerically in
\cite{Azuma:2004zq}. If the mass parameters of the potential are
related to the matrix size, the model becomes that introduced in 
\cite{Steinacker:2003sd}. The interpretation presented here is novel, 
as are the results on the entropy and critical behaviour 
and the extension to the full model.

A short description of the results obtained in this article is given
in \cite{emergent-geometry-prl}. This article is organised as
follows. In section $2$ we review the fuzzy sphere and its geometry.
In section $3$ we derive several theoretical predictions in the fuzzy
sphere phase including the critical behaviour of the model.  In
section $4$ we discuss the non-perturbative phase structure (phase
diagram) and Monte Carlo numerical results for various observables. 
We conclude in section $5$ with some discussion and speculations.

\section{The fuzzy sphere}
The ordinary round unit sphere, ${ S}^2$, can be defined as the two
dimensional surface embedded in flat three dimensional space
satisfying the equation 
$\sum_{a=1}^3n_a^2=1$ with $\vec{n}{\in}{R}^3$. 
One can use the $n_a$ as a nonholonomic coordinate system for
the sphere. In this coordinate system a general function can be
expanded as $f(\vec{n})=\sum_{l=0}^{\infty}f_{lm}Y_{lm}(\vec{n})$,
where $Y_{lm}$ are the standard spherical harmonics.  The basic 
derivations are provided by the $SO(3)$ generators 
defined by 
${\cal L}_a=-i{\epsilon}_{abc}n_b{\partial}_c$ 
and the Laplacian is 
${\cal L}^2={\cal L}_a{\cal L}_a$ with eigenvalues $l(l+1)$,
$l=0,...,\infty $.  Following Fr\"{o}hlich and Gaw\c{e}dzki
\cite{FroehlichGawedzki} (or Connes \cite{connes} for spin geometry) 
the geometry of a Riemannian manifold can be encoded in a spectral triple. 
For the ordinary sphere the triple is 
$(C^{\infty}({\bf S}^2),{\cal H},{\cal L}^2)$ 
where $C^{\infty}({ S}^2)$ is the algebra
of all functions $f(\vec{n})$ on the sphere and
${\cal H}$ is the infinite dimensional Hilbert space of square
integrable functions. 
Such a spectral triple is precisely the data that enters 
the scalar field action on the manifold. So a scalar field action 
which includes an appropriate Laplacian can specify the geometry. 

The fuzzy sphere can be viewed as a particular deformation of the
above triple which is based on the fact that the sphere is the
coadjoint orbit $SU(2)/U(1)$ \cite{HoppeMadore},
\begin{eqnarray}
g{\sigma}_3g^{-1}=n_a{\sigma}_a~,~g{\in}SU(2)~,~\vec{n}{\in}S^2 ,
\end{eqnarray}
and is therefore a symplectic manifold which can 
be quantized in a canonical fashion by simply quantizing the volume form 
\begin{eqnarray}
{\omega}=sin{\theta}d\theta{\wedge} d{\phi}=\frac{1}{2}{\epsilon}_{abc}n_adn_b{\wedge}dn_c.
\end{eqnarray}
The result of this quantization is to replace the algebra $C^{\infty}(S^2)$ 
by the algebra of $N{\times}N$ matrices $Mat_N$. $Mat_N$ becomes the 
$N^2-$dimensional Hilbert space structure ${H}_N$ when supplied 
with inner product 
$(f,g)=\frac{1}{N}Tr(f^{+}g)$ where 
$f,g{\in}Mat_{N} $. The spin $\frac{N-1}{2}$ IRR of $su(2)$ has both a 
left and a right action on this this Hilbert space. For the left
action the generators are $L_a$ and satisfying 
$[L_a,L_b]=i{\epsilon}_{abc}L_c ,~{\rm and}~ \sum_a L_a^2=c_2=\frac{N^2-1}{4}$.
The spherical harmonics $Y_{lm}(\vec{n})$ become the 
canonical $su(2)$ polarization tensors $\hat{Y}_{lm}$ and form a basis for 
$H_N$.  These are defined by
\begin{eqnarray}
[L_a,[L_a,\hat{Y}_{lm}]]=l(l+1)\hat{Y}_{lm}~,~[L_{\pm},\hat{Y}_{lm}]=\sqrt{(l{\mp}m)(l{\pm}m+1)}\hat{Y}_{lm{\pm}1}~,~[L_3,\hat{Y}_{lm}]=m\hat{Y}_{lm}.
\end{eqnarray}
and satisfy
\begin{eqnarray}
\hat{Y}_{lm}^{\dagger}=(-1)^m\hat{Y}_{l-m}~,~\frac{1}{N}Tr\hat{Y}_{l_1m_1}\hat{Y}_{l_2m_2}=(-1)^{m_1}{\delta}_{l_1l_2}{\delta}_{m_1,-m_2},
\end{eqnarray}
and the completeness relation 
\begin{equation}
\sum_{l=0}^{N-1}\sum_{m=-l}^{l}\hat{Y}_{lm}^{\dagger}\hat{Y}_{lm}={\bf 1}
\end{equation}
The ``coordinates functions'' on the fuzzy sphere 
${ S}^2_N$ are defined to be proportional to 
$\hat{Y}_{1\mu}$ tensors (as in the continuum) and satisfy
\begin{eqnarray}
x_1^2+x_2^2+x_3^2=1~,~
[x_a,x_b]=\frac{i}{\sqrt{c_2}}{\epsilon}_{abc}x_c,~{\rm where}~x_a=\frac{L_a}{\sqrt{c_2}}.
\end{eqnarray}
``Fuzzy'' functions on ${ S}^2_N$ are elements of the
matrix algebra while derivations are inner and given by the
generators of the adjoint action of $su(2)$ defined by 
$\hat{\cal L}_a{\phi}:=[L_a,{\phi}]$.
A natural choice of the Laplacian on the fuzzy sphere is therefore 
given by the Casimir operator 
\begin{eqnarray}
\hat{\cal L}^2=[L_a,[L_a,..]].
\end{eqnarray} 
Thus the algebra of matrices $Mat_{N}$ with $N=L+1$ decomposes under the
action of  $su(2)$ as 
$\frac{L}{2}{\otimes}\frac{L}{2}=0{\oplus}1{\oplus}2{\oplus}..{\oplus}L$, 
with the first $\frac{L}{2}$ standing for the left action
while the other $\frac{L}{2}$ stands for the right action of $su(2)$. 
It is not difficult to convince ourselves that this Laplacain has a
cut-off spectrum with eigenvalues $l(l+1)$ where $l=0,1,...,L$. 
Given the above discussion we see that 
a general fuzzy function (or element of the algebra) 
on ${ S}^2_N$ can be expanded 
in terms of polarization tensors as follows 
${f}=\sum_{l=0}^{L}\sum_{m=-l}^{l}{f}_{lm}\hat{Y}_{lm}$. 
The continuum limit is given by $L{\longrightarrow}{\infty}$. 
Therefore the fuzzy sphere can be described as a sequence of triples 
$(Mat_{N},H_N,\hat{\cal L}^2)$ with a well defined limit given 
by the triple  $(C^{\infty}({\bf S}^2),{\cal H},{\cal L}^2)$. 
The number of degrees of freedom in the function algebra of 
${\bf S}^2_N$ is $N^2$ and the noncommutativity parameter 
is $\theta=\frac{2}{\sqrt{N^2-1}}$.

\section{Theoretical predictions}

\subsection{Gauge action}
It has been shown in \cite{Grosse:1992bm,CarowWatamura:1997qw,CarowWatamura:1998jn,Grosse:2000gd} 
that the differential calculus 
on the fuzzy sphere is three dimensional and
as a consequence, treating gauge fields as one-forms,
a generic gauge field, $\vec{A}$ has $3$ components. Each component 
$A_a$, $a=1,2,3$, is an element of $Mat_{N}$ and the 
$U(1)$ gauge symmetry of the commutative sphere will become a $U(N)$ 
gauge symmetry on the fuzzy sphere with gauge transformations implemented 
as
$A_a{\longrightarrow}UA_aU^{\dagger}+U[L_a,U^{\dagger}]$ where $U\in U(N)$.  
In this approach to gauge fields on the fuzzy sphere, 
${\bf S}^2_N$, it is difficult to split the vector field $\vec{A}$ in a
gauge-covariant fashion into a tangent gauge field and a normal
scalar field. However, we can write a gauge-covariant expression for
the normal scalar field as
$\Phi=\frac{1}{2}(x_aA_a+A_ax_a+\frac{A_a^2}{\sqrt{c_2}})$. In the
commutative limit, $N\rightarrow\infty$, we have $A_a\rightarrow {\cal A}_a$ 
and 
$\Phi\rightarrow\varphi=n_a{\cal A}_a$ and the splitting into gauge field and 
scalar field becomes trivial being implemented by 
simply writing ${\cal A}_a=n_a{\varphi}+a_a$, with $n_a a_a=0$,
where $\vec{n}$ is the unit vector on ${\bf S}^2$, 
$\varphi=\vec{n}\cdot \vec{{\cal A}}$ is
the normal gauge-invariant component of $\vec{{\cal A}}$ and $\vec{a}$ is the
tangent gauge field.  

For this formulation of gauge field theory on the fuzzy sphere 
the most general action (up to quartic power in
$A_a$) on ${{\bf S}}^2_N$ is then 
\begin{eqnarray}
\label{matrix-action-A_a}
\fbox{$S_N[A]=\frac{1}{4g^2N}TrF_{ab}^2-\frac{1}{2g^2N}{\epsilon}_{abc}Tr\left[\frac{1}{2}F_{ab}A_c-\frac{i}{6}[A_a,A_b]A_c\right]+\frac{2m^2}{g^2N}Tr\Phi^2+\frac{{\rho}}{g^2N}Tr \Phi.$}\label{main0}
\end{eqnarray}
In above $Tr{\bf 1}=N$ and $F_{ab}=i[L_a,A_b]-i[L_b,A_a]+{\epsilon}_{abc}A_c+i[A_a,A_b]$ is the covariant curvature, and $S_N[0]=0$.  The limit $m^2{\longrightarrow}\infty $ 
gives a large mass to the scalar component and effectively projects it out of
the spectrum of small fluctuations. 

The associated continuum action $S_{\infty}$ is then at most quadratic
in the field and as a consequence the theory is largely trivial,
consisting of a gauge field and a scalar field that have a mixing in
their joint propagator.  Indeed we can show that
\begin{eqnarray}
S_{\infty}=\frac{1}{4g^2}\int
\frac{d{\Omega}}{4{\pi}} \bigg[(f_{ab})^2-4{\epsilon}_{abc}f_{ab}n_c\varphi
-2({\cal L}_a\varphi)^2+4(1+2m^2){\varphi}^2+4\rho {\varphi}\bigg],\label{higgs}
\end{eqnarray}
where $f_{ab}$ is the curvature of the tangent field $a_a$ and 
$f_{ab}=i{\cal L}_aa_b-i{\cal
L}_ba_a+{\epsilon}_{abc}a_c$. As one can immediately
see this theory consists of a $2$-component gauge field $a_a$ that 
mixes with a scalar field $\varphi$, i.e. the propagator mixes the two fields. 
In the following we will primarily be interested in the case with $\rho=0$.
We see also that the presence of the scalar field means that the geometry
is completely specified, in that all the ingredients of the spectral triple
are supplied by this field. In contrast a two dimensional gauge theory on its
own would not be sufficient to specify the geometry.

\subsection{Matrix model}
We introduce $\tilde{\alpha}^4=\frac{1}{g^2}=\beta$ 
where $\beta$ can be interpreted as an inverse temperature 
and we can rewrite the above gauge action 
(\ref{matrix-action-A_a}) (shifted by constants
and dropping the subscript $N$)  
in terms of 
$D_a=L_a+A_a$ as follows:
\begin{eqnarray}
S^{(0)}[D]&=&\frac{1}{4g^2N}TrF_{ab}^2-\frac{1}{2g^2N}{\epsilon}_{abc}Tr\left[\frac{1}{2}F_{ab}A_c-\frac{i}{6}[A_a,A_b]A_c\right]-\frac{\tilde{\alpha}^4c_2}{6}\nonumber\\
&=&\frac{\tilde{\alpha}^4}{N}\bigg[-\frac{1}{4}Tr[D_a,D_b]^2+\frac{2i}{3}{\epsilon}_{abc}TrD_aD_bD_c\bigg].
\label{Action-S0}
\end{eqnarray}
 \begin{eqnarray}
V[D]&=&\frac{2m^2}{g^2N}Tr\Phi^2+\frac{{\rho}}{g^2N}Tr\Phi-\frac{m^2
  c_2}{2g^2}+\frac{\rho\sqrt{c_2}}{2g^2}\nonumber\\ &=&\frac{\tilde{\alpha}^4}{N}\bigg[\frac{m^2}{2c_2}
  Tr(D_a^2)^2+\left(\frac{{\rho}}{2\sqrt{c_2}}-m^2\right)Tr
  (D_a^2)\bigg].
\end{eqnarray}
The complete action functional is then:
\begin{eqnarray}
\fbox{$S[D]\equiv {S}^{(0)}[D]+V[D]= S_N[A]-\frac{\tilde{\alpha}^4c_2}{6}-\frac{\tilde{\alpha}^4m^2c_2}{2}+\frac{\tilde{\alpha}^4\rho\sqrt{c_2}}{2},$}
\label{main11}
\end{eqnarray}
and the constants are chosen so that $S[0]=0$.
The action takes the rather simple form
\begin{equation}
\fbox{$S[D]=\frac{\tilde{\alpha}^4}{N}Tr\big[-\frac{1}{4}[D_a,D_b]^2+\frac{2i}{3}{\epsilon}_{abc}D_aD_bD_c+\frac{m^2}{2c_2}(D_a^2)^2-\mu D_a^2\big]$}
\label{Action-simpleform}
\end{equation}
where $\mu=m^2-\frac{{\rho}}{2\sqrt{c_2}}$.  It is invariant under
unitary transformations $U(N)$ and global rotations $SO(3)$. Extrema
of the model are given by the reducible representations of $SU(2)$ and
commuting matrices.  For sufficiently small $\rho$ and with
$c_2=(N^2-1)/4$ the classical absolute minima of the model is given by
the irreducible representation of $SU(2)$ of dimension $N$.  Small
fluctuations around this background can then be seen to have the
geometrical content of a Yang-Mills and scalar multiplet on a
background fuzzy sphere as described in the previous section.  As we
will see, for small enough coupling or low temperature, these
configurations also give the ground state of the fluctuating system.
For very small negative values of the parameter $\mu$, and for $m=0$
there is a local minimum at $D_a=0$ and a global minimum at $D_a\sim
L_a$, separated by a barrier.  As $\mu$ is made more negative 
the difference in energy (or Euclidean action) between the two extrema 
becomes less and eventually for $\mu=-\frac{2}{9}$ the
two minima become degenerate, one occurring at $D_a=0$ while the
other occurs at $D_a=\frac{2}{3}L_a$ and they are separated by a barrier
whose maximum occurs at $D_a=\frac{1}{3}L_a$.

In this special case (i.e. $m=0$ and $\mu =-\frac{2}{9}$ or equivalently 
$\rho=\frac{4}{9}\sqrt{c_2}$) the action 
takes the form
\begin{equation}
S[D]=\frac{\tilde{\alpha}^4}{N}\bigg(\frac{i}{2}Tr[D_a,D_b]+\frac{1}{3}{\epsilon}_{abc}D_c\bigg)^2.
\end{equation}
and we see that the configurations $D_a=0$ and $D_a=\frac{2}{3}L_a$
both give zero action, however there is a unique configuration with
$D_a=0$ while there is an entire $SU(N)$ manifold of configurations
$D_a=\frac{2}{3}U L_a U^{\dagger}$ which are equivalent. The classical
model has a first order transition at $m=0$ for $\mu=-\frac{2}{9}$ and
the classical ground state switches from $D_a=\phi L_a$, where
$\phi=(1+\sqrt{1+4\mu})/2$, for $\mu > -2/9$ to $D_a=0$ for 
$\mu < -2/9$.  The quantity $\phi$ is therefore a useful order parameter for
this transition. However, one would expect fluctuations to have a
significant effect on this classical picture.

In fact both theoretical and numerical studies show that 
in the fluctuating theory, the fuzzy sphere phase only 
exists for $\mu>-2/9$ \cite{Azuma:2005bj}.

\subsection{Quantization and Observables for small $m^2$}

The quantum version of the model is taken to be 
that obtained by functional integration with respect to the gauge field.
This amounts to integration over the three Hermitian 
matrices $D_a$ with Dyson measure and the partition function $Z$ is given by
\begin{eqnarray}
&&Z=\int dD_a e^{-{S}[D]}=e^{-\frac{3N^2}{4}\log {{\alpha}^4}}\int dX_ae^{-N \hat{S}[X]}~,~X_a=\alpha D_a~,~\tilde{\alpha}=\alpha \sqrt{N}\nonumber\\
&&\hat{S}[X]=-\frac{1}{4}Tr[X_a,X_b]^2+\frac{2i\alpha}{3}{\epsilon}_{abc}TrX_aX_bX_c+\frac{m^2}{2c_2}
Tr(X_a^2)^2+\left(\frac{\rho {\alpha}^2}{2\sqrt{c_2}}-m^2\alpha^2\right)Tr (X_a^2).\nonumber\\
\end{eqnarray}
The latter form of the expression (in terms of $X_a$) allows us 
to take $\tilde\alpha=0$ and we see 
that this limit is equivalent to removing all but the leading 
commutator squared term. Also the quartic term proportional to $m^2$ survives. However if $m^2$ and $\rho$ are scaled appropriately 
with $\tilde\alpha$ only the Chern-Simons 
term ${\epsilon}_{abc}TrX_aX_bX_c$ is removed.

The set of gauge equivalent configurations is parameterized by 
the $SU(N)$ group manifold which is compact, so there is no need to gauge fix 
and the functional integral is well defined, being an ordinary integral over 
${\bf R}^{3N^2}$. However,  the volume of the gauge group diverges 
in the limit $N\rightarrow\infty$ and to make contact with the commutative 
formulation it is convenient to gauge fix in the standard way. 

In the background field gauge formulation we separate the 
field as  $ X_a=\tilde\alpha D_a+Q_a$. The action is 
invariant under $D_a{\longrightarrow}D_a$,
$Q_a{\longrightarrow}UQ_aU^{\dagger}+U[D_a,U^{\dagger}]$. 

Following the standard Faddeev-Popov procedure \cite{faddeev} and taking 
the background field configuration to be $D_a=\phi L_a$ one finds,
keeping $m^2$ fixed as the $N\rightarrow\infty$ limit is taken, that 
$F=-\ln Z$ is given by 
\begin{eqnarray}
\fbox{$\frac{F}{N^2}=\frac{3}{4}\log\tilde{\alpha}^4+\frac{\tilde{\alpha}^4}{2}
\bigg[\frac{{\phi}^4}{4}-\frac{{\phi}^3}{3}
+m^2\frac{{\phi}^4}{4}-\mu\frac{\phi^2}{2}\bigg]+
\log\tilde{\alpha}{\phi}, $}\label{formula1}
\end{eqnarray}
with $\mu=m^2-\frac{\rho}{2\sqrt{c_2}}$. 

The most notable feature of this expression is that the entire 
fluctuation contribution is summarized in the logarithmic term 
$\log(\tilde{\alpha}\phi)$.  From (\ref{formula1}) we see that 
the effective potential for the order parameter $\phi$ is
\begin{equation}
\label{V_eff}
\frac{V_{\rm eff}}{2c_2}=\tilde{\alpha}^4\left[\frac{\phi^4}{4}
-\frac{\phi^3}{3}+m^2\frac{\phi^4}{4}-\mu\frac{\phi^2}{2}\right]+\log \phi^2
\end{equation}
The effective potential is not bounded below at $\phi=0$ due to the 
$\ln\phi$ term. However, our analysis assumes the existence of a fuzzy sphere
ground state and so the effective potential can only be trusted in this phase.
It has a local  minimum for $\phi$ positive and sufficiently large 
$\tilde\alpha$ and in this regime the fuzzy sphere configuration exists, 
for lower values of $\tilde\alpha$ our numerical study indicates that 
the model is indeed in a different phase.

Setting $\phi\frac{\partial F}{\partial\phi}=0$ (or equivalently 
$\phi\frac{\partial V_{\rm eff}(\phi)}{\partial\phi}=0$) gives
\begin{equation}
\label{equ-for-phi}
\frac{\tilde{\alpha}^4}{2}
\bigg[{\phi}^4-{\phi}^3
+m^2{\phi}^4-\mu{\phi^2}\bigg]+1=0 , 
\end{equation}
the solution of which specifies $\phi$. 
We define  the average of the action, which will be one of the principal 
observables of our numerical study, as
\begin{equation}
{\cal S}=<S>/N^2
=\tilde{\alpha}^4\frac{d}{d\tilde{\alpha}^4}\bigg(\frac{F}{N^2}\bigg) .
\end{equation}  
Then a direct computation and use of eq. 
(\ref{equ-for-phi}) yields 
\begin{eqnarray}
\label{average_action}
\fbox{${\cal S}=\frac{3}{4}-\frac{\tilde{\alpha}^4{\phi}^3}{24}-\frac{\tilde{\alpha}^4\mu{\phi}^2}{8}.$}\label{u}
\end{eqnarray}
We can also compute the expected radius of the sphere
\begin{equation}
\label{sphere_expected_radius}
\frac{1}{R}=\frac{<Tr D_a^2>}{Nc_2}=
-\frac{2}{\tilde{\alpha}^4}\frac{dF}{d \mu}=\phi^2
\end{equation}
This can also be calculated directly in perturbation theory as
\begin{eqnarray}
\nonumber\\
\frac{1}{R}&=&{\phi}^2 +\frac{1}{c_2\tilde{\alpha}^4{\phi}^2}Tr_3 Tr_{N^2}(\frac{1}{{\cal L}_c^2{\delta}_{ab}+4m^2x_ax_b}).
\end{eqnarray}
Hence, we conclude that the expected inverse radius of the 
fuzzy sphere is given by
\begin{eqnarray}
\fbox{$\frac{1}{R}={\phi}^2.$}
\end{eqnarray}
Also
\begin{eqnarray}
\frac{<Tr((D_a^2)^2)>}{2N^3c_2}
=\frac{1}{\tilde{\alpha}^4}\frac{\partial}{{\partial}m^2}\bigg(\frac{F}{N^2}\bigg)=\frac{{\phi}^4}{8} ,  
\end{eqnarray}
which yields
\begin{eqnarray}
\fbox{$\frac{<Tr(D_a^2)^2>}{2N^3c_2}=\frac{\phi^4}{8}.$}
\end{eqnarray}

Scaling $(D_a)_{ij}$ to $(1-\epsilon )(D_a)_{ij}$ in both the
action and measure amounts to a simple coordinate transformation 
which leaves the partition function invariant. However it also leads
to the  nontrivial identity
\begin{eqnarray}\label{wardid}
\frac{\tilde{\alpha}^4}{N}<K_m>=3N^2~,~K_m=Tr\bigg(-[D_a,D_b]^2 +2 i \epsilon_{abc} D_a D_b D_c-2 m^2D_a^2+\frac{2m^2}{c_2} (D_a^2)^2\bigg).\label{wa1}
\end{eqnarray}
Using this identity we can express 
\begin{eqnarray}
{\cal S}=\frac{3}{4}+\frac{\tilde{\alpha}^4}{6N^3}<i{\epsilon}_{abc}Tr
D_aD_bD_c>-\frac{\tilde{\alpha}^4}{2N^3}m^2<Tr D_a^2>.\label{wa2}
\end{eqnarray}
or equivalently in the form
\begin{eqnarray}
{\cal S}=1+\frac{\tilde{\alpha}^4}{12N^3}<Tr
[D_a,D_b]^2>-\frac{\tilde{\alpha}^4}{3N^3}m^2<Tr D_a^2>-\frac{\tilde{\alpha}^4}{6N^3c_2}m^2<Tr (D_a^2)^2>.\label{wa3}\nonumber\\
\end{eqnarray}
We define the Yang-Mills and Chern-Simons actions by
\begin{eqnarray}
\fbox{$
4{\rm YM}=-\frac{<Tr[D_a,D_b]^2>}{2Nc_2},\rm{ and }~3{\rm CS}=\frac{<i{\epsilon}_{abc}Tr D_aD_bD_c>}{Nc_2}.$}
\end{eqnarray} 
Then using the above results we find
\begin{eqnarray}
4{\rm YM}={\phi}^4+\frac{8}{\tilde{\alpha}^4}~ \rm{ and}~3{\rm CS}=-{\phi}^3.
\end{eqnarray} 
In the above we have extensively used the fact that 
$\phi$ must satisfy (\ref{equ-for-phi}) and taken $\rho=0$ i.e. $\mu=m^2$.

Another significant observable for our numerical study is the specific heat
$C_v$ defined as 
\begin{eqnarray}
C_v:=\frac{<S^2>-<S>^2}{N^2}=\frac{<S>}{N^2}-\tilde{\alpha}^4\frac{d}{d\tilde{\alpha}^4}\bigg(\frac{<S>}{N^2}\bigg).
\end{eqnarray}
A direct calculation yields
\begin{eqnarray}\label{cv-exact}
\fbox{$C_v=\frac{3}{4}+\frac{\tilde{\alpha}^5\phi}{32}(\phi +2m^2)\frac{d\phi}{d\tilde{\alpha}}.$}\label{cv}
\end{eqnarray}

Finally one can recover perturbation theory in the coupling 
$g^2=1/\tilde{\alpha}^4$ by expanding  in $1/\tilde{\alpha}^4$.
In particular the one-loop predictions are obtained by 
using the solution of (\ref{equ-for-phi}) 
expanded to first order which is
$\tilde{\alpha}{\longrightarrow}\infty$ given by
\begin{eqnarray}
\phi =1-\frac{2}{1+2m^2}\frac{1}{\tilde{\alpha}^4}
+O(\frac{1}{\tilde{\alpha}^8}).
\end{eqnarray}
In the next section we will look at the solution of (\ref{equ-for-phi}) in more detail and the consequences for the transition.

\subsection{Phase Transitions}
Let us now examine the predictions for the quantum transition as
determined by $F$ given in (\ref{formula1}) or the effective potential 
(\ref{V_eff}) which we repeat here for convenience 
\begin{eqnarray}
\fbox{$\frac{V_{\rm eff}}{2c_2}=\tilde{\alpha}^4
\bigg[\frac{\phi^4}{4}-\frac{\phi^3}{3}+m^2\frac{\phi^4}{4}
-\mu\frac{\phi^2}{2}\bigg]+\log\tilde{\phi}^2.$}\label{veff}
\end{eqnarray}
But first let us review the classical case.  The only difference between the 
full quantum potential (\ref{V_eff}) and 
the corresponding classical potential is the quantum induced logarithm
of $\phi$, which as we will see plays a crucial role. The extrema of
the classical potential occur at
\begin{equation}
\bar{\phi}=(1+m^2)\phi=\left\{0,~ \frac{1-\sqrt{1+4t}}{2}, ~ 
\frac{1+\sqrt{1+4t}}{2}\right\}.
\end{equation} 
where $t=\mu(1+m^2)$. 
The first and last expressions are local minima
and the middle one is the maximum of the barrier between them. For 
$\mu$ positive the global minimum is the third expression, i.e. the 
largest value of $\phi$.  When written in terms of $\bar{\phi}$
we see the potential takes the same form as that for $m=0$ and
we can read off that if $\mu$ is sent negative then 
this minimum  becomes degenerate with that at $\phi=0$ 
at $t=\mu(1+m^2)=-\frac{2}{9}$ and the maximum height of the barrier
is given by $\frac{\tilde\alpha^4}{324 (1+m^2)^3}$. 
When $\mu=m^2$ the minimum is clearly $\phi=1$, i.e $D_a=L_a$ for all $m$
and this is separated from the local minimum at $\phi=0$ by a barrier.
The classical transition therefore has the same character as that of
the $m=0$ model and the transition is 1st order and 
occurs only when $\rho$ is tuned to a critical value.

\begin{figure}
\begin{center}
\includegraphics[width=5.8cm,angle=-90]{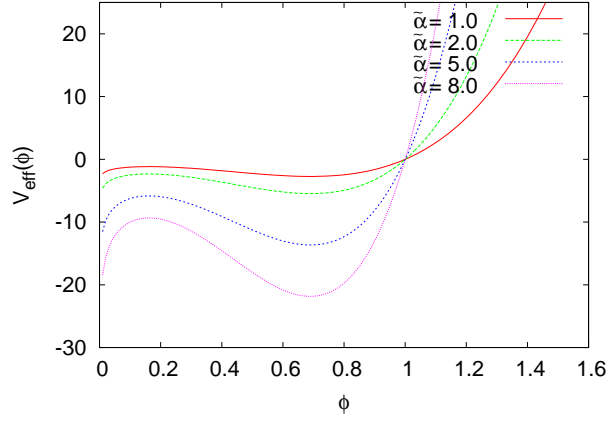}
\caption{{The effective potential for $m^2=20$. The minimum corresponds
with the fuzzy sphere solution, as we lower the coupling constant the
minimum disappears and the fuzzy sphere collapses.}}\label{figVeff}
\end{center}
\end{figure}

Let us now consider the effect of the fluctuation induced 
logarithm of $\phi$. The potential is plotted in figure \ref{figVeff}
for different values of $\tilde\alpha$ and $\mu=m^2=20$.
The condition $V^{'}_{\rm eff}(\phi)=0$ 
gives us extrema of the model. For large enough $\tilde{\alpha}$  
(or low enough temperature) and large enough $m$ and $\mu$ it 
admits four real solutions two for positive $\phi$ and two for negative 
$\phi$. The largest of the positive $\phi$ solutions can be identified
with the least free energy and therefore the ground state of the system 
in this phase of the theory. It will
determine the actual radius of the sphere. 
The second positive solution is the local maximum (figure
\ref{figVeff}) of $V_{\rm eff}(\phi)$ and will determine the height of the 
barrier in the effective potential. As the coupling is decreased
(or the temperature increased) these two solutions merge and
the barrier disappears. This is the critical point of the model
and it has no classical counterpart since, in the classical case, the
barrier between the two minima never disappears. For smaller couplings
than this critical coupling $\tilde\alpha_*$ the fuzzy sphere
solution $D_a= {\phi}L_a$ no longer exists and the effective potential 
cannot be relied on. This is in accord with our numerical simulations 
which indicate that as the matrix size is increased the radius as defined in 
(\ref{sphere_expected_radius}) appears to go to zero.

The condition when the barrier disappears is
\begin{eqnarray}
\label{barrier_disappears}
V^{''}_{\rm eff}=\frac{\tilde{\alpha}^4}{2}\big[3{\phi}^2-2 \phi +3m^2{\phi}^2-\mu\big]-\frac{1}{{\phi}^2}=0.
\end{eqnarray}
Solving both (\ref{equ-for-phi}) and (\ref{barrier_disappears}) 
yields $4(1+m^2)\phi^2_*-3\phi_*-2\mu=0$ and therefore the critical values
\begin{eqnarray}
\fbox{$\bar{\phi}_{*}=\frac{3}{8}(1+\sqrt{1+\frac{32t}{9}}).$}
\end{eqnarray}
\begin{eqnarray}\label{critline}
\fbox{$g_{*}^2=\frac{1}{\tilde{\alpha}_*^4}
=\frac{{\phi}_*^2({\phi}_*+2\mu)}{8}.$}
\end{eqnarray}
where as defined earlier $\bar{\phi}_*=\phi/(1+m^2)$ and $t=\mu(1+m^2)$.
Setting both $m^2=0$ and $\mu=0$ yields
\begin{eqnarray}
\fbox{$
{\phi}_{*}=\frac{3}{4}~,~\tilde{\alpha}_*^4=(\frac{8}{3})^3.$}\label{cv0}
\end{eqnarray}
while setting $m=0$ alone leads to no significant simplification 
and we still have $\phi_*=\bar\phi_*$ with $t=\mu$.

If we take $\mu$ negative as in the discussion of the last section 
we see that $g_*$ goes to zero at $t=-1/4$ and the critical 
coupling $\tilde{\alpha}_* $ is sent to infinity and therefore
for $t<-\frac{1}{4}$ the model has no fuzzy sphere phase.
This case arises when 
\begin{equation}
\rho > 2\sqrt{c_2}(m^2+\frac{1}{4(1+m^2)}).
\end{equation}
However, in the region $-\frac{1}{4}<\mu <-\frac{2}{9}$ the action 
(\ref{Action-simpleform}) is completely positive. It is therefore
not sufficient to consider only the configuration $D_a=\phi L_a$,
but rather all $SU(2)$ representations must be considered. 
Furthermore for large $\tilde{\alpha}$ the ground state will be dominated
by those representations with the smallest Casimir. This means that
there is no fuzzy sphere solution for $\mu<-\frac{2}{9}$. A result that
we also observe in simulations and in agreement with the result 
of \cite{Azuma:2005bj}.
We therefore see that the classical transition described above is
significantly affected by fluctuations and in particular
the fuzzy sphere phase dissapears when $\rho$ is increased to
the special value of $\frac{4\sqrt{c_2}}{9}$.

The other limit of interest is the limit $\mu=m^2{\longrightarrow}\infty$. 
In this case 
\begin{eqnarray}
\fbox{$
{\phi}_{*}=\frac{1}{\sqrt{2}}~,~\tilde{\alpha}^4_{*}=\frac{8}{m^2}.$}\label{cvinfy}
\end{eqnarray}
This means that the phase transition is located at a smaller value of
the coupling constant $\tilde{\alpha}$ as $m$ is increased.  In other
words the region where the fuzzy sphere is stable is extended to lower
values of the coupling or higher temperatures. These
results agree nicely with numerical data. 

As we cross the critical value of $\tilde\alpha$, a rather exotic phase
transition occurs where the geometry disappears as the temperature is
increased. The fuzzy sphere phase has the background geometry of a two
dimensional spherical non-commutative manifold which macroscopically
becomes a standard commutative sphere for $N\rightarrow\infty$.  The
fluctuations are then of a $U(1)$ gauge theory which mixes with a
scalar field on this background.  In the high temperature phase, which we 
call a matrix phase, the order parameter $\phi$ is not well 
defined and the fluctuations are around diagonal matrices so  
the model is a pure matrix one
corresponding to a zero dimensional Yang-Mills theory in the large $N$
limit. The fuzzy sphere phase occurs for 
$\tilde{\alpha}>\tilde{\alpha}_{*}$ while the matrix phase occurs
for $\tilde{\alpha}<\tilde{\alpha}_{*}$.

\subsection{Predictions from the effective potential}

Since $\phi=0$ is not a solution of equation (\ref{equ-for-phi}) 
the extremal equation  can be rearranged and when expressed in terms
of $\bar\phi$ takes the form
\begin{eqnarray}
\label{Pofphi}
P(\bar\phi)\equiv\bar{\phi}^4-\bar{\phi}^3-t\bar{\phi}^2+\frac{2}{a^4}=0.
\end{eqnarray}
where $a^4=\tilde{\alpha}^4/(1+m^2)^3$. By the substitution $\bar\phi=(1-4x)\tilde\phi +x$, $P(\bar\phi)$
can be brought to the form of a 
\begin{eqnarray}
\tilde{\phi}^4-\tilde{\phi}^3-\lambda \tilde{\phi}+\frac{2}{{\tilde{\beta}^4}}=0.
\end{eqnarray}
The new parameters $\lambda$ and $\tilde\beta$ are given by
\begin{eqnarray}
&&\lambda =\frac{x}{(1-4x)^3} \big[x+\frac{4t}{3}\big]~,\\
&&\frac{2}{\tilde{\beta}^4}=\frac{1}{(1-4x)^4}\bigg[\frac{2(1+m^2)^3}{\tilde{\alpha}^4}-\frac{1}{4}\bigg(x+\frac{m^2(1+m^2)}{3}\bigg)\bigg(x+\frac{5m^2(1+m^2)}{3}\bigg)\bigg]~,
\end{eqnarray}
and the shift $x$ must take one of the two values
\begin{eqnarray}
x\longrightarrow x_{\pm}=\frac{1}{4}(1\pm \sqrt{1+\frac{8t}{3}}).
\end{eqnarray}
We choose $x=x_{-}$ since $x_{-}(0)=0$ and so this case allows us
to easily recover the case with $\mu=m^2=0$ and for $m^2{\longrightarrow}0$ 
we get
\begin{eqnarray}
x_{-}=-\frac{m^2}{3}+O(m^4)~,~\lambda =O(m^4)~,~\frac{2}{\tilde{\beta}^4}=\frac{2}{\tilde{\alpha}^4}(1-\frac{7}{3}m^2+O(m^4))
\end{eqnarray}
The two positive solutions of (\ref{Pofphi}), $\bar\phi_{\pm}$, 
for general values of $m^2$ and $\mu$  are then given by
\begin{eqnarray}
&&(1+m^2){\phi}_{\pm}=\frac{1}{4}\bigg[1+(1-4x)\sqrt{1+{\delta} }{\pm}(1-4x)\sqrt{2-{\delta}+\frac{2}{\sqrt{1+{\delta}}}(1+8\lambda)}\bigg]\nonumber\\
&&{\delta}=4W^{\frac{1}{3}}\bigg[\bigg(1+\sqrt{1-V}\bigg)^{\frac{1}{3}}+\bigg(1-\sqrt{1-V}\bigg)^{\frac{1}{3}}\bigg]\nonumber\\
&&V=\frac{1}{W^2}\bigg(\frac{8}{3\tilde{\beta}^4}-\frac{\lambda}{3}\bigg)^3~,~W=\frac{1}{\tilde{\beta}^4}+\frac{{\lambda}^2}{2}.\label{form1}
\end{eqnarray}
We rewrite the above solution (\ref{form1}) as follows,   
(collecting definitions here for completeness)
\begin{eqnarray}
&&t=\mu(m^2+1)~,~a^4=\frac{\tilde{\alpha}^4}{(1+m^2)^3}\nonumber\\
&&q=1+\frac{8t}{3}-\frac{a^4t^3}{27}~,~p=q^2-\frac{(\frac{8}{3}+\frac{a^4t^2}{9})^3}{a^4}.
\end{eqnarray} 
Then we can show that
\begin{eqnarray}
&&(1-4x)^3(1+8\lambda)=1+4t~,~W=\frac{q}{(1+\frac{8t}{3})^3a^4}~,~V=1-\frac{p}{q^2}.
\end{eqnarray}
and
\begin{eqnarray}
\delta =\frac{4d}{1+\frac{8t}{3}}~,~d=a^{-\frac{4}{3}}\bigg(\big(q+\sqrt{p}\big)^{\frac{1}{3}}+\big(q-\sqrt{p}\big)^{\frac{1}{3}}\bigg).
\end{eqnarray}
Substituting one finds the relatively simple form
\begin{eqnarray}\label{phisol}
\fbox{$\bar{\phi}_{\pm}=(1+m^2){\phi}_{\pm}=\frac{1}{4}+\frac{1}{2}\sqrt{\frac{1}{4}+\frac{2t}{3}+d}\pm
  \frac{1}{2}\sqrt{\frac{1}{2}+\frac{4t}{3}-d+\frac{1+4t}{4\sqrt{\frac{1}{4}+\frac{2t}{3}+d}}}.$}
\end{eqnarray}
For completeness the remaining two solutions of the quartic $P(\bar\phi)=0$ 
are given by 
\begin{eqnarray}
\fbox{$\bar{\phi}^n_{\pm}=(1+m^2){\phi}^n_{\pm}=\frac{1}{4}-\frac{1}{2}\sqrt{\frac{1}{4}+\frac{2t}{3}+d}\pm
  \frac{1}{2}\sqrt{\frac{1}{2}+\frac{4t}{3}-d-\frac{1+4t}{4\sqrt{\frac{1}{4}+\frac{2t}{3}+d}}}.$}
\end{eqnarray}
At the critical point $a=a_c$, $p$ becomes zero and 
the two solutions $\bar{\phi}_{+}$ and $\bar{\phi}_-$ are equal.  

\begin{figure}
\begin{center}
\includegraphics[width=5cm,angle=-90]{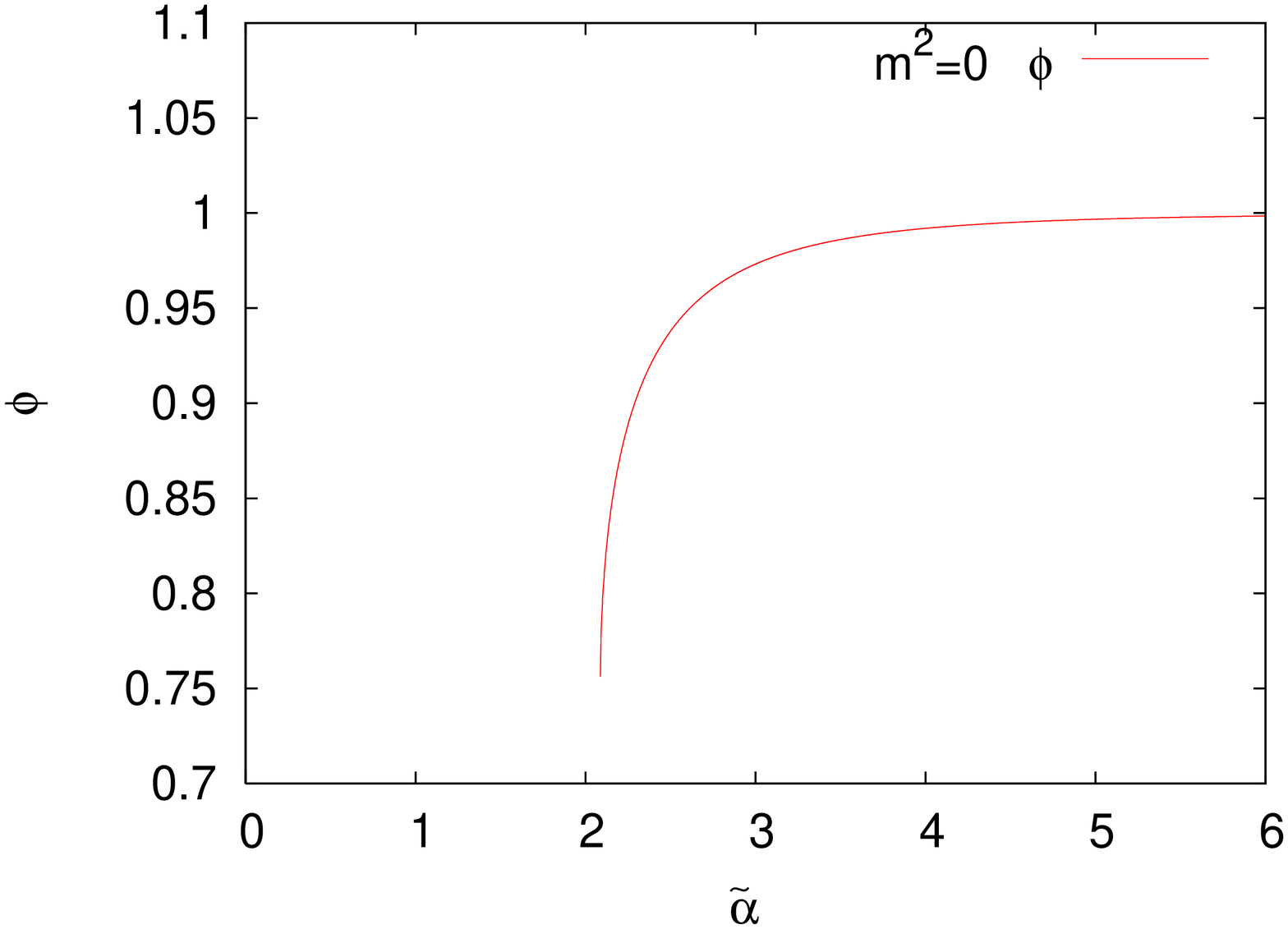}
\includegraphics[width=5cm,angle=-90]{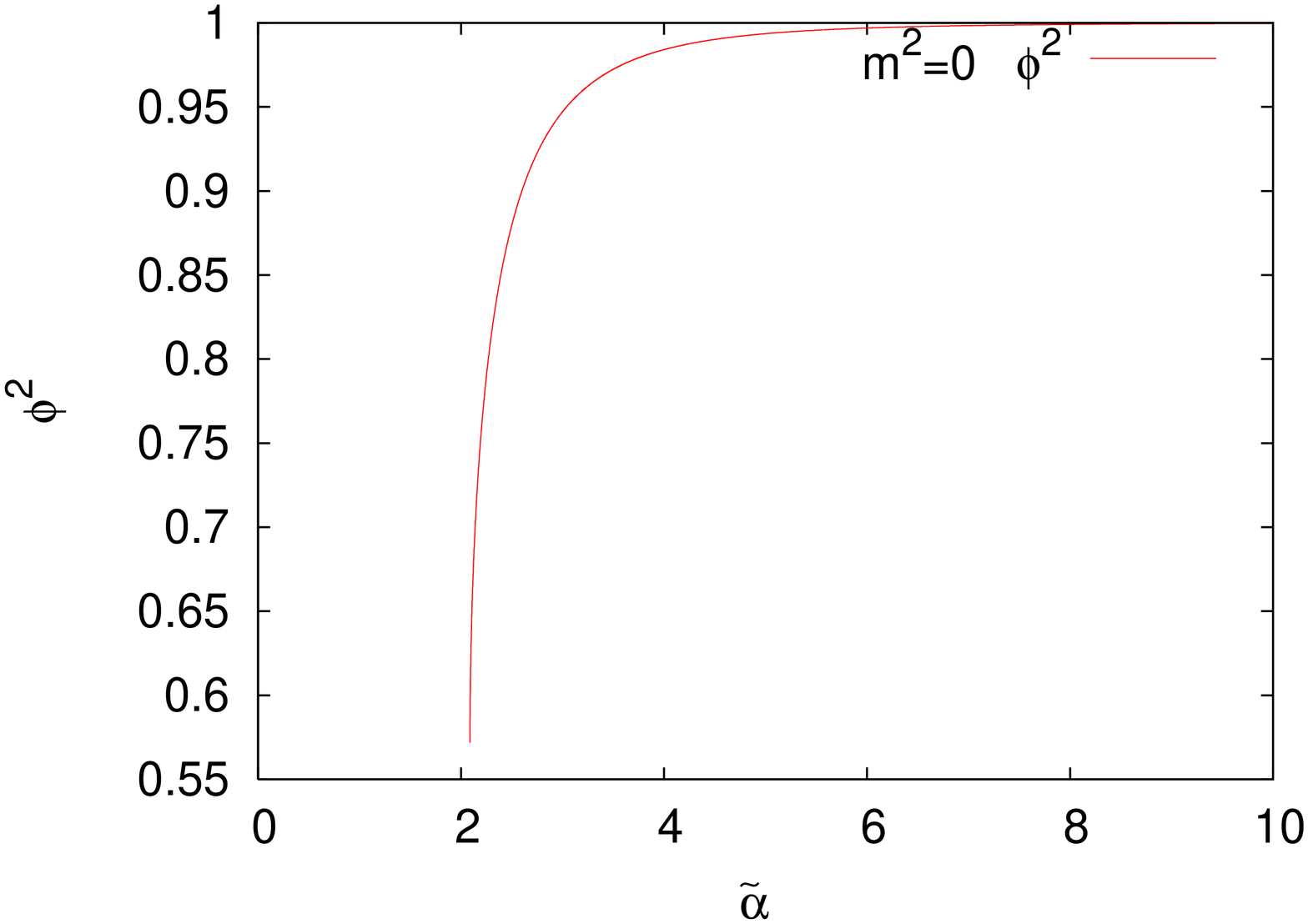} 
\includegraphics[width=5cm,angle=-90]{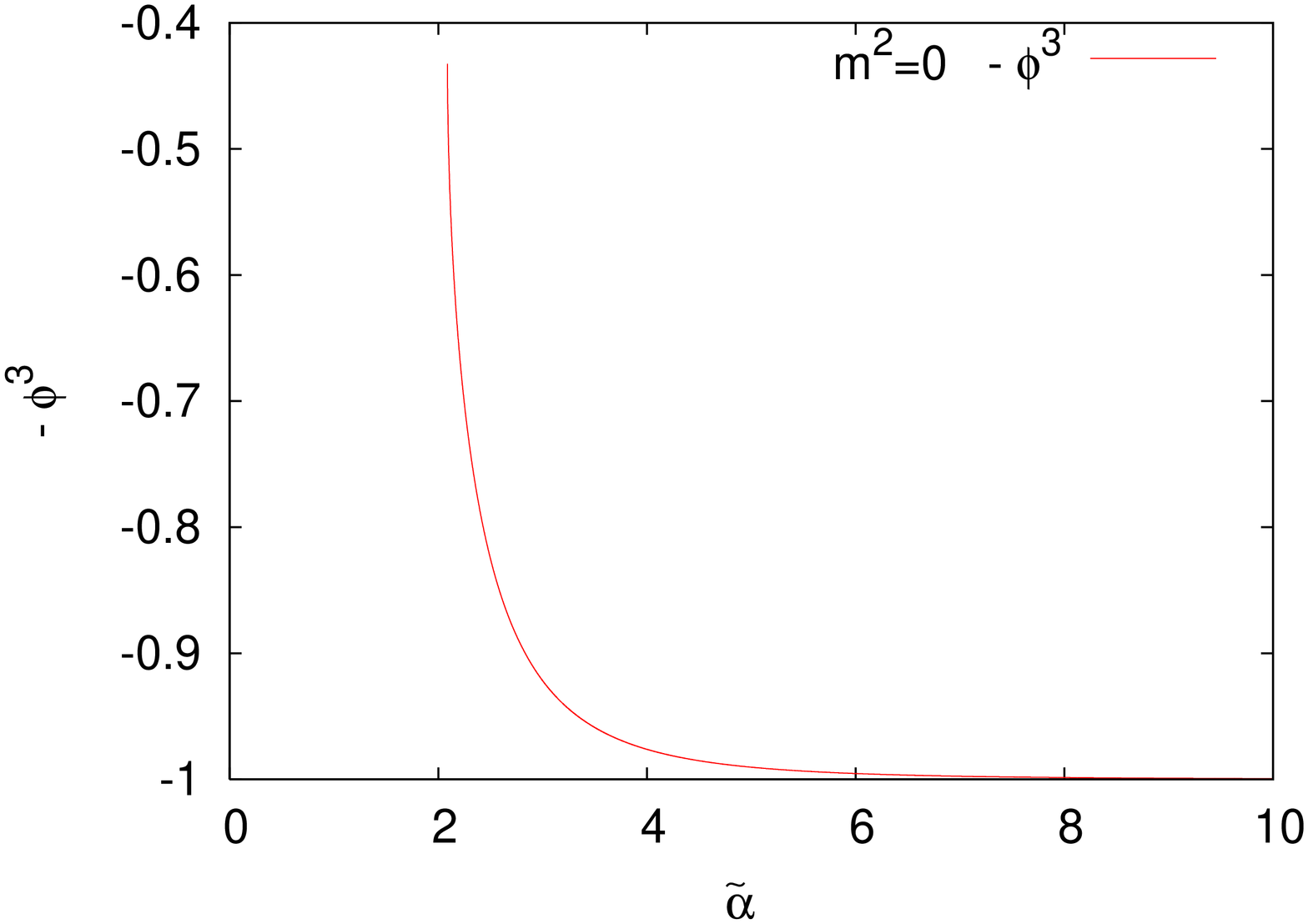}
 \includegraphics[width=5cm,angle=-90]{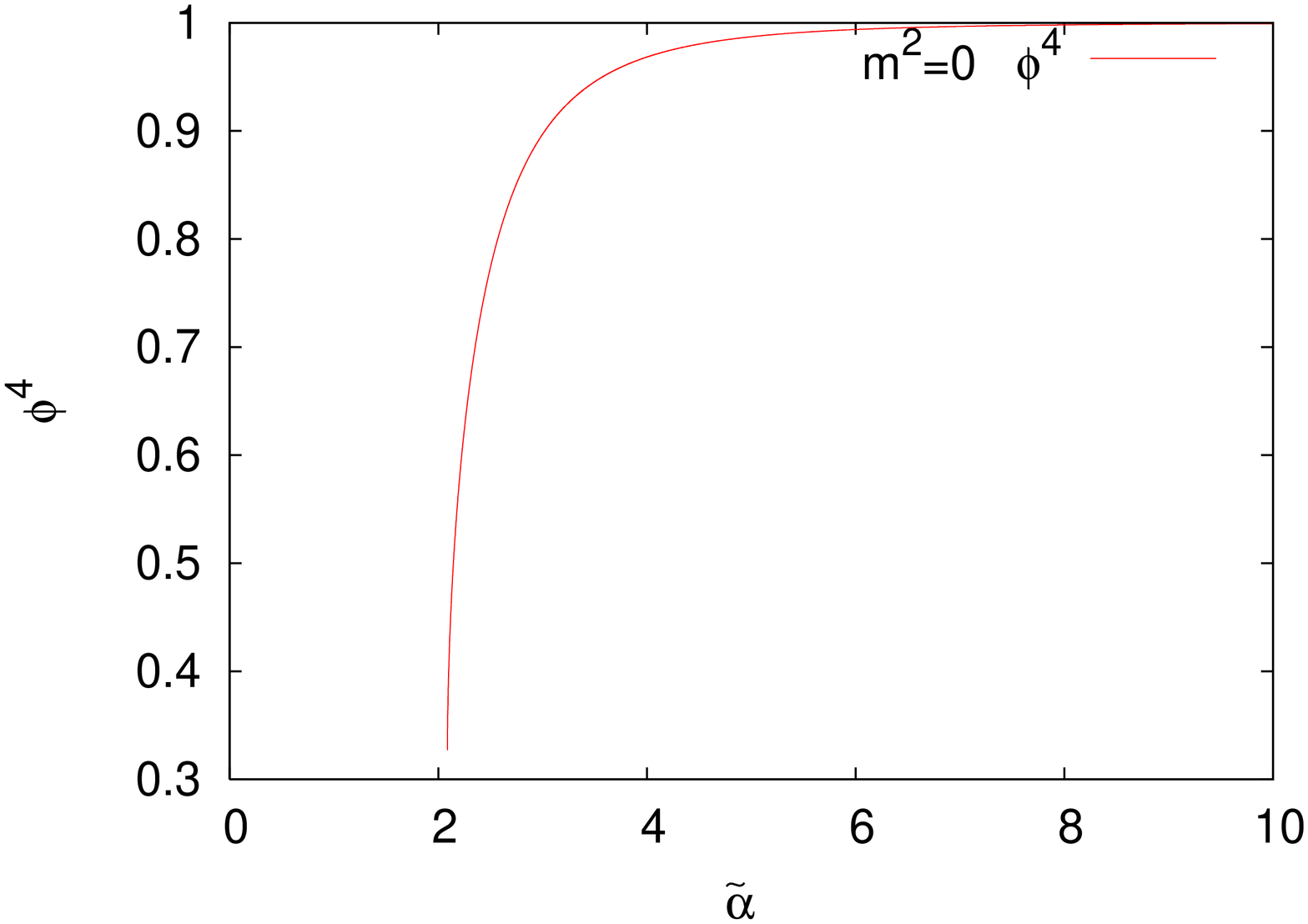} 
\caption{{The functions $\phi$, ${\phi}^2$, $-{\phi}^3$ and ${\phi}^4$
    for $m^2=0$. $\phi$ corresponds with $\phi_{+}$; the minimum of
    the effective potential (eq.\ref{veff}). }}\label{figphim0}
\end{center}
\end{figure}

\begin{figure}
\begin{center}
\includegraphics[width=5cm,angle=-90]{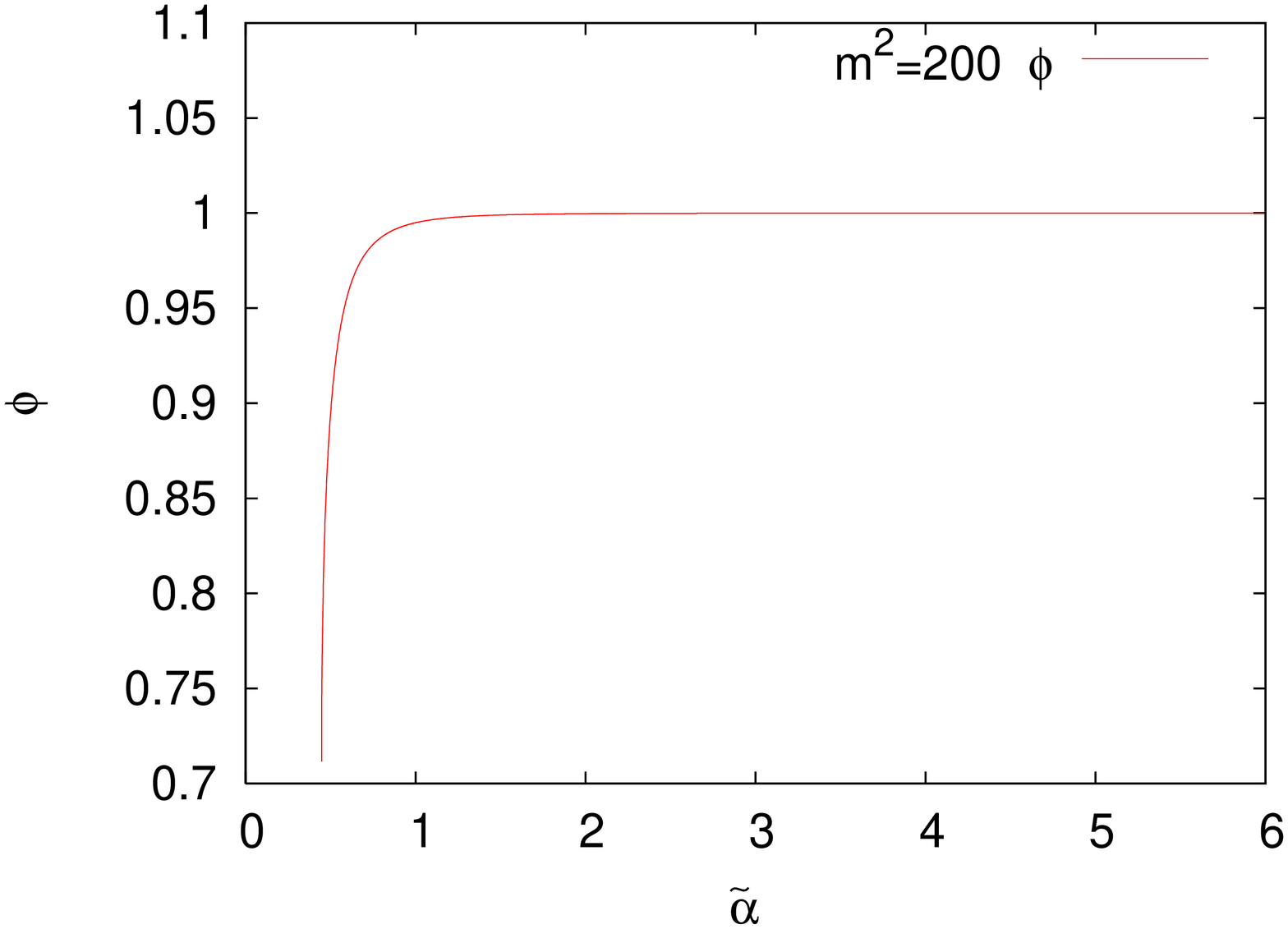}
\includegraphics[width=5cm,angle=-90]{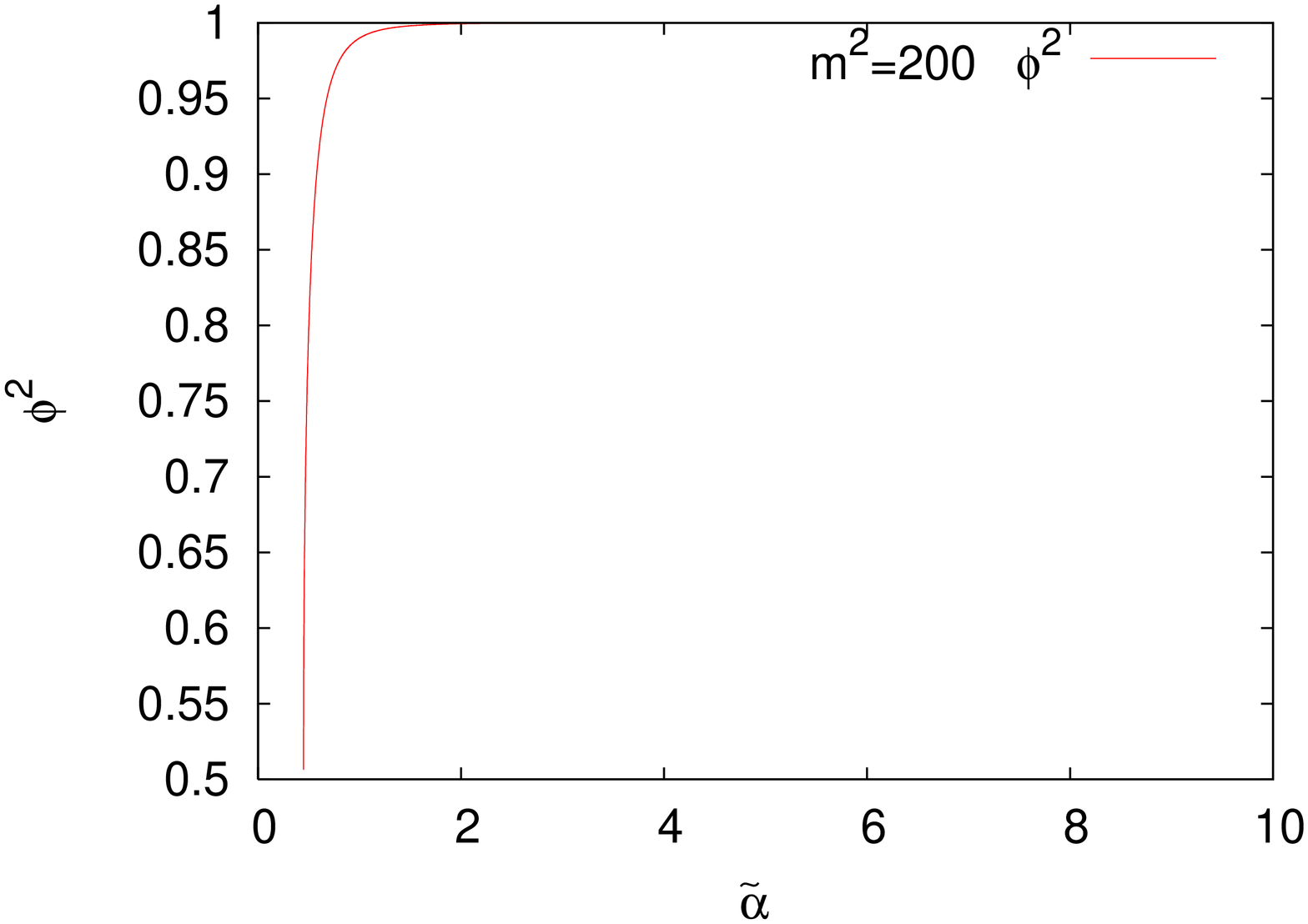} 
\includegraphics[width=5cm,angle=-90]{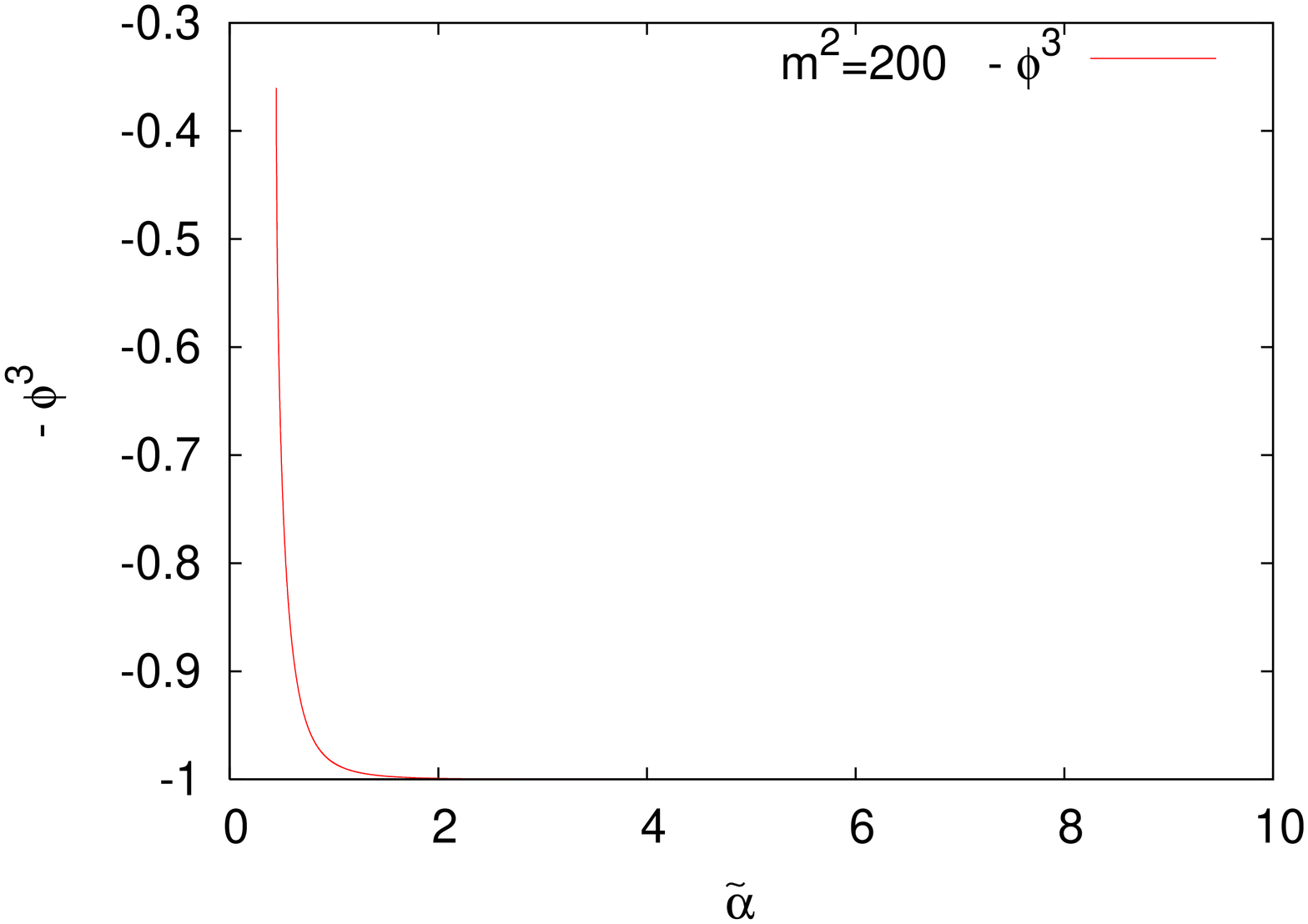}
 \includegraphics[width=5cm,angle=-90]{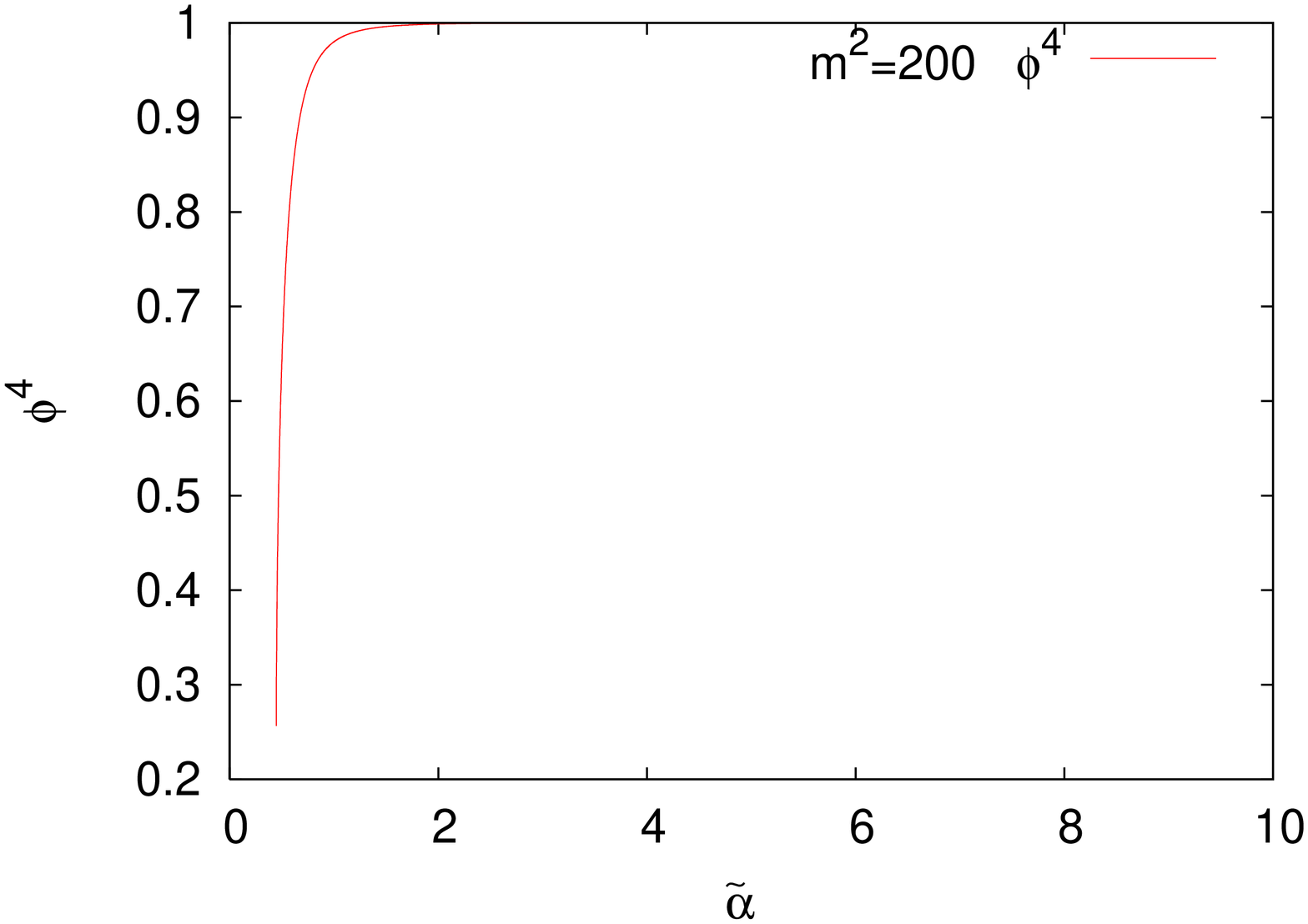} 
\caption{{The functions $\phi$, ${\phi}^2$, $-{\phi}^3$ and ${\phi}^4$ for $m^2=200$.}}\label{figphim200}
\end{center}
\end{figure}

In the fuzzy sphere phase the ground state of the system is given by
$\phi_+$ and the barrier  maximum is at $\phi_-$.  The minimum
${\phi}_{+}$ together with the powers ${\phi}_{+}^2$, ${\phi}_{+}^3$
and ${\phi}_+^4$ are plotted in figures \ref{figphim0} and
\ref{figphim200} for $m^2=0~{\rm and}~200$. 

In the specific heat we will need the derivative of the minimum with
respect to $\tilde{\alpha}$. This is given by
\begin{eqnarray}
\frac{d\bar{\phi}_{-}}{d\tilde{\alpha}}&=&\frac{1}{4}\frac{dd}{d\tilde{\alpha}}\bigg[-\frac{1+4t}{8}\frac{1}{\sqrt{\frac{1}{2}+\frac{4t}{3}-d+\frac{1+4t}{4\sqrt{\frac{1}{4}+\frac{2t}{3}+d}}}}\frac{1}{(\frac{1}{4}+\frac{2t}{3}+d)^{\frac{3}{2}}}\nonumber\\
&-&\frac{1}{\sqrt{\frac{1}{2}+\frac{4t}{3}-d+\frac{1+4t}{4\sqrt{\frac{1}{4}+\frac{2t}{3}+d}}}}+\frac{1}{\sqrt{\frac{1}{4}+\frac{2t}{3}+d}}\bigg].
\end{eqnarray}
\begin{eqnarray}
\frac{dd}{d\tilde{\alpha}}&=&-\frac{4(1+m^2)^3a^{\frac{8}{3}}}{3\tilde{\alpha}^5}\bigg[1+\frac{8t}{3}+\frac{(1+\frac{8t}{3})q}{\sqrt{p}}-\frac{4(\frac{8}{3}+\frac{t^2a^4}{9})^2}{a^4\sqrt{p}}\bigg]\frac{1}{(q+\sqrt{p})^{\frac{2}{3}}}\nonumber\\
&-&\frac{4(1+m^2)^3a^{\frac{8}{3}}}{3\tilde{\alpha}^5}\bigg[1+\frac{8t}{3}-\frac{(1+\frac{8t}{3})q}{\sqrt{p}}+\frac{4(\frac{8}{3}+\frac{t^2a^4}{9})^2}{a^4\sqrt{p}}\bigg]\frac{1}{(q-\sqrt{p})^{\frac{2}{3}}}.
\end{eqnarray}

\subsection{Critical behaviour }
For $\mu=m^2=0$ we have $x=x_{-}=0$ and  $\lambda =0$ and the solution
\begin{eqnarray}
\phi_+=\tilde{\phi}_+=&&\bar{\phi}_+ =\frac{1}{4}\bigg[1+\sqrt{1+{\delta} }+\sqrt{2-{\delta}+\frac{2}{\sqrt{1+{\delta}}}}\bigg]\nonumber\\
&&{\delta}=\frac{4}{\tilde{\alpha}^{\frac{4}{3}}}\bigg[\bigg(1+\sqrt{1-\frac{\tilde{\alpha}_*^4}{\tilde{\alpha}^4}}\bigg)^{\frac{1}{3}}+\bigg(1-\sqrt{1-\frac{\tilde{\alpha}_*^4}{\tilde{\alpha}^4}}\bigg)^{\frac{1}{3}}\bigg]\nonumber\\
&&\tilde{\alpha}_*^4=(\frac{8}{3})^{3}.
\end{eqnarray}
Expanding near the critical point we have 
$\delta=3 -\frac{16}{3}\epsilon +O({\epsilon}^2)$ and 
we obtain thus the expression
\begin{eqnarray}
\fbox{$\phi =\frac{1}{4}\bigg[3+\sqrt{6\epsilon}-\frac{4\epsilon}{3}+O({\epsilon}^{\frac{3}{2}})\bigg]~,~\epsilon=\frac{\tilde{\alpha}-\tilde{\alpha}^{*}}{\tilde{\alpha}^{*}}.$}\label{sq1}
\end{eqnarray}
Substituting into (\ref{average_action}) near the critical point we obtain
the expression for the scaled average action
\begin{eqnarray}
\fbox{${\cal S}=\frac{5}{12}-\frac{1}{3^{\frac{1}{8}}2^{\frac{5}{8}}}\sqrt{\tilde{\alpha}-\tilde{\alpha}_*}-\frac{7}{3^{\frac{5}{4}}2^{\frac{5}{4}}}(\tilde{\alpha}-\tilde{\alpha}_*)+O((\tilde{\alpha}-\tilde{\alpha}_*)^{\frac{3}{2}})~,$}
\end{eqnarray}
and the specific heat is then given by
\begin{eqnarray}
\fbox{$C_v=\frac{29}{36}+\frac{1}{2^{\frac{11}{8}}3^{\frac{7}{8}}}\frac{1}{\sqrt{\tilde{\alpha}-\tilde{\alpha}_*}} +O((\tilde{\alpha}-\tilde{\alpha}_*)^{\frac{1}{2}}).$}\label{sq2}
\end{eqnarray}
This gives a divergent specific heat with critical exponent
\begin{eqnarray}
\fbox{$\alpha=\frac{1}{2}~.$}
\end{eqnarray}

If instead we consider $\mu=0$ but $m\neq0$ so that $t=0$ then the above
expressions remain essentially the same provided we substitute $a$ for $\tilde\alpha$.
The critical behaviour remains the same for this case with the critical
inverse temperature 
\begin{equation}
\beta_c=\tilde{\alpha}^4_*=(\frac{8}{3})^{3}(1+m^2)^3~,
\end{equation}
so that increasing $m^2$ with $\mu=0$ sends the critical temperature lower
and so the region of stability of the fuzzy sphere solution is reduced.

\subsection{The generic case}

At the critical point the coupling $a$ takes the value $a_*$ 
and the order parameter $\phi$ takes the value
${\phi}_*$. At this critical point the two solutions 
$\phi_{+}$ and $\phi_{-}$ merge.  This is more easily determined by 
requiring the additional equation 
\begin{eqnarray}
&&\frac{2(1+m^2)^3}{\tilde{\alpha}^4}{\phi}^2 \frac{d^2V_{\rm eff}}{d{\phi}^2}=Q(\bar{\phi})\equiv3\bar{\phi}^4-2\bar{\phi}^3-t\bar{\phi}^2-\frac{2}{{a}^4}=0.
\end{eqnarray}
Putting the two equations $P(\bar{\phi})=0$ and $Q(\bar{\phi})=0$
together we obtain
\begin{eqnarray}
P(\bar{\phi})+Q(\bar{\phi})=4\bar{\phi}^4-3\bar{\phi}^3-2t\bar{\phi}^2=\bar{\phi}\frac{dP(\bar{\phi})}{d\bar{\phi}}.
\end{eqnarray}
In other words $\frac{dP(\bar{\phi})}{d\bar{\phi}}=0$ at the critical
point. Expanding around the critical 
\begin{eqnarray}
\bar{\phi} =\bar{\phi}_*+\sigma.
\end{eqnarray}
and using
$P(\bar{\phi}_*)=P^{'}(\bar{\phi}_*)=0$ we obtain
\begin{eqnarray}
P(\bar{\phi})={\sigma}^4+(4\bar{\phi}_*-1){\sigma}^3+\frac{3\bar{\phi}_*+4t}{2}{\sigma}^2+\frac{2}{a^4}-\frac{2}{a_*^4}=0.
\end{eqnarray}
For small $\sigma$, treating ${\sigma}^4$ and ${\sigma}^3$ perturbatively
we get 
\begin{eqnarray}
{\sigma}&=&
\sqrt{\frac{2}{3\bar{\phi}_*+4t}\bigg(\frac{2}{a_*^4}-\frac{2}{a^4}\bigg)}
-\frac{(4\bar{\phi}_*-1)}{(3\bar{\phi}_*+4t)^2}\bigg(\frac{1}{a_*^4}-\frac{1}{a^4}\bigg)
+...
\nonumber\\
&=&
\frac{4}{a_*^{\frac{5}{2}}}\frac{1}{\sqrt{3\bar{\phi}_*+4t}}\sqrt{a-a_*}
-\frac{(16\bar{\phi}_*-1)}{a_*^5(3\bar{\phi}_*+4t)^2}(a-a_*)
+...
\end{eqnarray}
Hence
\begin{eqnarray}
\fbox{$\bar\phi=\bar{\phi}_*+
\frac{4}{a_*^{\frac{5}{2}}}\frac{1}{\sqrt{3\bar{\phi}_*+4t}}\sqrt{a-a_*}+...$}
\end{eqnarray}
or equivalently to leading order we have
\begin{eqnarray}
\fbox{$\phi={\phi}_*+
\frac{4}{{\tilde\alpha}_*^{\frac{5}{2}}}\frac{1}{\sqrt{3{\phi}_*+4\mu}}\sqrt{\tilde{\alpha}-\tilde{\alpha}_*}+...$}\label{sq3}
\end{eqnarray}

The average action ${\cal S}$ near the critical point can be computed using
this expression of $\phi$ in equation (\ref{u}). 
The result is that 
\begin{eqnarray}
{\cal S}={\cal S}_*
-a_c^4\frac{\bar{\phi}_*(\bar{\phi}_*+2t)}{\sqrt{3\bar{\phi}_*+4t}}
\sqrt{\frac{a-a_c}{a_c}}+...
\end{eqnarray}
Where 
\begin{eqnarray}
{\cal S}_*=\frac{3}{4}-\frac{(\bar{\phi}_*+3t)}{3(\bar{\phi}_*+2t)}
\end{eqnarray}
which interpolates between ${\cal S}_*=\frac{5}{12}$ for $t=0$ and 
${\cal S}_*=\frac{1}{4}$ for large $t$. 

In order to compute
the specific heat we need the derivative
\begin{eqnarray}
\frac{d\phi}{d\tilde{\alpha}}&=&
\frac{2}{\tilde{\alpha}_*^{\frac{5}{2}}}\frac{1}{\sqrt{3{\phi}_*+4\mu}}\frac{1}{\sqrt{\tilde{\alpha}-\tilde{\alpha}_*}}+...
\end{eqnarray}
The divergent term in the specific heat is still given by a square
root singularity. From (\ref{cv}) we get
\begin{eqnarray}
\fbox{$C_v=C_v^B+\frac{{\phi}_*({\phi}_*+2\mu)}{16\sqrt{3{\phi}_*+4\mu}}\frac{\tilde{\alpha}_*^{\frac{5}{2}}}{\sqrt{\tilde{\alpha}-\tilde{\alpha}_*}}+...$}\label{sq4}
\end{eqnarray}
where the background constant contribution to the specific heat $C_v^B$ 
is given by 
\begin{eqnarray}
C_v^B=\frac{3}{4}+\frac{(3+4t)\bar{\phi}_*+2t}{8(3\bar{\phi}+4t)^2}
\end{eqnarray}
If we set $\mu=0$ in (\ref{sq3}) and (\ref{sq4}) we recover
(\ref{sq1}) and (\ref{sq2}). 
Let us recall that the critical value
$\tilde{\alpha}_*$ can be given by the formula
\begin{eqnarray}
\tilde{\alpha}_*^4=\frac{8}{{\phi}_*^2({\phi}_*+2\mu)}.\label{sq5}
\end{eqnarray}
Then  (\ref{sq4}) can  be put in the form
\begin{eqnarray}
\fbox{$C_v=C_v^B+\frac{1}{8\sqrt{1+\frac{\tilde{\alpha}_*^4{\phi}_*^3}{16}}}\frac{\sqrt{\tilde{\alpha}_*}}{\sqrt{\tilde{\alpha}-\tilde{\alpha}_*}}+...$}
\end{eqnarray}
The prediction here is that the critical exponent of the 
specific heat for this model 
is given precisely by
\begin{eqnarray}
\alpha=\frac{1}{2}
\end{eqnarray}
Specializing to the case $\mu=m^2$ we see the coefficient of the 
singularity for any small $m^2$ (i.e the amplitude) is
\begin{eqnarray}
c(m^2)=\frac{\sqrt{\tilde{\alpha}_*}}{8\sqrt{1+\frac{\tilde{\alpha}_*^4{\phi}_*^3}{16}}}.
\end{eqnarray}
If we  extrapolate these results to large $m^2$ 
where we know that ${\phi}_*\longrightarrow
1/\sqrt{2}$ and $\tilde{\alpha}_*^4\longrightarrow 8/m^2$  we get 
\begin{eqnarray}
\fbox{$C_v=\frac{3}{4}+\frac{1}{32\sqrt{2}m^2}+\frac{1}{2^{\frac{21}{8}}m^\frac{1}{4}}\frac{1}{\sqrt{\tilde{\alpha}-\tilde{\alpha}_*}}+...$}
\end{eqnarray}
The coefficient of the singularity and the critical value $\tilde\alpha_*$
become very small and vanish when
$m^2\longrightarrow \infty$. For comparative purposes we can compute the ratio
\begin{eqnarray}
\frac{c(m^2)}{c(0)}=\bigg[\frac{1}{8}\big(\frac{3}{2}\big)^{7}\frac{1}{m^2}\bigg]^{\frac{1}{8}}.
\end{eqnarray}
For $m^2=200$ we get the ratio $c(200)/c(0)=0.57$ which is not yet very
small. As we will see below our data (see figure \ref{cvmlarge}) 
in the critical region for large $m$ is not precise enough to 
confirm or rule out the presence of a singularity.


\section{Numerical results}
Let us now turn to the numerical simulations. 
A fully nonperturbative study of this model is done in
\cite{O'Connor:2006wv,emergent-geometry-prl}. For the model with $m^2=0$ see also
\cite{Azuma:2004zq,Azuma:2004ie}. In Monte Carlo simulations we use the Metropolis algorithm
and the action (\ref{main11}) with $N$ in the range $N=8$ to $104$ and
$m^2$ in the range $m^2=0$ to $1000$.  The errors were estimated using the
binning-jackknife method.  We measure the radius of the sphere $R$
(the order parameter) defined by $Nc_2/R=<Tr D_a^2>$, the average
value of the action $<S>$ and the specific heat
$C_{v}=\frac{<(S-<S>)^2>}{N^2}$ as functions of $\tilde{\alpha}$ for
different values of $N$ and $m^2$. We also measure the eigenvalue
distributions of several operators.

\subsection{The theory with $m^2=0$}

\begin{figure}
\begin{center}
\includegraphics[width=5.8cm,angle=-90]{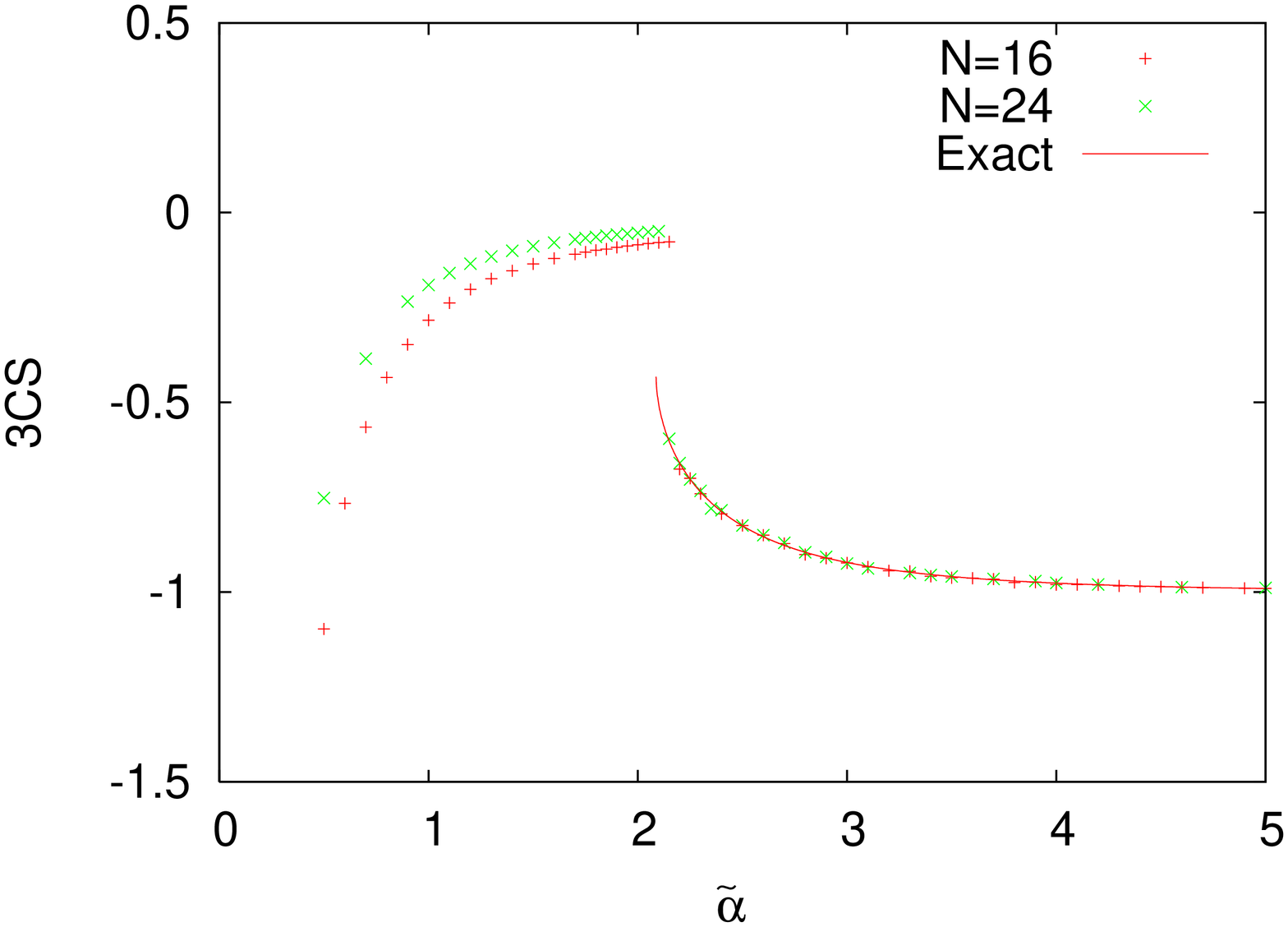}
\includegraphics[width=5.8cm,angle=-90]{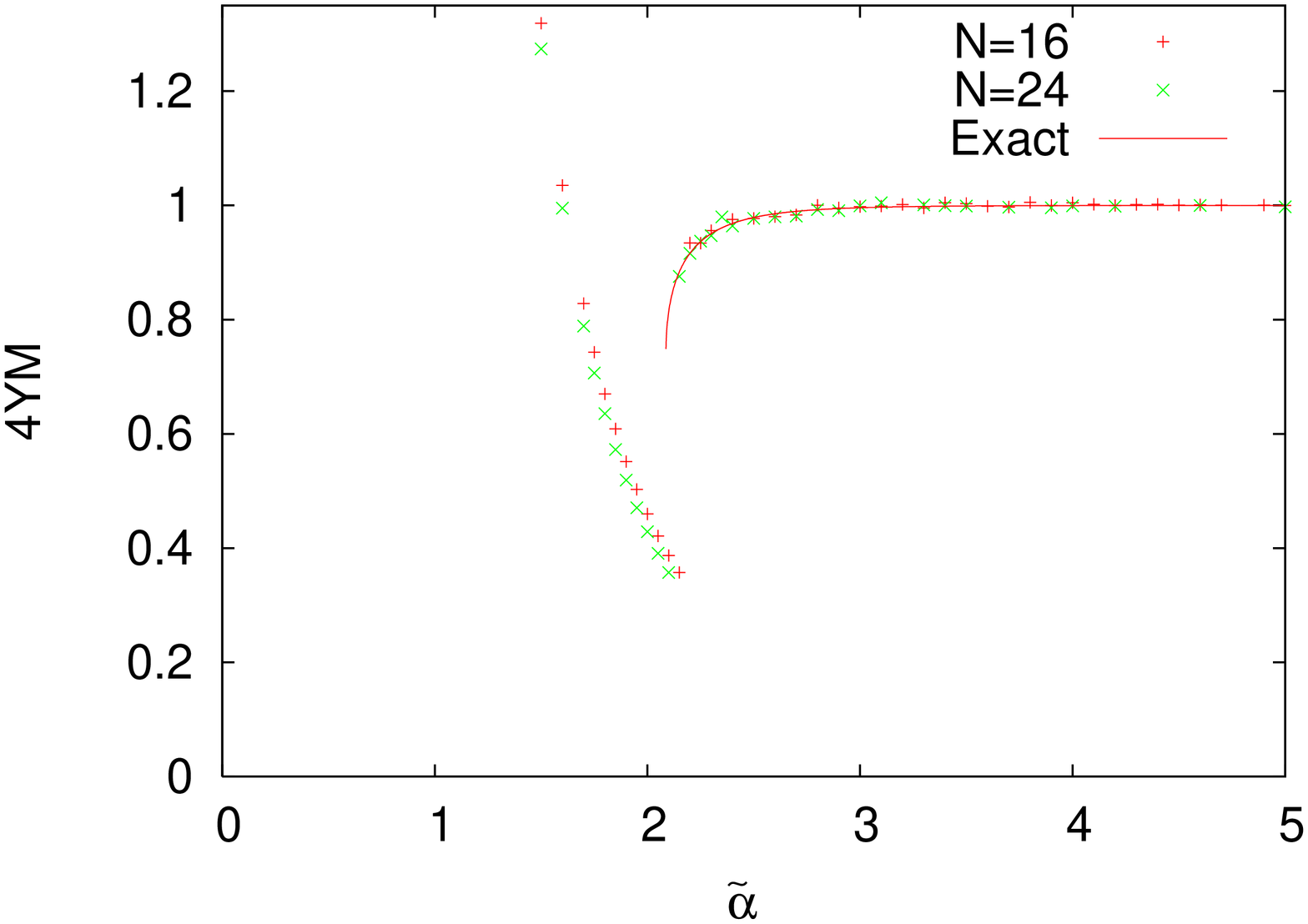}\\
\includegraphics[width=5.8cm,angle=-90]{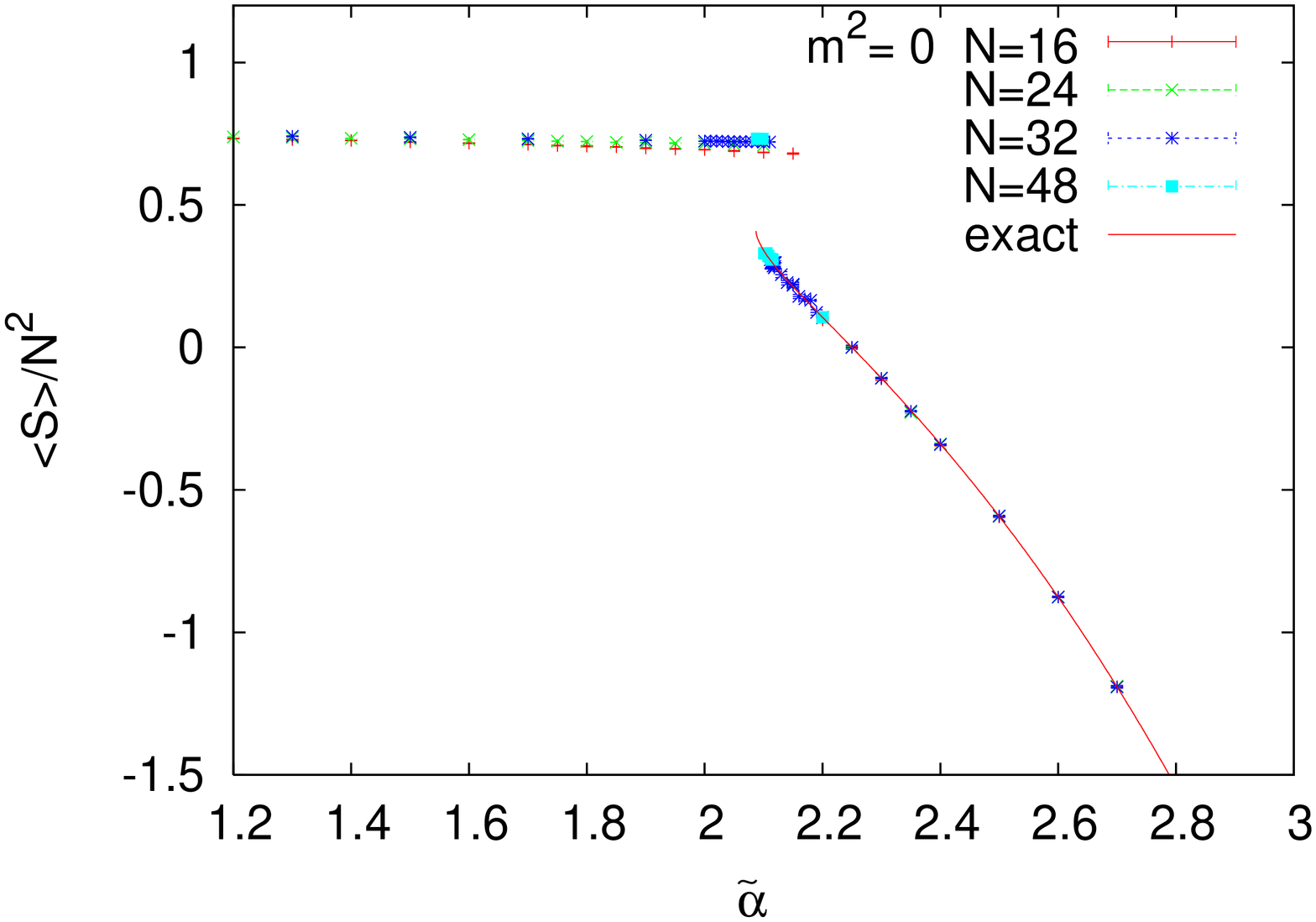}
\caption{The observables $CS$, $YM$ and $\frac{<S>}{N^2}$ for $m^2=0$ as a
  function of the coupling constant for different matrix sizes $N$. The
  solid line corresponds to the theoretical prediction using the local
  minimum (\ref{phisol}) of the effective potential.}\label{obsm0}
\end{center}
\end{figure}

\begin{figure}
\begin{center}
\includegraphics[width=5.8cm,angle=-90]{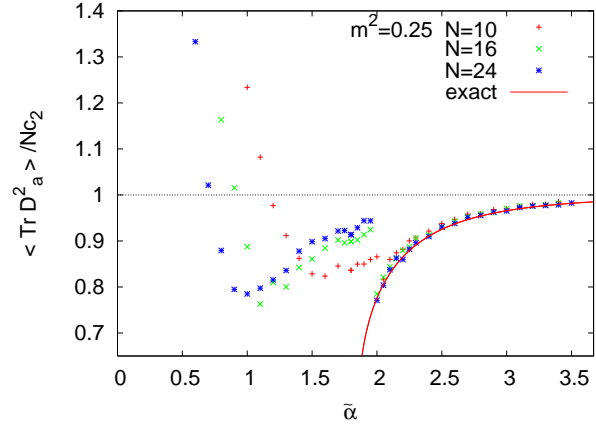}
\caption{The inverse radius for $m^2=0.25$ against $\tilde{\alpha}$ for $N=10,16,24$. The solid line corresponds to
  $\phi^2$, given by eq.(\ref{phisol}).}\label{radm0.75}
\end{center}
\end{figure}

\begin{figure}
\begin{center}
\includegraphics[width=7cm,angle=-90]{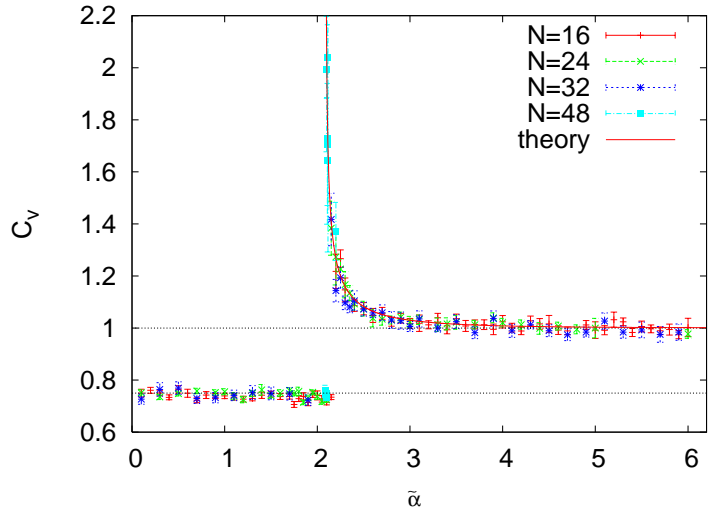}
\caption{The specific heat for $m^2=0$ as a function of the coupling
  constant for $N=16,24,32$,$48$. The curve corresponds with the
  theoretical prediction given by eq. (\ref{cv-exact}) for
  $m^2=0$.}\label{figcvm0}
\end{center}
\end{figure}

For $m^2=0$ we observe that the expectation values 
${\cal S}$, ${\rm YM}$, ${\rm CS}$ (defined above)
are all discontinuous at $\tilde{\alpha}_s=2.1\pm 0.1$ (figure
\ref{obsm0}). This is where the transition occurs. Indeed this agrees
with the theoretical value $\tilde{\alpha}_*=2.087$. The theory also
predicts the behaviour of these actions in the fuzzy sphere
phase. There clearly exists a latent heat and hence we are dealing
with a $1$st order transition which terminates at some value of
$m^2$. The radius is also discontinuous at the critical point (figure
\ref{radm0.75}) whereas the specific heat is discontinuous and
divergent (figure \ref{figcvm0}).  Near the critical point we compute
a divergent specific heat with critical exponent $\alpha=1/2$
(equation (\ref{sq2})). The fit for data fixing $\alpha=0.5$ gives the
critical value $\tilde{\alpha}_c=2.125\pm 0.007$ again with good
agreement with the theory. From the matrix side the specific heat
seems to be a constant equal to $0.75$ and therefore the critical
exponent is zero. The inverse radius in the matrix phase goes through
a minimum and then rise quickly and sharply to infinity.

\begin{center}
\begin{tabular}{|c|c|}
\hline
fuzzy sphere ($\tilde{\alpha}>\tilde{\alpha}_*$ )& matrix phase ($\tilde{\alpha}<\tilde{\alpha}_*$)\\
$R=1$ & $
R=0$\\
$C_v=1$  & $C_v=0.75$  \\
\hline
\end{tabular}
\end{center}


\subsection{The action, radius and specific heat for $m^2\neq 0$}

\begin{figure}
\begin{center}
\includegraphics[width=5.8cm,angle=-90]{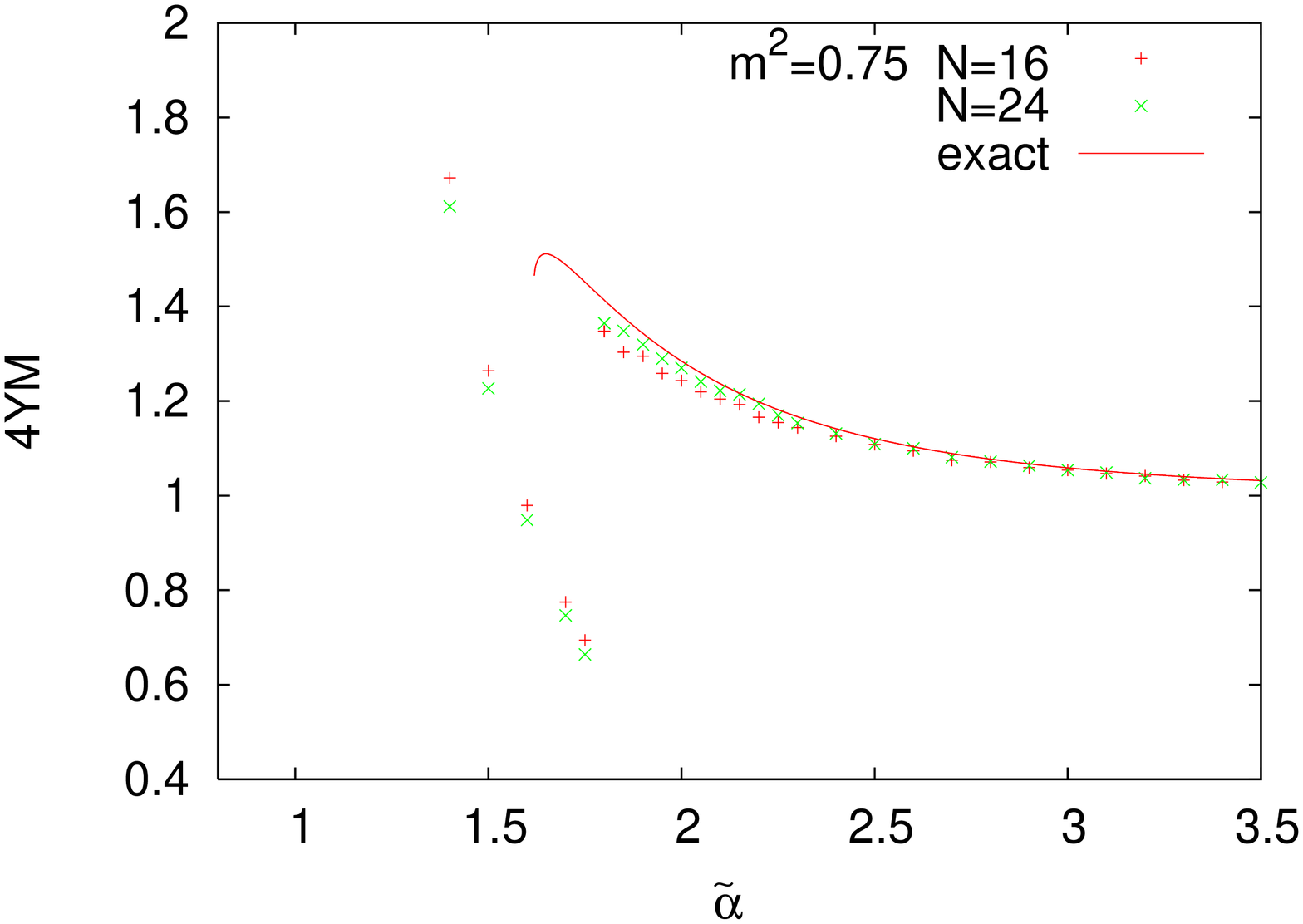}
\includegraphics[width=5.8cm,angle=-90]{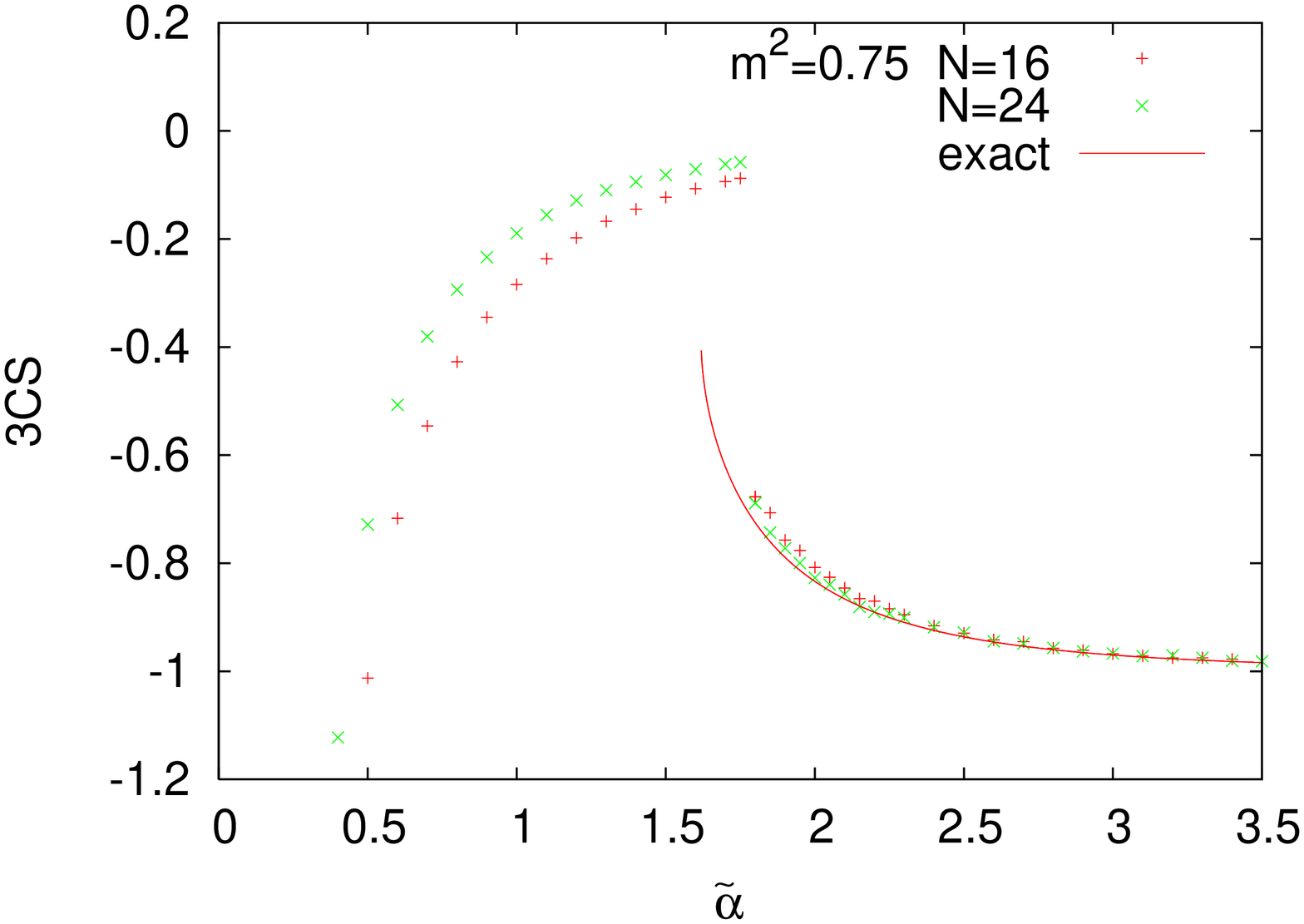}
\includegraphics[width=5.8cm,angle=-90]{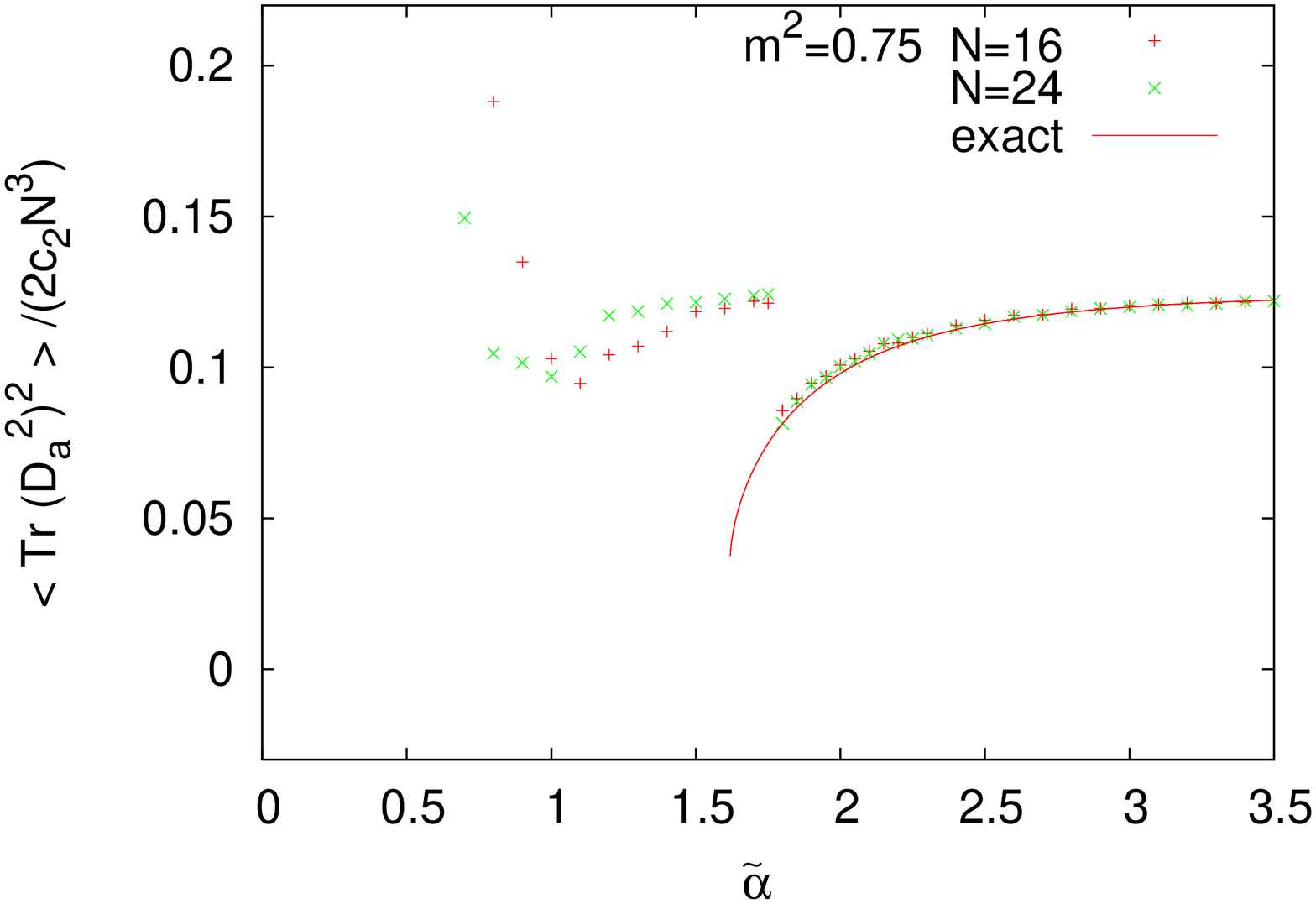}
\includegraphics[width=5.8cm,angle=-90]{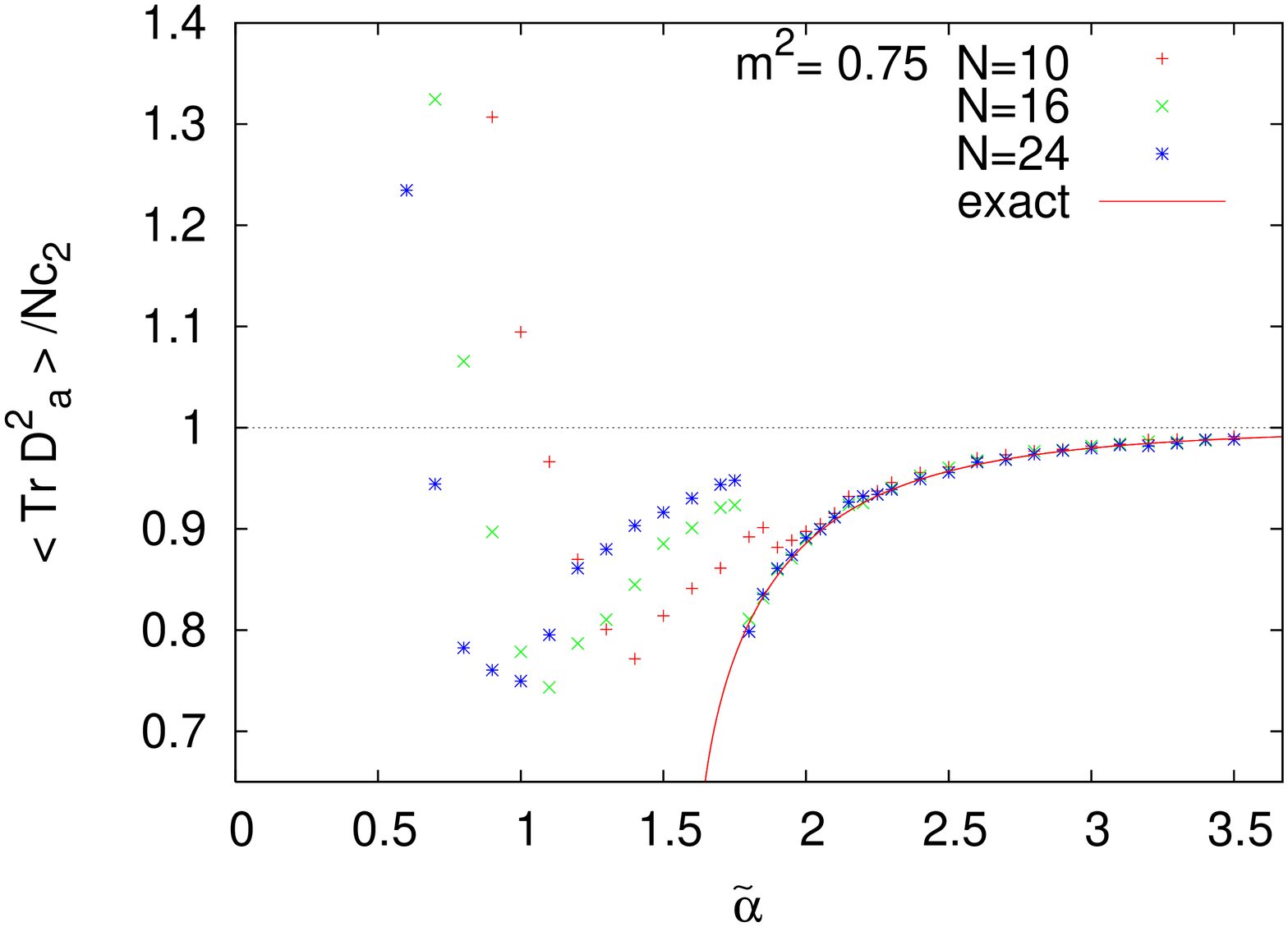}
\includegraphics[width=5.8cm,angle=-90]{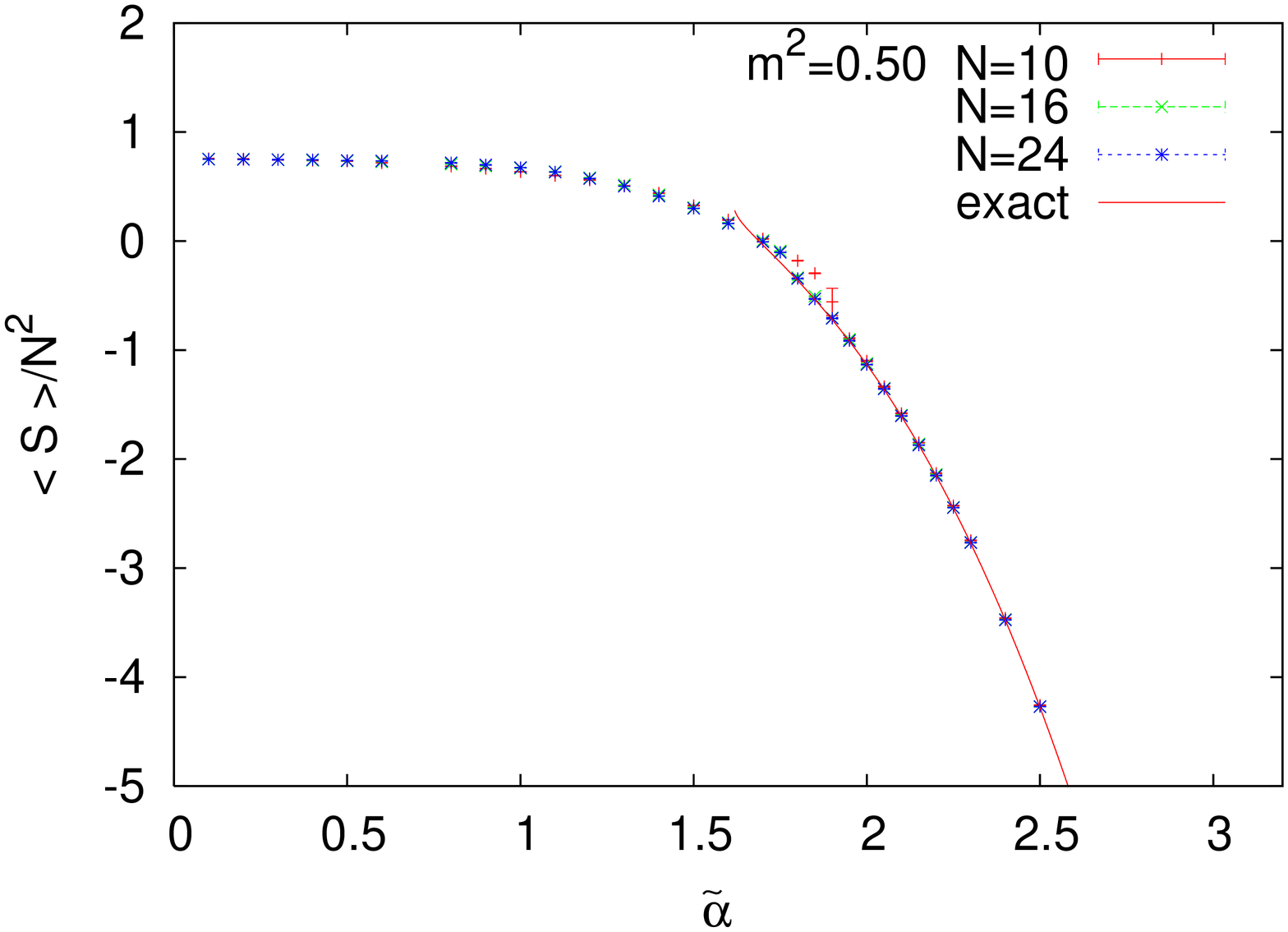}
\includegraphics[width=5.8cm,angle=-90]{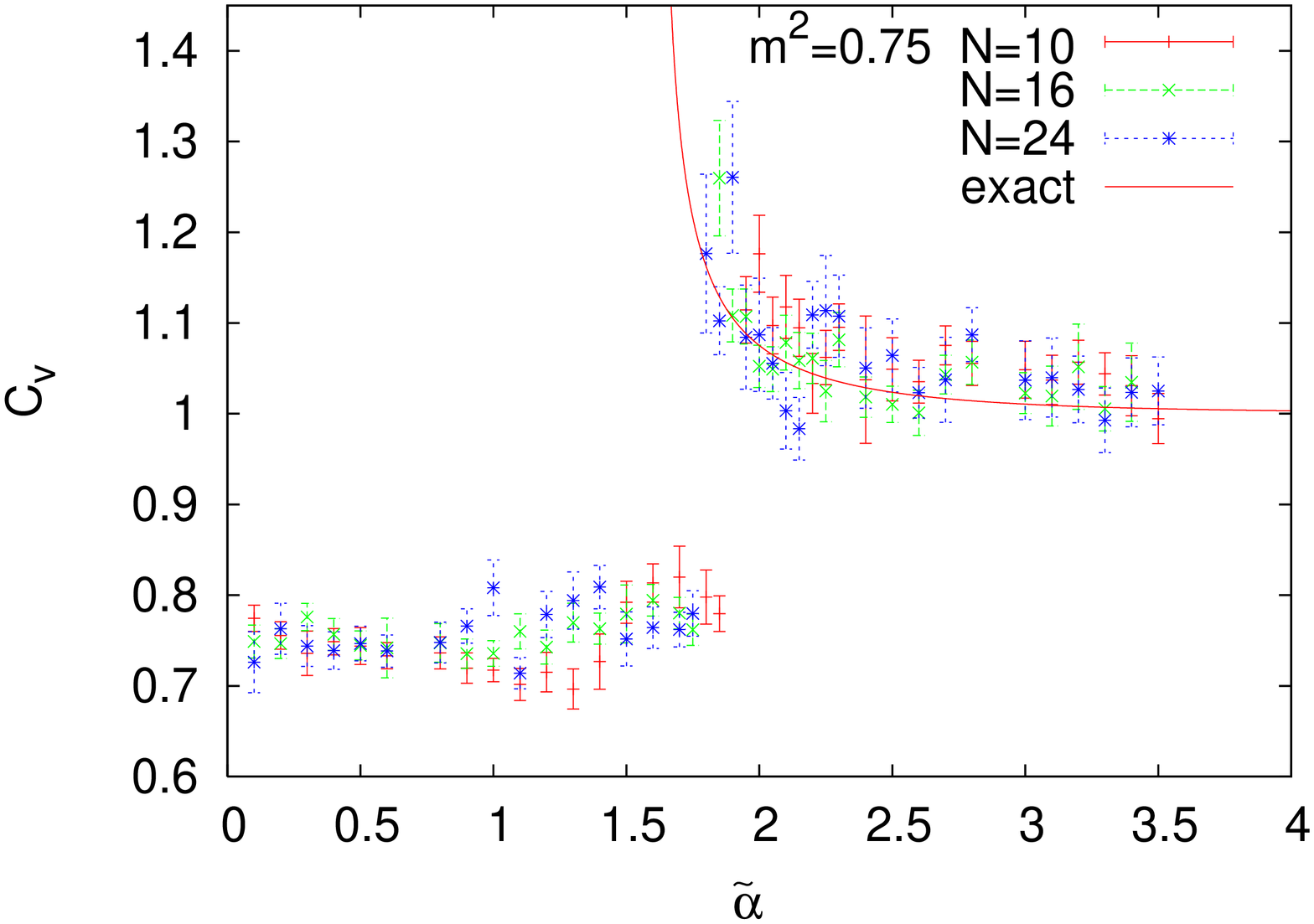}
\caption{Different observables for $m^2=0.75$ plotted against
  $\tilde{\alpha}$ for different matrix sizes. Again the solid lines represent the theoretical
  predictions using the local minimum of the effective potential.}\label{obsm0.75}
\end{center}
\end{figure}

For small values of $m^2$ we determine the critical value
$\tilde{\alpha}_s$ as the point of discontinuity in 
${\cal S},{\rm YM},{\rm CS}$, $Tr(D_a^2)^2$ and the radius $Tr D_a^2$. This
is where the divergence in $C_v$ occurs. For example for $m^2=0.75$
the action looks continuous but its parts are all discontinuous with a
jump. The radius is also discontinuous with a jump. The specific heat
is still divergent in this case (figure \ref{obsm0.75}). This is
still $1$st order.

\begin{figure}
\begin{center}
\includegraphics[width=5.8cm,angle=-90]{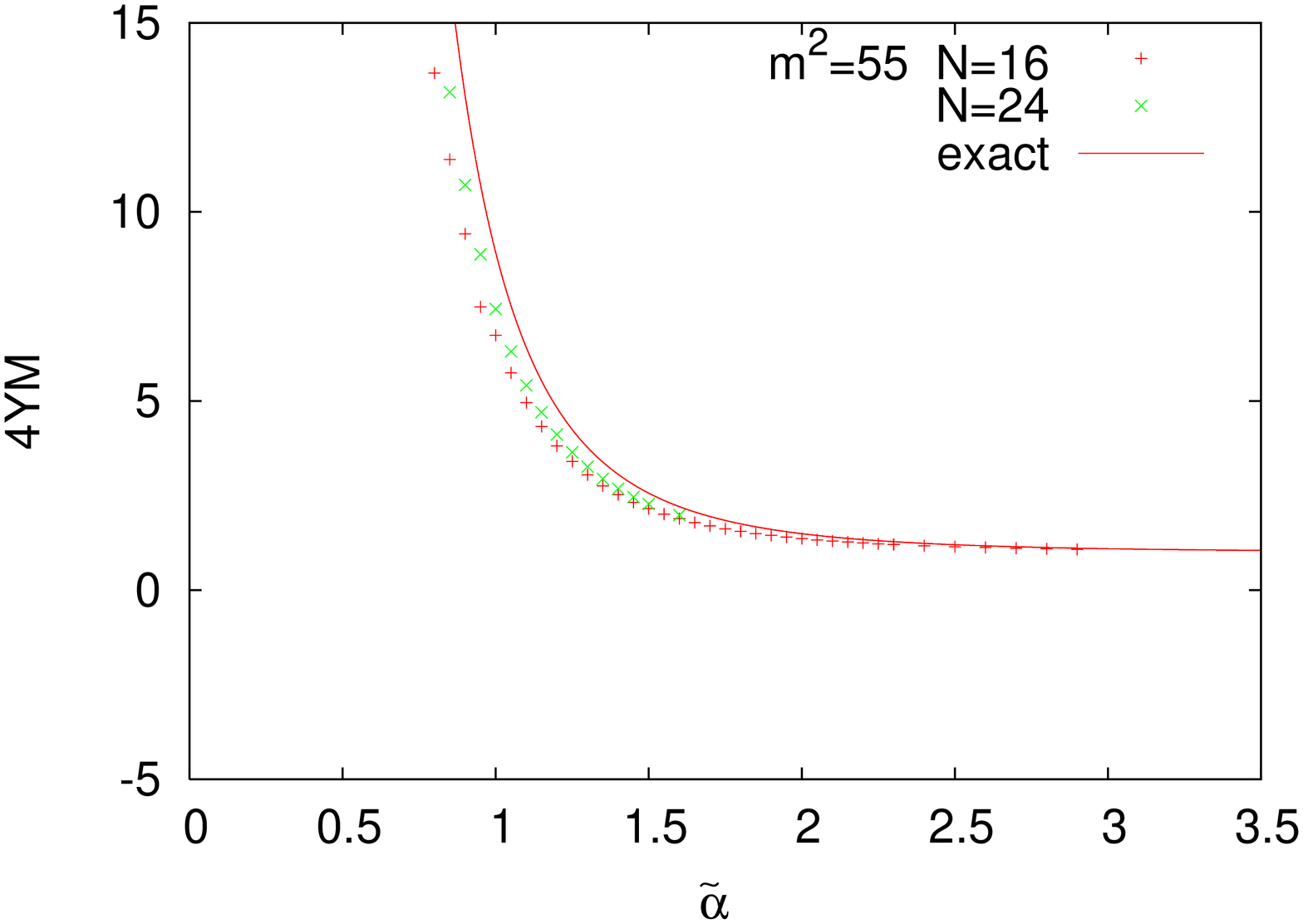}
\includegraphics[width=5.8cm,angle=-90]{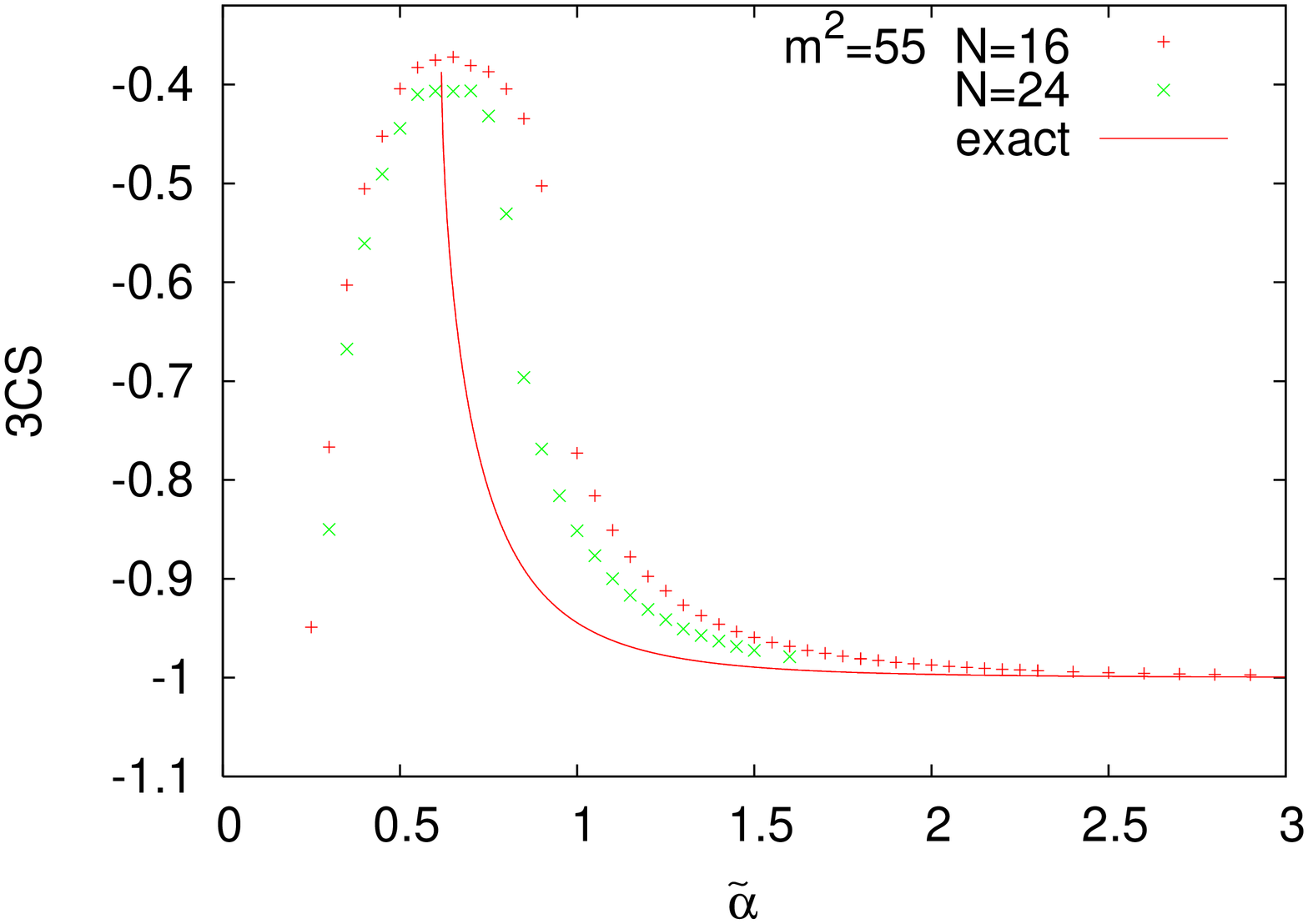}
\includegraphics[width=5.8cm,angle=-90]{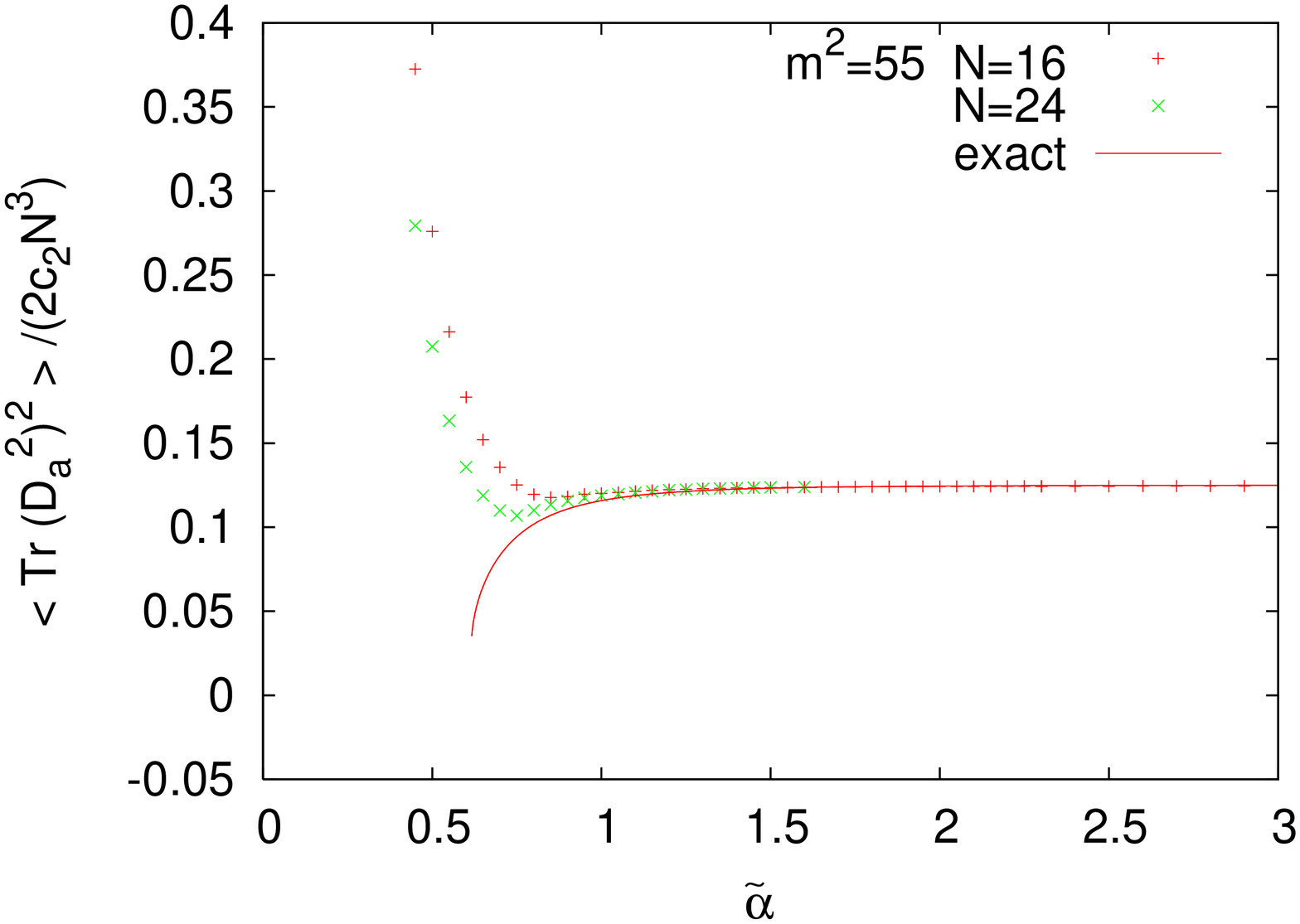}
\includegraphics[width=5.8cm,angle=-90]{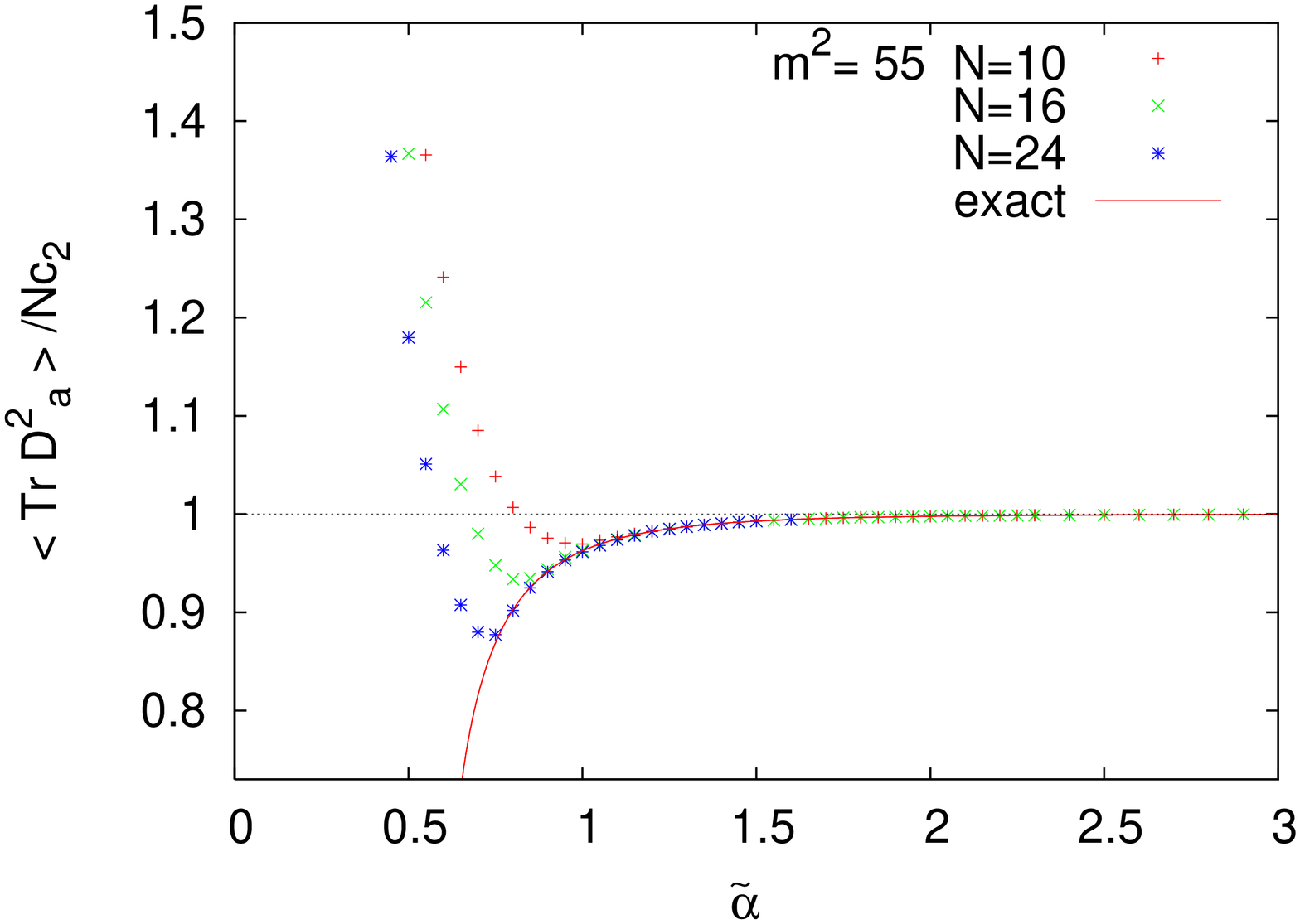}
\includegraphics[width=5.8cm,angle=-90]{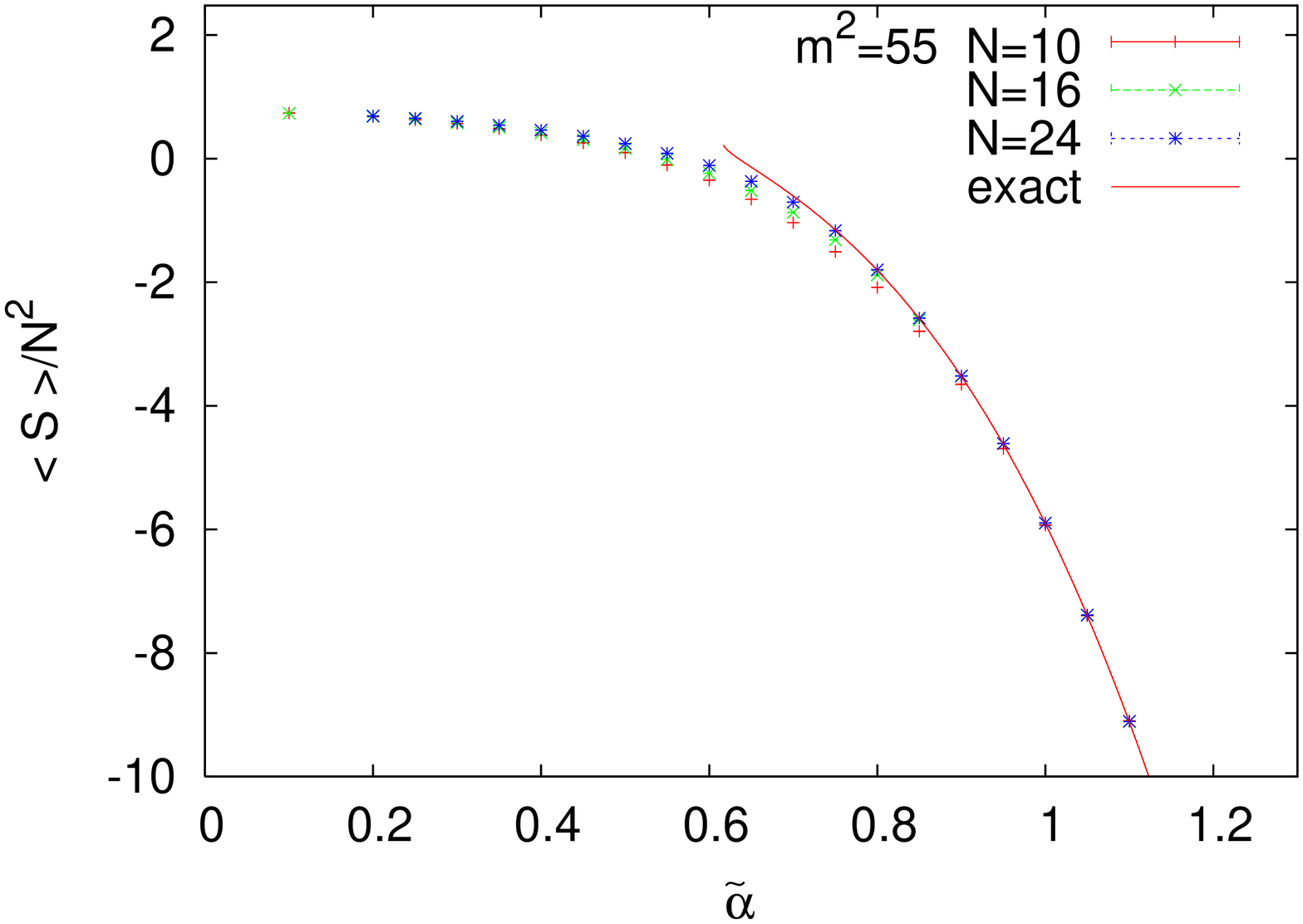}
\includegraphics[width=5.8cm,angle=-90]{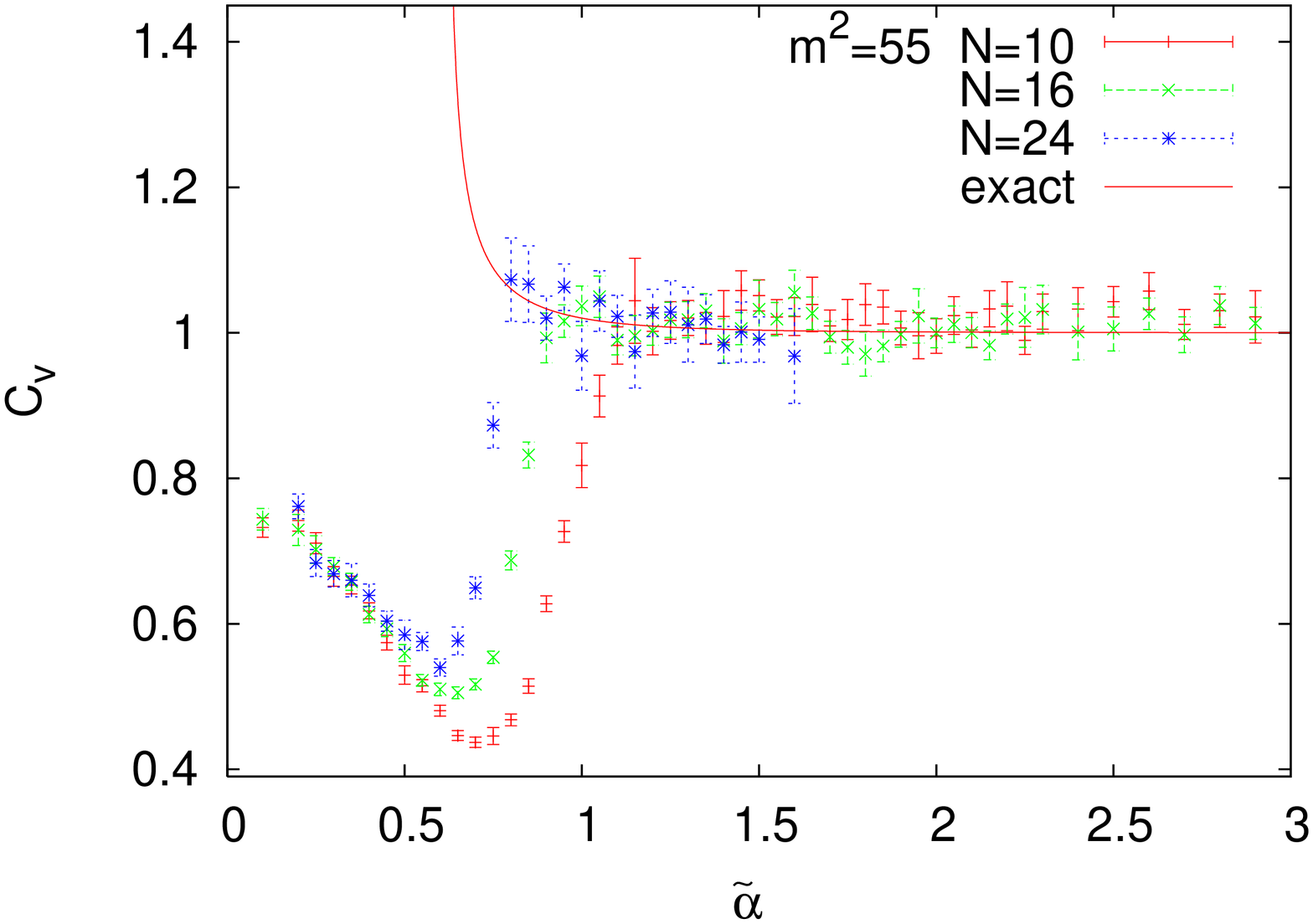}
\caption{Different observables for $m^2=55$ as functions of the
  coupling constant for different matrix sizes.}\label{obsm55}
\end{center}
\end{figure}

For $m^2=55$ the action and its parts become continuous. We find in
particular that the Chern-Simons and the radius are becoming continuous
around $m^2=40-50$.  These are the two operators which are associated
with the geometry. The specific heat seems now to be continuous (figure
\ref{obsm55}). This looks like a $3$rd order transition.

\subsection{The limit $m^2\longrightarrow \infty$ and specific heat}

\begin{figure}
\begin{center}
  \includegraphics[width=5cm,angle=-90]{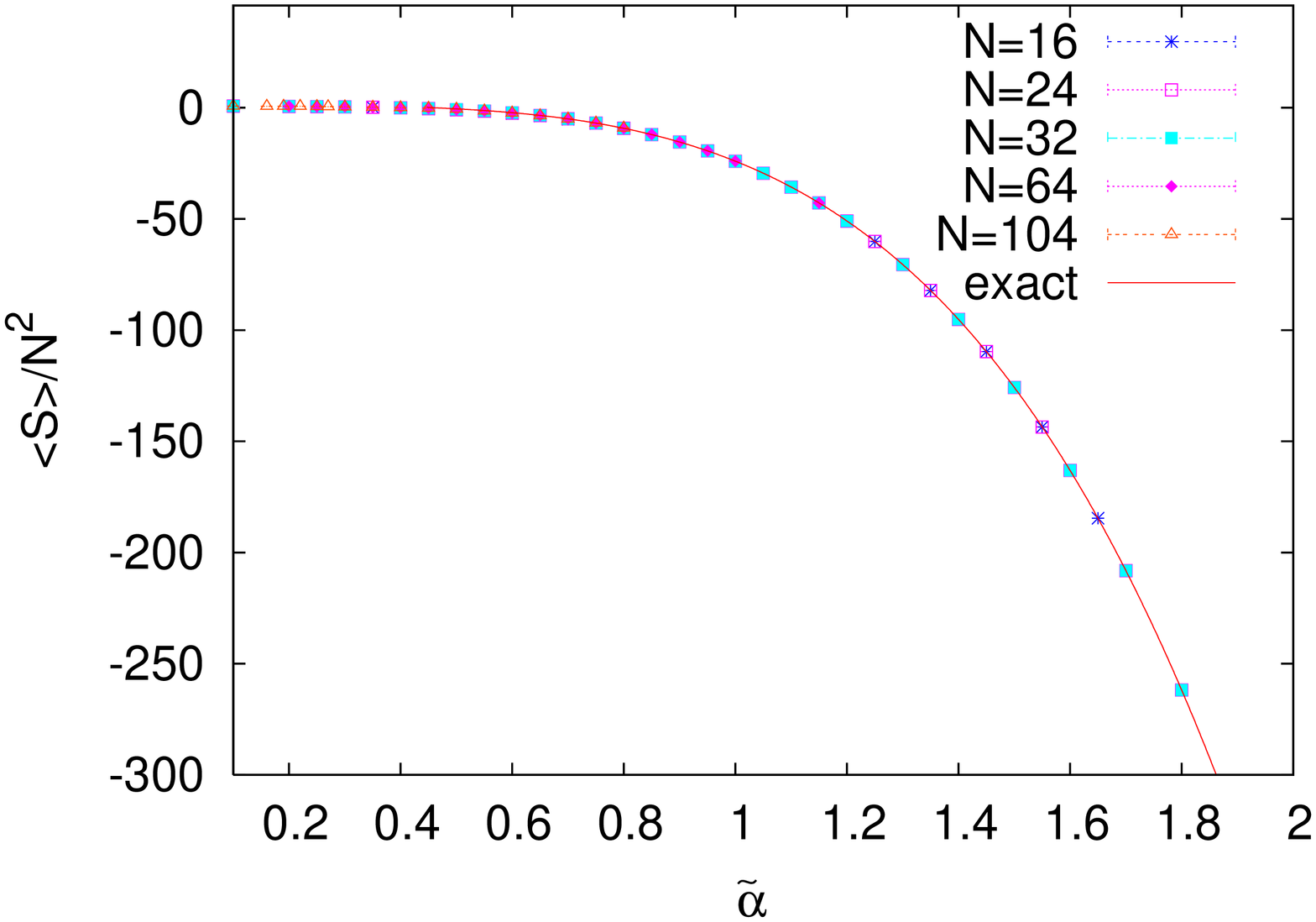}
  \includegraphics[width=5cm,angle=-90]{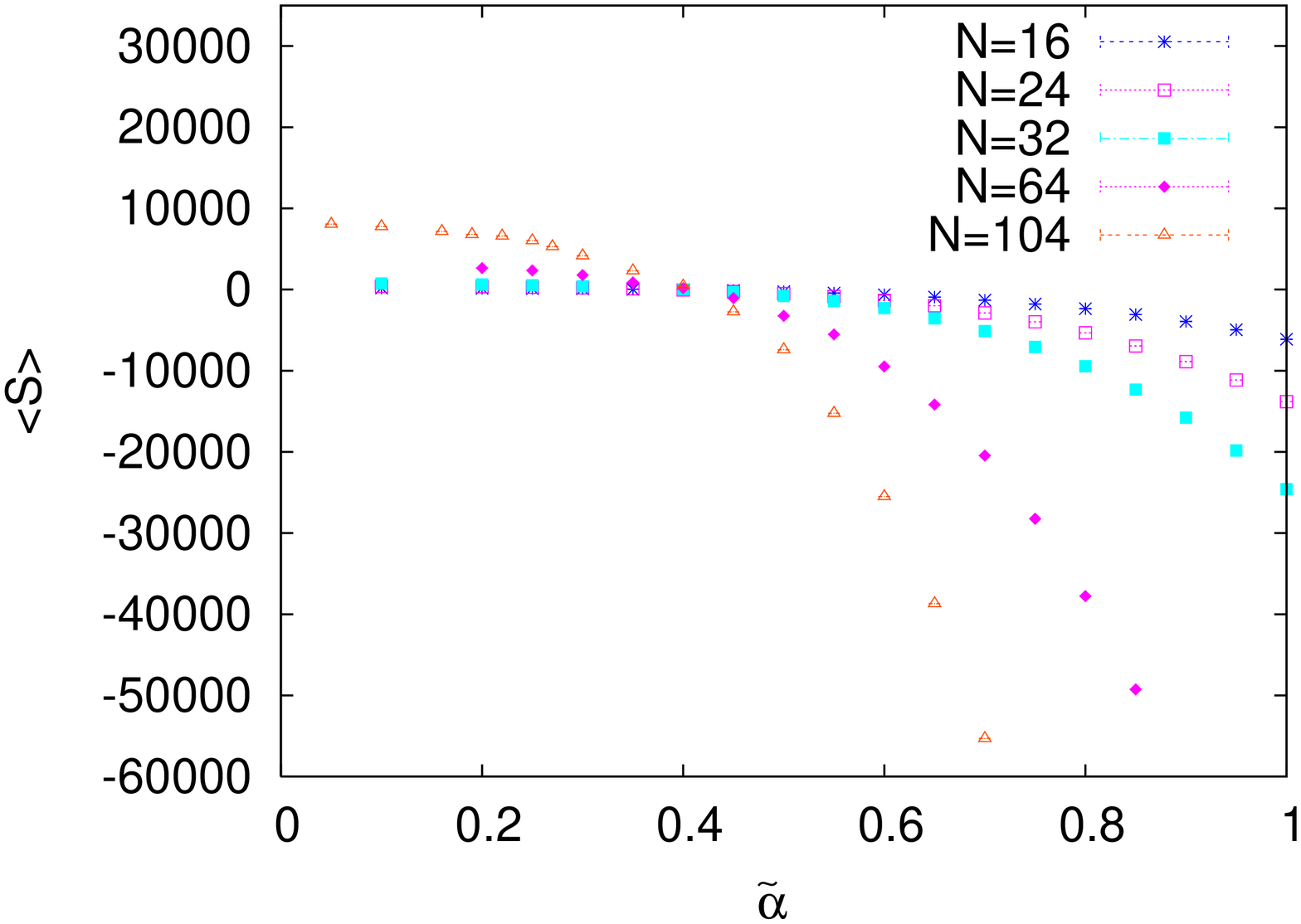}
\caption{The observables $\frac{<S>}{N^2}$ (left) and $<S>$ (right)
  for $m^2=200$ plotted as functions of $\tilde{\alpha}$ for
  $N=16,24,32,64,104$. The value of $\tilde{\alpha}$ at which the
  curves $<S>$ for different values of $N$ cross is defined as
  $\tilde{\alpha}_s$.}\label{Savm200}
\end{center}
\end{figure}

In this case we measure the critical value $\tilde{\alpha}_s$ as
follows. We observe that different actions $<S>$ which correspond to
different values of $N$ (for some fixed value of $m^2$) intersect at
some value of the coupling constant $\tilde{\alpha}$ which we define
$\tilde{\alpha}_s$ (figure \ref{Savm200}). This is the critical
point. For example we find for $m^2=200$ the result
$\tilde{\alpha}_s=0.4\pm 0.1$. The theoretical value is
$\tilde{\alpha}_*=0.44$. For large $m^2$ the theoretical critical
value $\tilde{\alpha}_*$ is given by equation (\ref{cvinfy}). The
measured value $\tilde{\alpha}_s$ tends to be smaller than this
predicted value.

\begin{figure}
\begin{center}
\includegraphics[width=5.8cm,angle=-90]{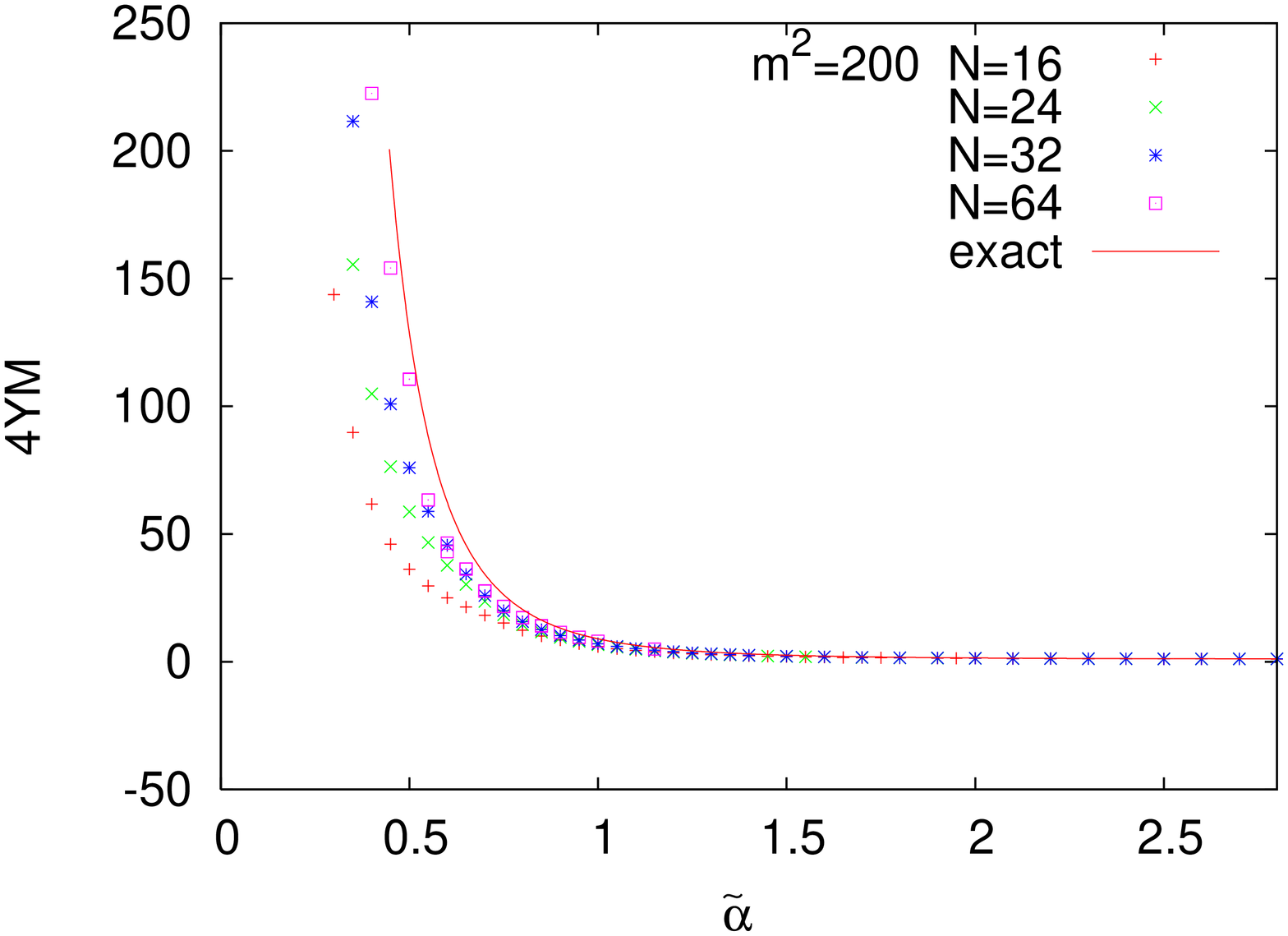}
  \includegraphics[width=5.8cm,angle=-90]{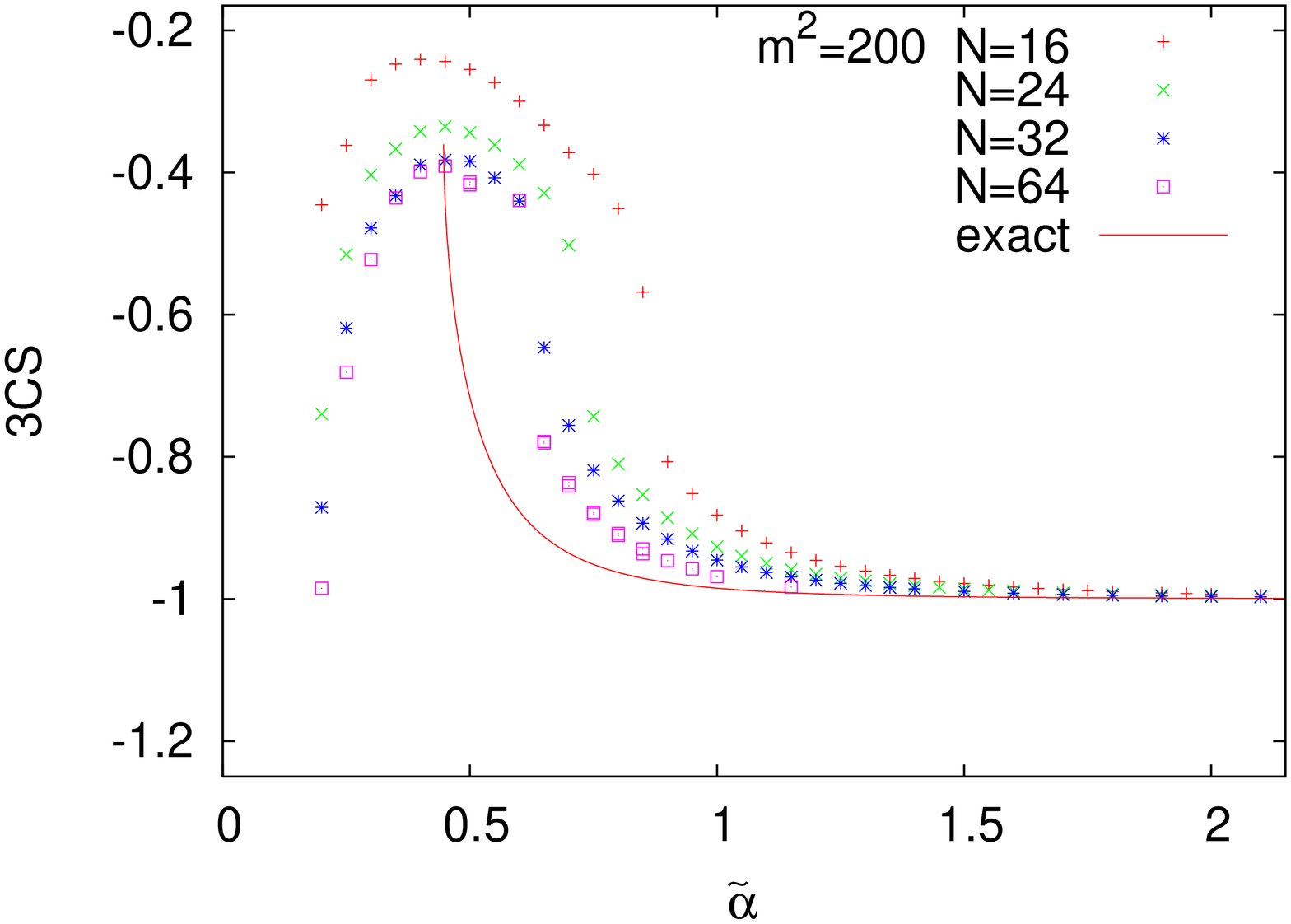}
 \includegraphics[width=5.8cm,angle=-90]{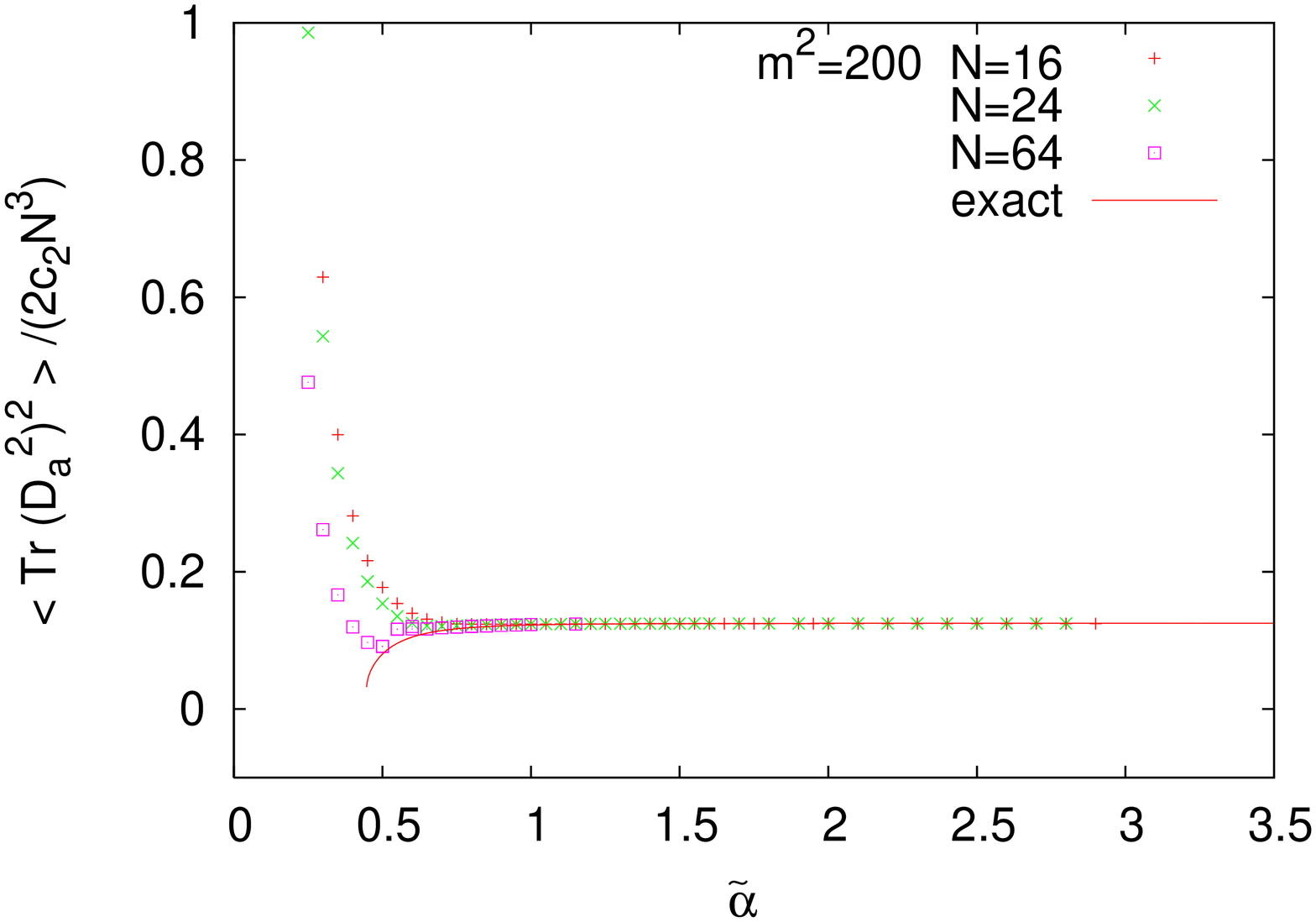}
\includegraphics[width=5.8cm,angle=-90]{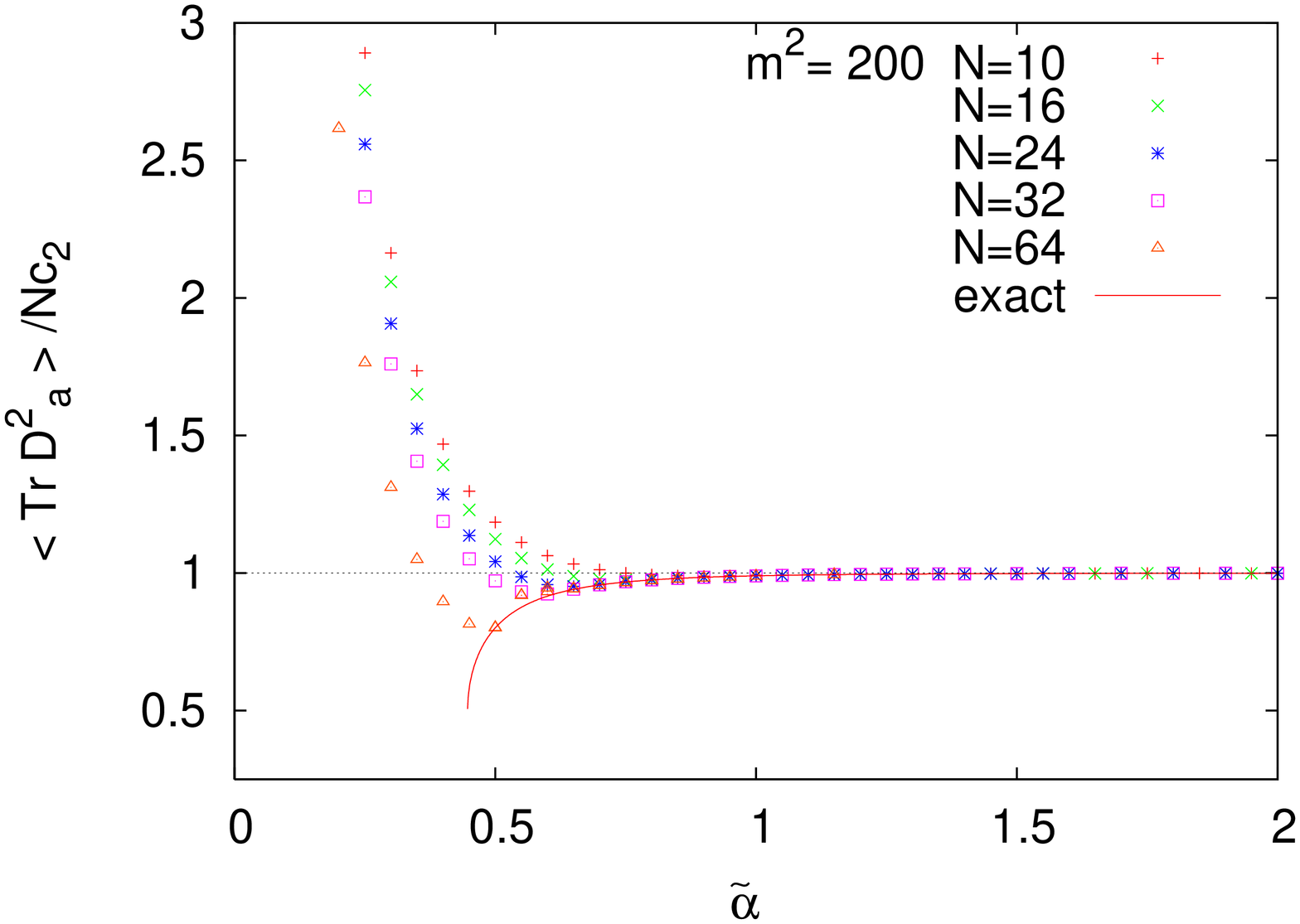}
\caption{Other observables for $m^2=200$ as functions of the coupling
  constant for $N=16,24,32,64$. The solid lines correspond to the theoretical
  predictions. We observe that the data tend to
  approach the theoretical prediction as $N$ is increased.}\label{obsm200}
\end{center}
\end{figure}

\begin{figure}
\begin{center}
\includegraphics[width=5.8cm,angle=-90]{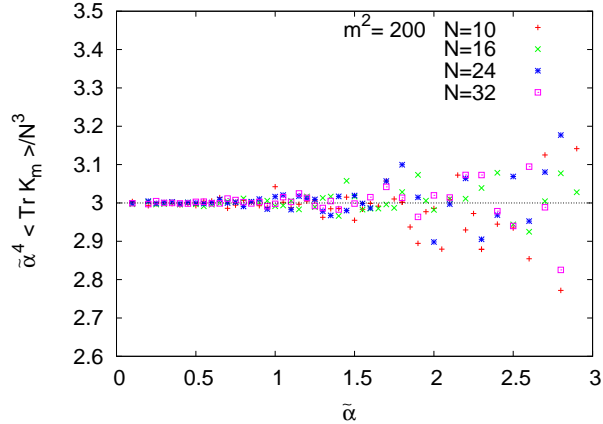}
\caption{The Ward identity (\ref{wardid}) for $m^2=200$ as a function of the coupling
  constant $\tilde{\alpha}$ for $N=10,16,24,32$.}\label{figward}
\end{center}
\end{figure}

The quantities ${\cal S}$, ${\rm YM}$, ${\rm CS}$, 
$\langle Tr(D_a^2)^2\rangle$ and the radius are 
all continuous across the transition point in this regime (figure
\ref{obsm200}). We observe that near the critical point the numerical
results approach the theoretical curves as we increase $N$. We also
checked the Ward identities (\ref{wa1}),(\ref{wa2}) and (\ref{wa3})
(figure \ref{figward}).

\begin{figure}
\begin{center}
\includegraphics[width=5.8cm,angle=-90]{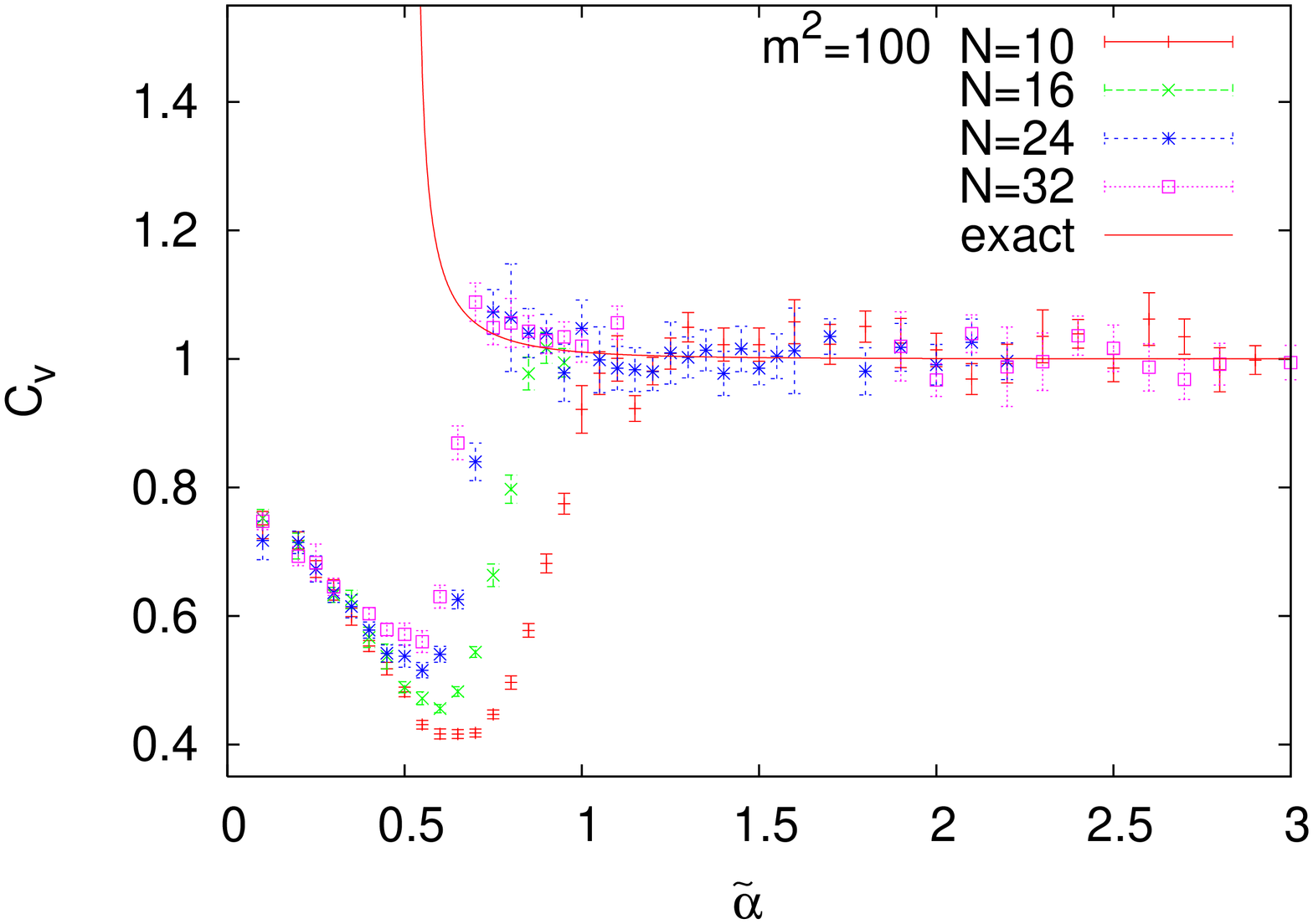}   
\includegraphics[width=5.8cm,angle=-90]{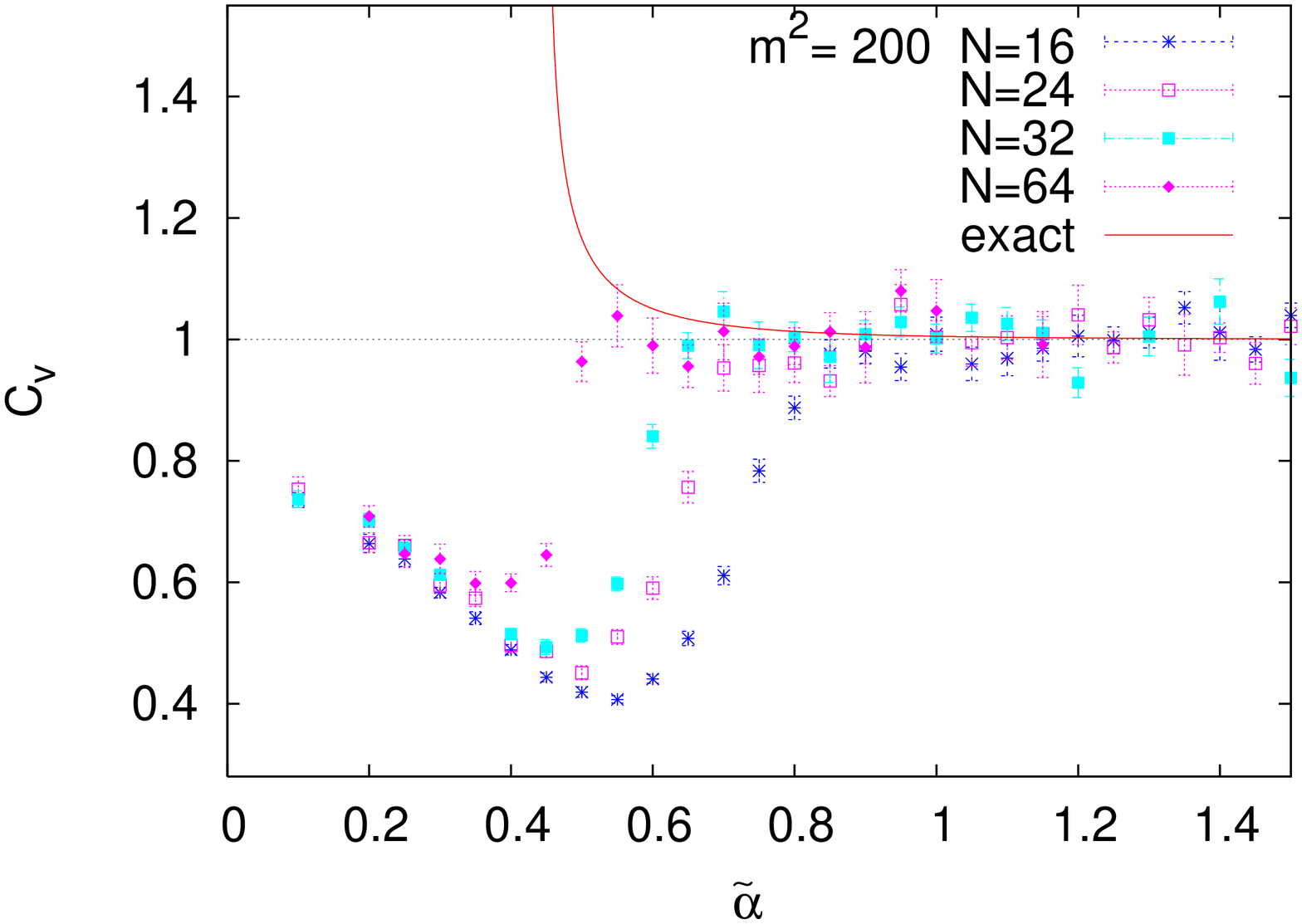} 
\caption{The specific heat for $m^2=100~{\rm and}~200$ for different matrix
  sizes. In the region of the fuzzy sphere the value of the specific
  heat is $1$. In the region of the matrix phase the value of the
  specific heat tends to take the constant value $0.75$ as the size of
  matrix $N$ increases.}\label{cvmlarge}
\end{center}
\end{figure}

\begin{figure}
\begin{center}
\includegraphics[width=5.8cm,angle=-90]{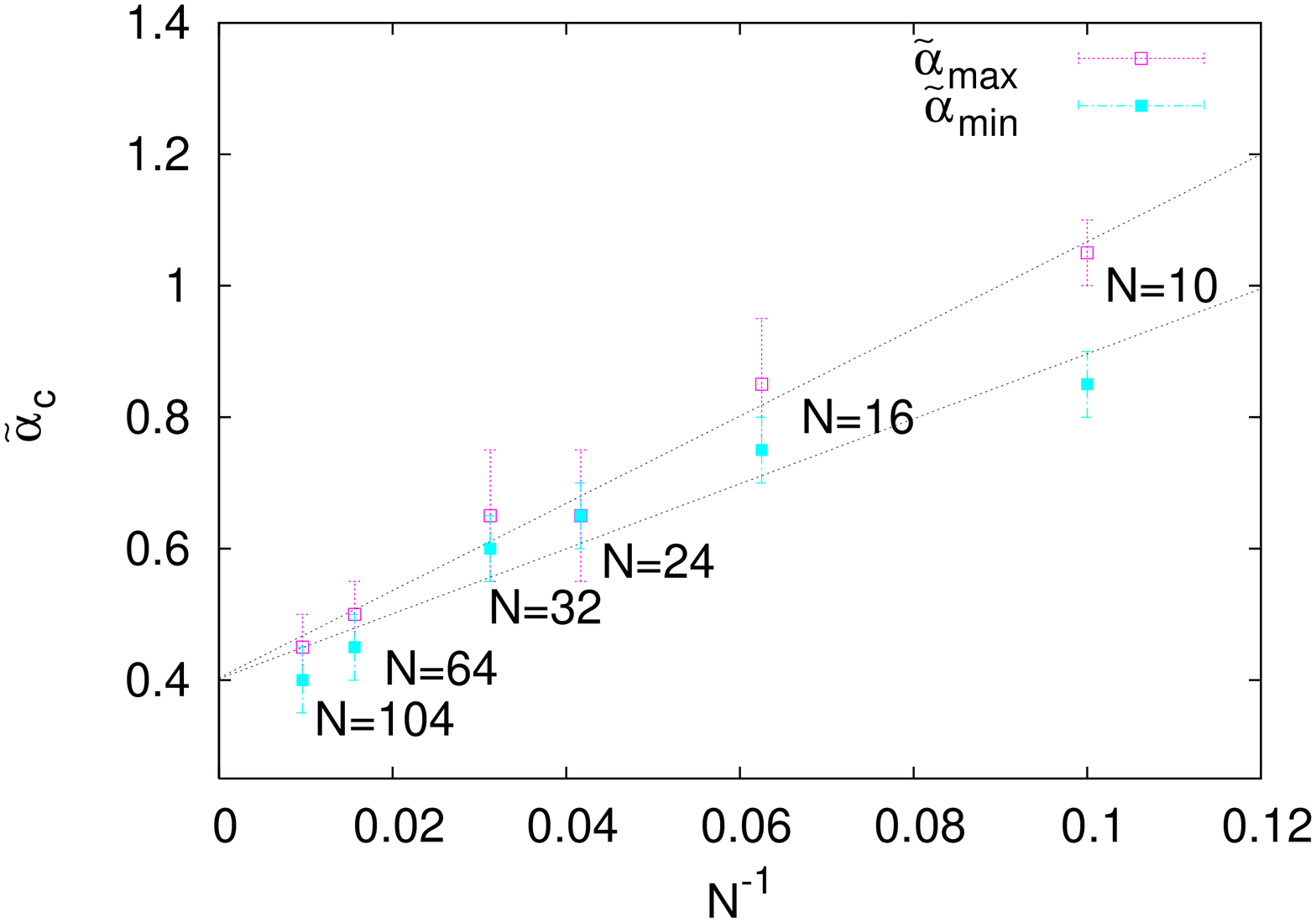}
\caption{By extrapolating the measured values of $\tilde{\alpha}_{\rm
    max}$ and $\tilde{\alpha}_{\rm min}$ to $N=\infty $ we obtain the
    critical value $\tilde{\alpha}_{c}$.}\label{figex}
\end{center}
\end{figure}

The specific heat in the fuzzy sphere phase is constant equal to $1$,
it starts to decrease at $\tilde{\alpha}_{\rm max}$, goes through a
minimum at $\tilde{\alpha}_{\rm min}$ and then goes up again to the
value $0.75$ when $\tilde{\alpha}\longrightarrow 0$ (figure
\ref{cvmlarge}). The values $\tilde{\alpha}_{\rm max,min}$ decrease
with $N$ while the minimum value of $C_v$ increases. Extrapolating the
$\tilde{\alpha}_{\rm max}$ and $\tilde{\alpha}_{\rm min}$ to
$N=\infty$ (figure \ref{figex}) we obtain our estimate for
the critical coupling
$\tilde{\alpha}_c$ which agree with $\tilde{\alpha}_s$ within
errors. The {matrix}-to-${S}^2_N$ phase transition looks then $3$rd
order. However it could be that for large $N$ the specific heat
becomes discontinuous at the critical point with a jump, i.e the
transition is discontinuous with $2$nd order fluctuations. 

Thus it seems that the specific heat in the regime of large values of
$m^2$ is such that $i)$ $\tilde{\alpha}_{\rm max, min}$ approach
$\tilde{\alpha}_s$ in the limit $N\longrightarrow \infty$ and $ii)$
that the specific heat becomes constant in the matrix phase and equal to
$C_v=0.75$. There remains the question of whether or not the specific
heat has critical fluctuations at the critical point for large $m^2$.

The theory still predicts a transition
(equation (\ref{sq4})) with critical fluctuations and a divergence 
in the specific heat with a critical exponent
$\alpha=1/2$ and with a very small amplitude (the coefficient of the
square root singularity).  There is possibly some evidence for
this even for $m^2=100$ but none for $m^2=200$. To resolve this question 
we need to go very near the critical point and simulate with bigger $N$.

\subsection{The phase diagram}

\begin{figure}
\begin{center}
\includegraphics[width=8cm,angle=-90]{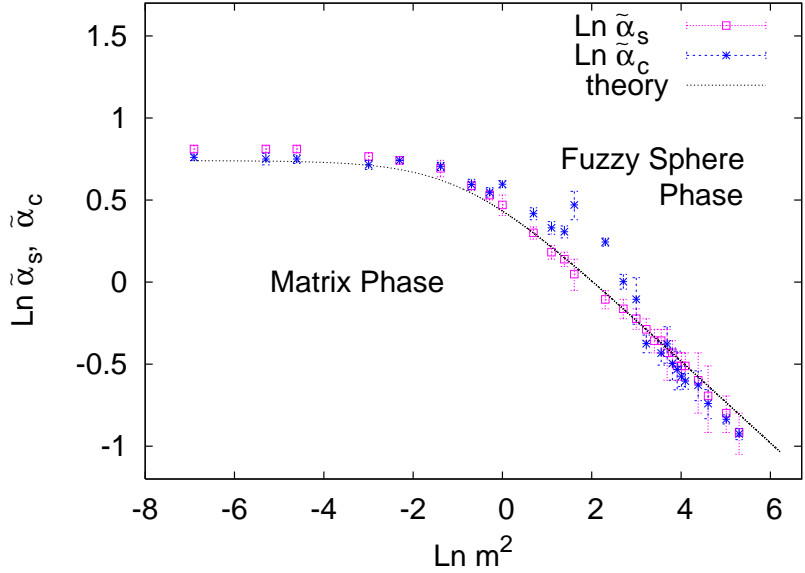}
\caption{The phase diagram shows the curve separating the geometrical
  and matrix phases of the model (\ref{Action-simpleform}) with
  $\mu=m^2$.  The critical curve is given by eq.(\ref{critline}) and
  describes (at least for small values of $m^2$) a line of exotic
  transition with a jump in the entropy yet with divergent critical
  fluctuations and a divergent specific heat with critical exponent
  $\alpha=0.5$, when approached from the fuzzy sphere side. The points
  which lay off the critical line in the middle mass region might
  suggest the existence of a multicritical point.}\label{phadia}
\end{center}
\end{figure}

Our phase diagram in terms of the parameters $\tilde{\alpha}$ and
$m^2$ which we have studied is given in figure \ref{phadia}.  We have
identified two different phases of the matrix model (\ref{main11}).
In the {\it geometrical} or {\it fuzzy sphere phase} we have a $U(1)$
gauge theory on ${ S}^2_N$; the geometry of the sphere and the
structure of the $U(1)$ gauge group are stable under quantum
fluctuations so the theory in the continuum limit is an ordinary
$U(1)$ on the sphere. In the {\it matrix phase} the fuzzy sphere
vacuum collapses under quantum fluctuations and there is no underlying
sphere in the continuum large $N$ limit.  In this phase the model
should be described by a pure matrix model without any background
spacetime geometry.  The transition in the Ehrenfest
classification would be labeled a first order transition, but this 
classification is not very helpful. The transition described here 
is a very exotic one with both a latent heat and a divergent specific
heat. We know of no other example of such a transition.  As we follow
this line of transitions the entropy jump or equivalently the latent
heat becomes zero at around $m^2\sim 40$ and remains zero for
larger $m^2$. Our theoretical analysis indicates that there is still a
divergent specific heat, however our numerical simulations are not
fine enough to determine whether this is so or not.

For large $m^2$ the transition (we expect) still has a divergent
specific heat as the transition is approached from the fuzzy sphere
side. This is the conclusion of our theoretical analysis and our
numerical results are consistent with this. But our numerical results
are not conclusive.  It may also be that the transition is even in
Ehernfest's classification a $3rd$ order, where there is a jump in the
specific heat with no divergence.  We could not determine the nature
of the transition in the large $m$ regime with any confidence from the
numerical data.  In all cases the fuzzy sphere-to-matrix theory
transitions are from a one-cut phase (the matrix phase) to a point or
discrete spectrum in the geometrical (fuzzy sphere) phase.

The specific heat in the fuzzy sphere phase takes the value $1$ where
the gauge field can only contribute the amount $1/2$. This can be
understood as follows. The high temperature limit of the
specific heat for any matrix model is governed by the largest
term and must go like $N_{total}/{degree}$ where $N_{total}$ is 
the total number of degrees of freedom and $degree$ is the degree
of the polynomial. This gives the limiting high temperature limit of
the specific heat be $\frac{3}{4}$ for all values of the parameters.
In the simple model with $m=0$ and $\mu=0$ this value is achieved
from the transition point onwards. 

\begin{figure}
\begin{center}
\includegraphics[width=5.8cm,angle=-90]{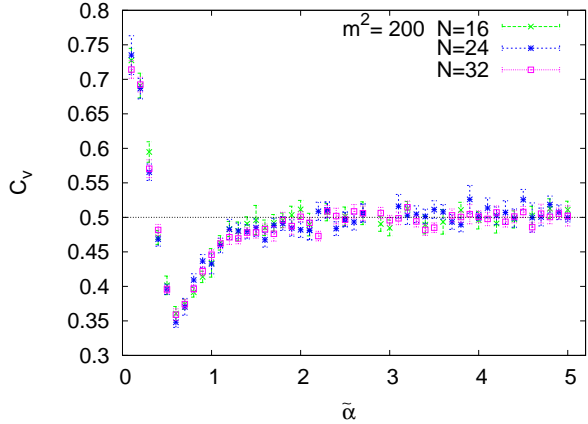}
\caption{The specific heat for the pure potential with $m^2=200$ as
  function of the coupling constant computed with respect to the
  Boltzmann weight $e^{-V}$.}\label{cvpote}
\end{center}
\end{figure}

For the full model in the large 
$m^2$ regime the effect of the potential $V$ should be dominant. This can be seen by remarking that in the strong-coupling limit
$m^2{\longrightarrow}\infty$, $\tilde{\alpha}{\longrightarrow}0$
keeping fixed $\tilde{\alpha}^4m^2$ the action $S$ reduces to
$V$. Note that if we consider $V$ alone with a measure given by $\int
[d{\Phi}^{'}]$ (with ${\Phi}^{'}=\sqrt{D_a^2}$) instead of $\int
[dD_a]$ then we will get the usual quartic potential dynamics with a
well known $3$rd order transition.  Here when we consider the model
given by the potential $V$ with the measure $\int [dD_a]$ we obtain
the specific heat given in figure \ref{cvpote}. In the region of
parameters corresponding to the matrix phase the specific heat shows
in this case a structure similar to that of the full model
$S$. However in the region of parameters corresponding to the fuzzy
sphere phase the specific heat is given now by ${C_{v}}=1/2$. Thus the
field $\Phi$ contributes only the amount $1/2$ to $C_v$. Indeed from
the eigenvalue distribution of the operator $\Phi$ computed in the
fuzzy sphere phase it is shown explicitly that $\Phi$ has Gaussian
fluctuations.  The behaviour of $C_v$ in the full model $S$ is thus a
non-trivial mixture of the behaviours in $S_0$ (first order
transition) and $V$ ($3$rd order transition) considered separately.

Given that the behaviour of the full model can be described as a 
non-trivial mixture of the model $S_0$ and the potential we
expect that the effect of adding the potential to the model is
to shift the transition temperature and provide a non-trivial background 
specific heat. The divergence of the specific heat arises
from the interplay of the two terms in $S_0$, i.e. between 
the the Chern-Simons and Yang-Mills terms. Given that this competition leads
to a divergence of the specific heat it should eventually emerge from the
background sufficiently close to the transition.

\subsection{The eigenvalue distributions for large $m^2$} 

\subsubsection{The low temperature phase (fuzzy sphere)}

\begin{figure}[htbp!]
\begin{center}
\includegraphics[width=5.8cm,angle=-90]{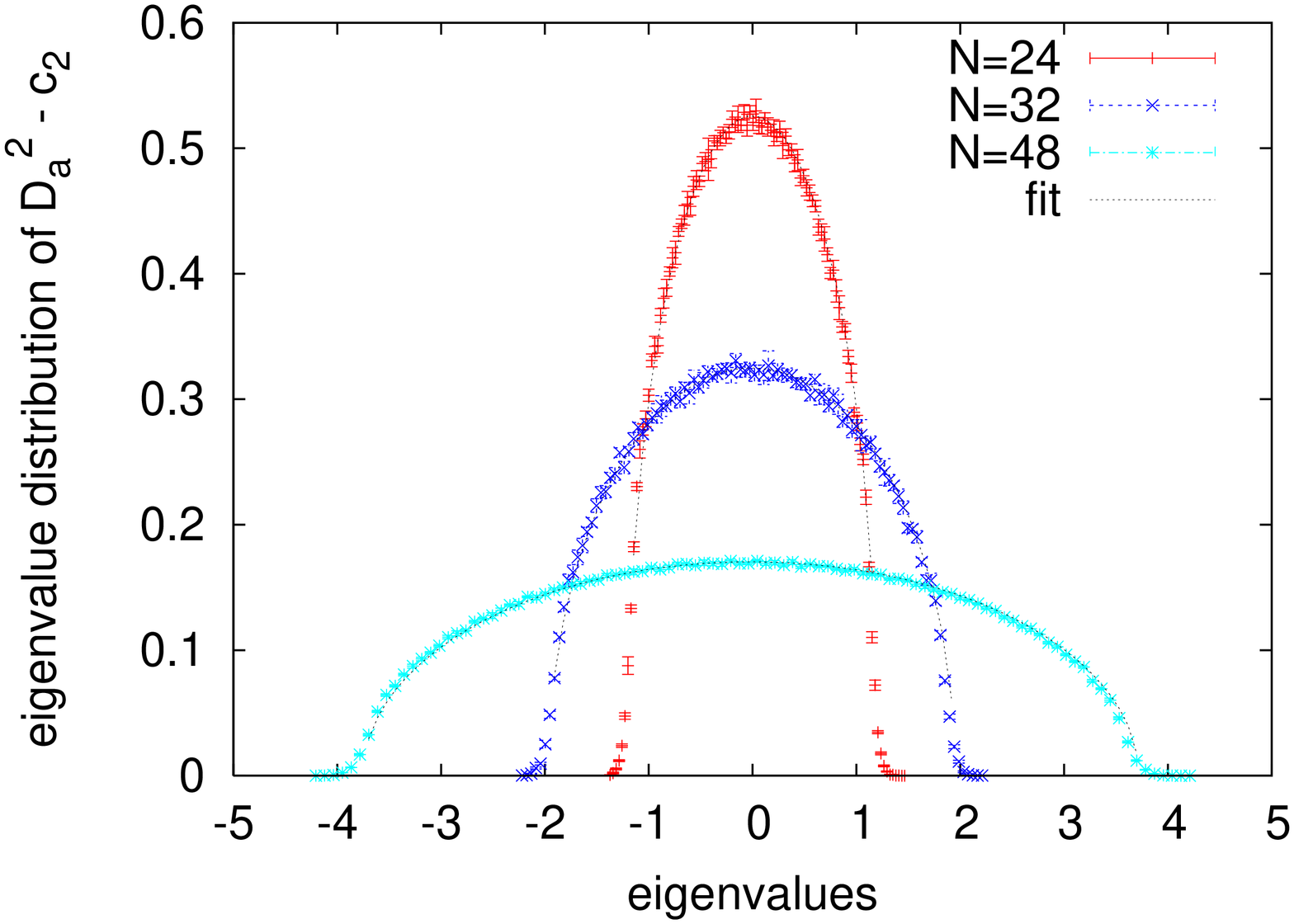}
\includegraphics[width=5.8cm,angle=-90]{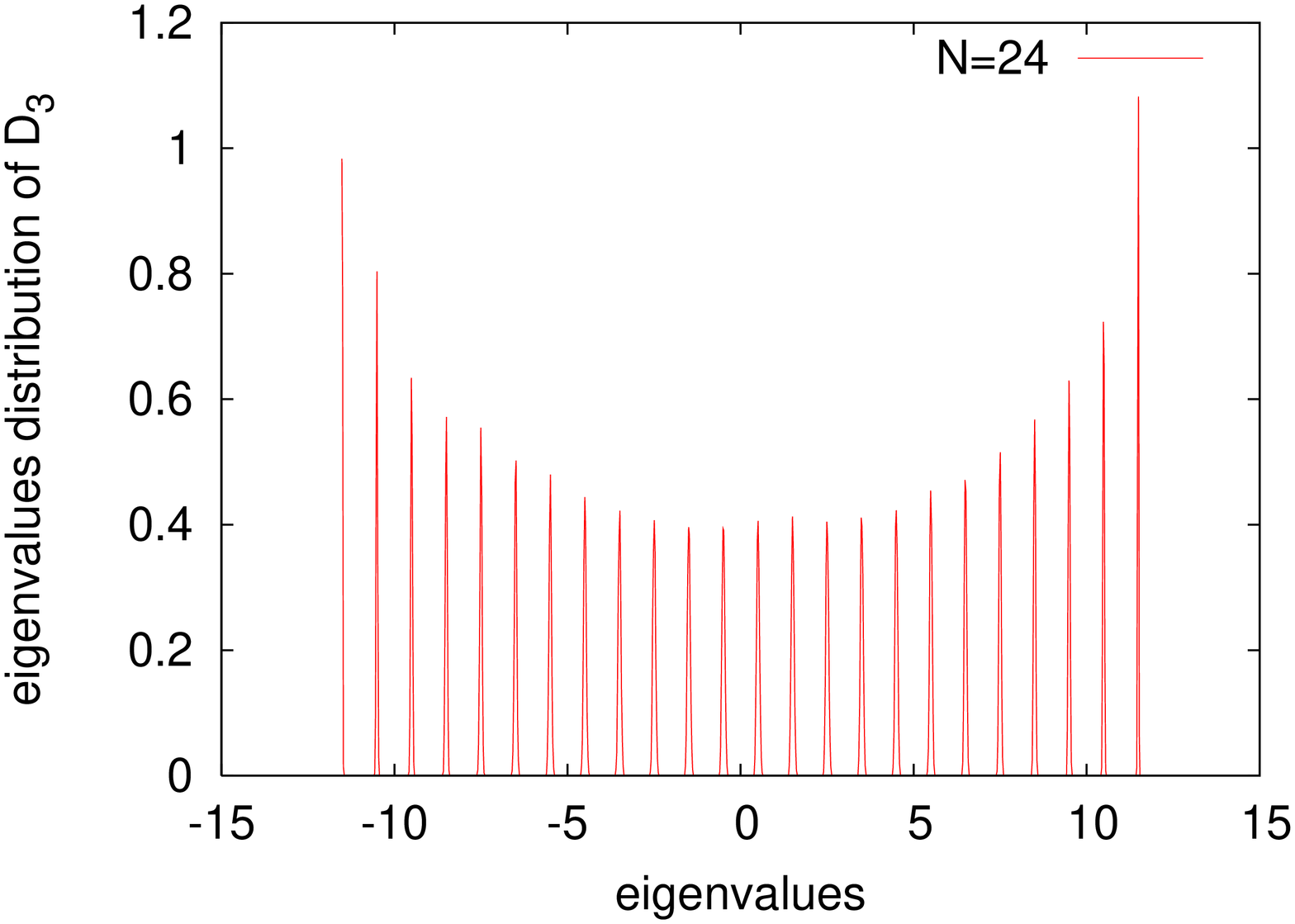}
\includegraphics[width=5.8cm,angle=-90]{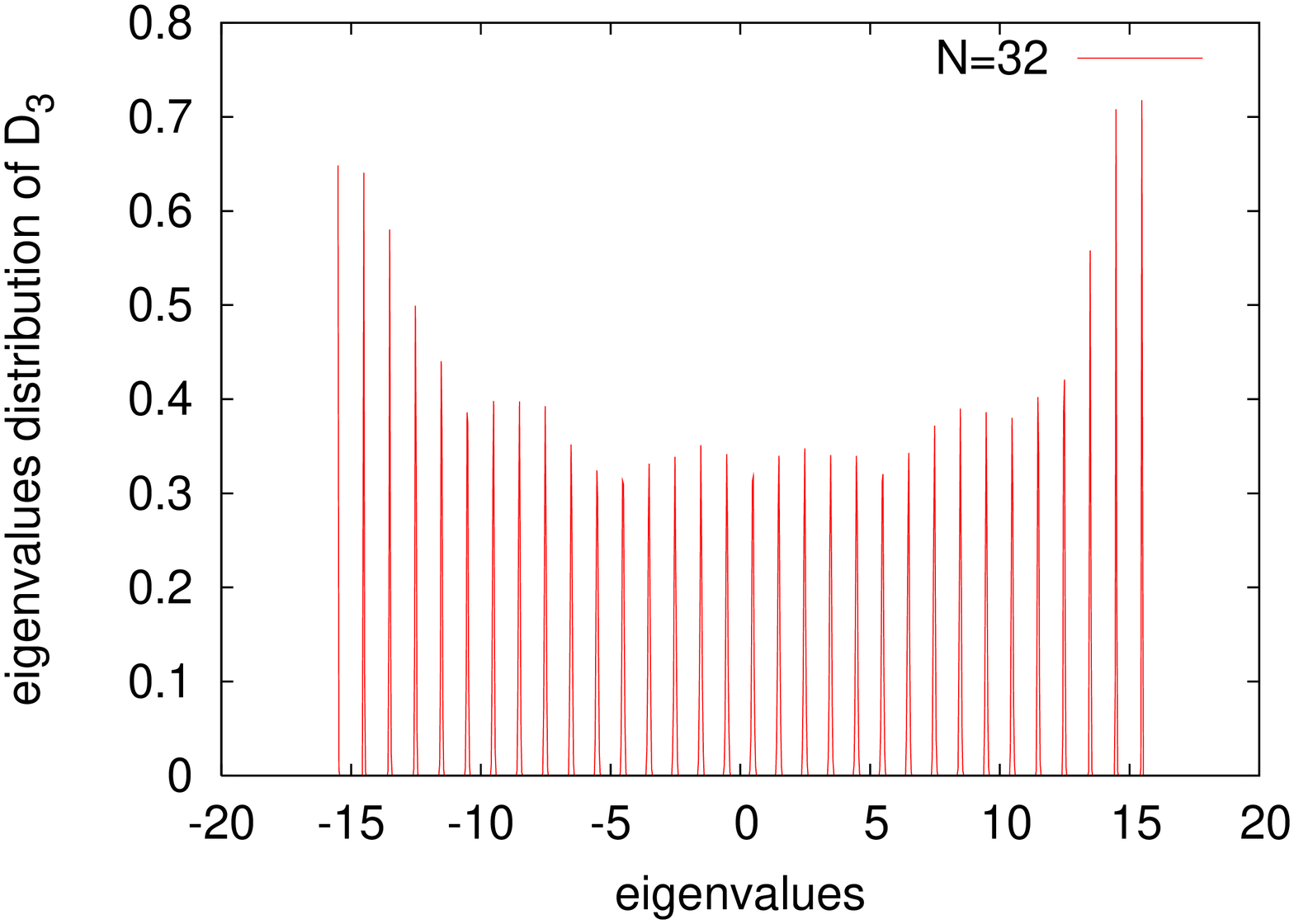} 
\includegraphics[width=5.8cm,angle=-90]{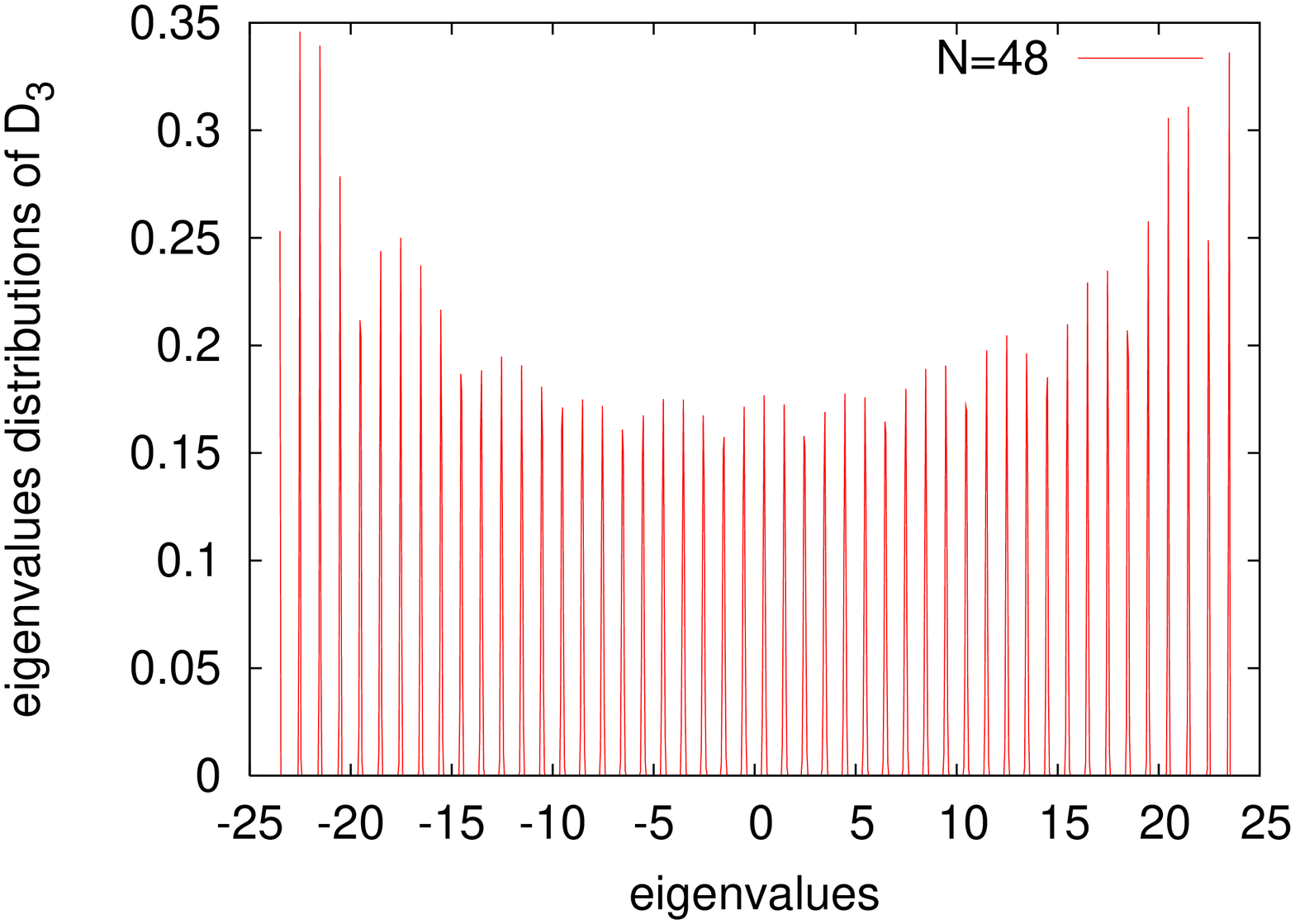} 
\caption{The eigenvalue distribution of $D_a^2-c_2$ (left) for
  $N=24,32,48$ and $D_3$ (right) for
  $N=24$, $N=32$ and $N=48$ in the fuzzy sphere phase with $m^2=200$ and $\tilde{\alpha}=5$. The fit for $D_a^2-c_2$ corresponds with
  the Wigner semi-circle law (\ref{WSClaw}).}\label{disfuzzy}
\end{center}
\end{figure}

Numerically we can check that the normal scalar field and the tangent
gauge field decouple from each other in the ``fuzzy sphere phase'' in
the limit $m^2{\longrightarrow}\infty$ by computing the eigenvalues of
the operators $D_a^2$ and $D_3$. See figure \ref{disfuzzy}. The fit
for the distribution of eigenvalues of $D_a^2-c_2$ is given by the
Wigner semi-circle law
\begin{eqnarray}\label{WSClaw}
\rho(x)=\frac{2}{a^2_{\rm eff} \pi}\sqrt{a^2_{\rm eff}-x^2}.
\end{eqnarray}
This distribution is consistent with the effective Gaussian  potential
\begin{eqnarray}\label{gaussian}
V_{\rm eff}^{\rm sphe}=\frac{2}{a^2_{\rm eff}}Tr (D_a^2-c_2)^2~,~a^2_{\rm eff} = \frac{4c_2N}{m^2
  \tilde{\alpha}_{\rm eff}^4}.
\end{eqnarray}
We find numerically
\begin{center}
\begin{tabular}{|c|c|c|}
\hline
$N$& $a^2_{\rm eff}$&$a^2$ \\
\hline
$24$ & $1.4632\pm~ 0.0016$& $0.1104$\\
\hline
$32$  & $3.8206\pm 0.0068$& $0.2619 $  \\
\hline
$48$&$13.9886\pm 0.0110$ & $0.8844$  \\
\hline
\end{tabular}
\end{center}
The parameter $a$ is the theoretical prediction given by $~a^2 = {4c_2N}/{m^2\tilde{\alpha}^4}$ which goes like $N^3$. The effective parameter $a_{\rm eff}^2$ is found to behave as
\begin{eqnarray}
a_{\rm eff}^2=(3.7645\pm 0.6557)\times 10^{-5}\times N^{3.3170\pm 0.0492}.
\end{eqnarray} 
This means that the renormalized value $\tilde{\alpha}_{\rm eff}$ of the gauge coupling constant is slowly decreasing with $N$. Equivalently the parameter $a_{\rm eff}$ yields
a small correction to the classical  potential $V$ which is linear in $\Phi$. Indeed we can show that

\begin{eqnarray}
Z&=&\int dD_ae^{-\frac{2}{a^2_{\rm eff}} Tr(D_a^2-c_2)^2}\nonumber\\
&=&e^{C_0}\int dD_ae^{-\frac{2}{a^2}
Tr(D_a^2-c_2)^2-\frac{4c_2}{a}(\frac{1}{a}-\frac{1}{a_{\rm eff}})Tr (D_a^2-c_2)
}\nonumber\\
C_0&=&-2Nc_2^2\bigg(\frac{1}{a}-\frac{1}{a_{\rm eff}}\bigg)^2-\frac{3}{2}N^2\log\frac{a}{a_{\rm eff}}.\label{arg}
\end{eqnarray}
This shows explicitly that  having $a_{\rm eff}\neq a$ means that there is an extra linear term in $\phi$ added to the classical potential.

\begin{figure}
\begin{center}
\includegraphics[width=5.8cm,angle=-90]{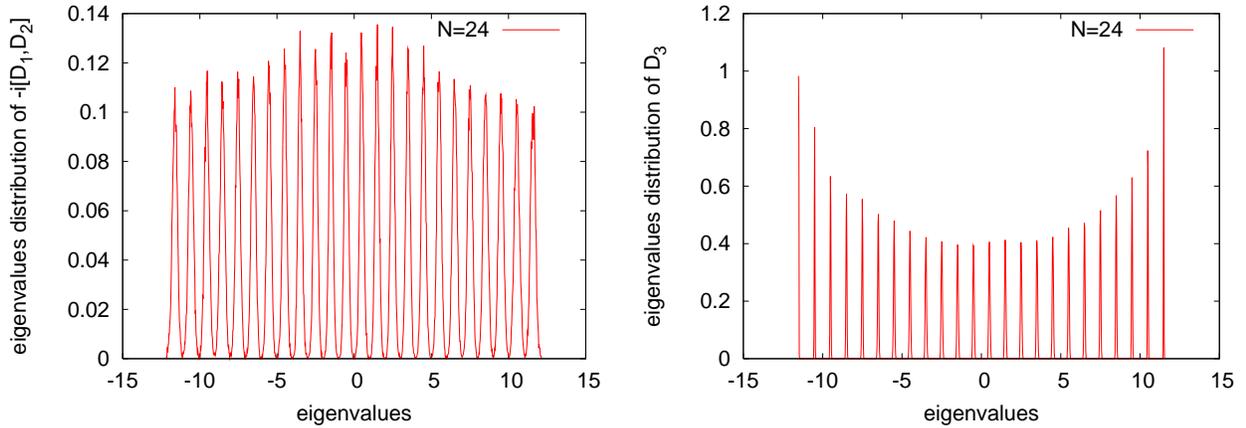}
\includegraphics[width=5.8cm,angle=-90]{evD3-Fuzzy-Sphere-a5.00-m200-BIG.ps}
\caption{{The eigenvalue distributions of $-i[D_1,D_2]$ and $D_3$ in the fuzzy sphere phase for $m^2=200$ and $\tilde{\alpha}=5$ .}}\label{comD}
\end{center}
\end{figure}

In the fuzzy sphere phase the field configurations $D_a$  are
thus given by (or are close to)  representations of
$SU(2)$ of spin $s=\frac{N-1}{2}$ as we can clearly see on  figure \ref{disfuzzy}. Indeed the eigenvalues of $D_3$ for $\tilde{\alpha}=5$ and
$m^2=200$ are found to lie within the range $-\frac{N-1}{2},\cdots, 0,\cdots
,\frac{N-1}{2}$ as expected for $N=24$ and $N=32$. The eigenvalues of the commutator $-i[D_1,D_2]$ are also found to lie in the range  $-\frac{N-1}{2},\cdots, 0,\cdots
,\frac{N-1}{2}$ (figure \ref{comD}).

\subsubsection{The high temperature phase (matrix phase)}
\begin{figure}[htbp!]
\begin{center}
\includegraphics[width=4.8cm,angle=-90]{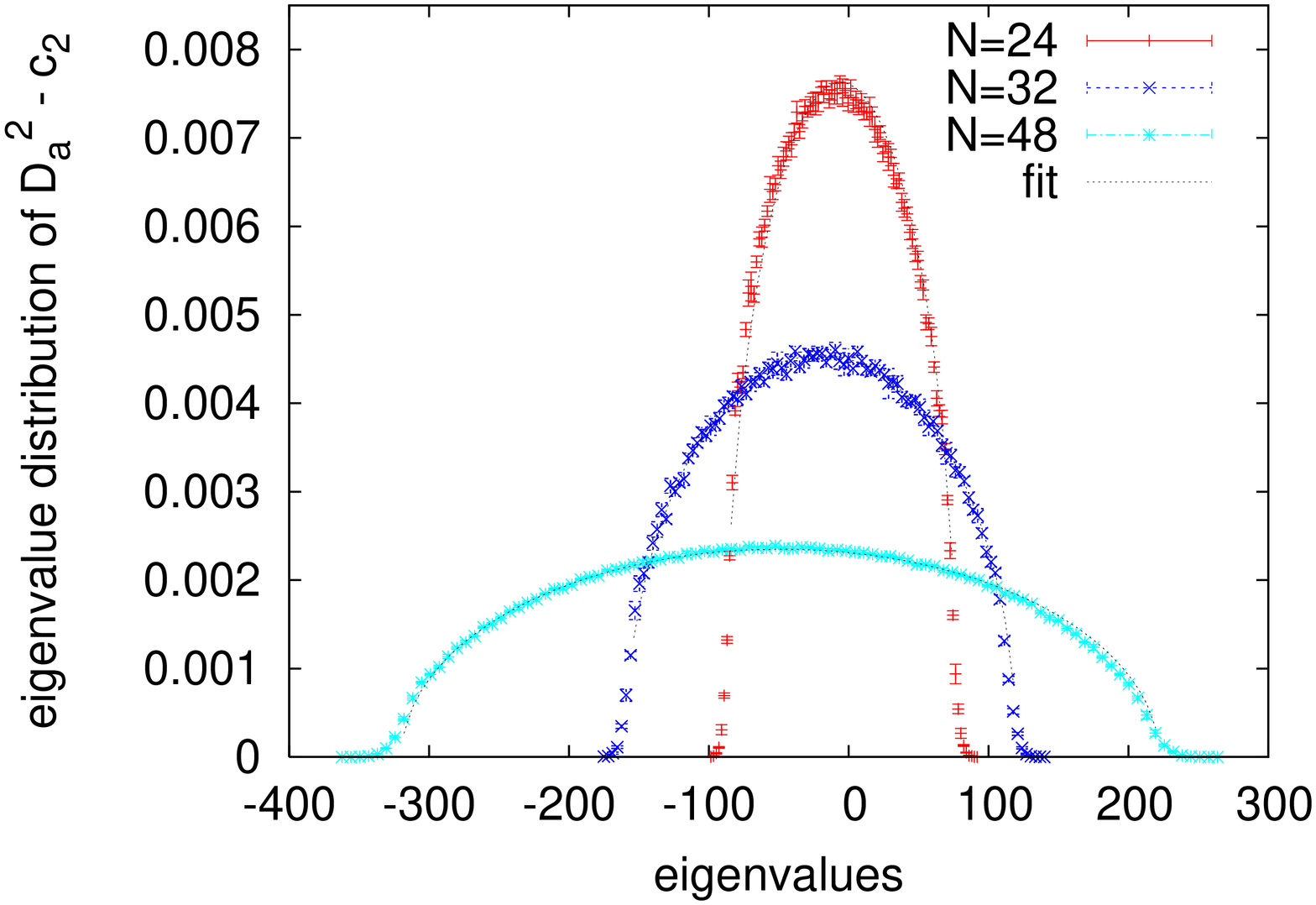}
\includegraphics[width=4.8cm,angle=-90]{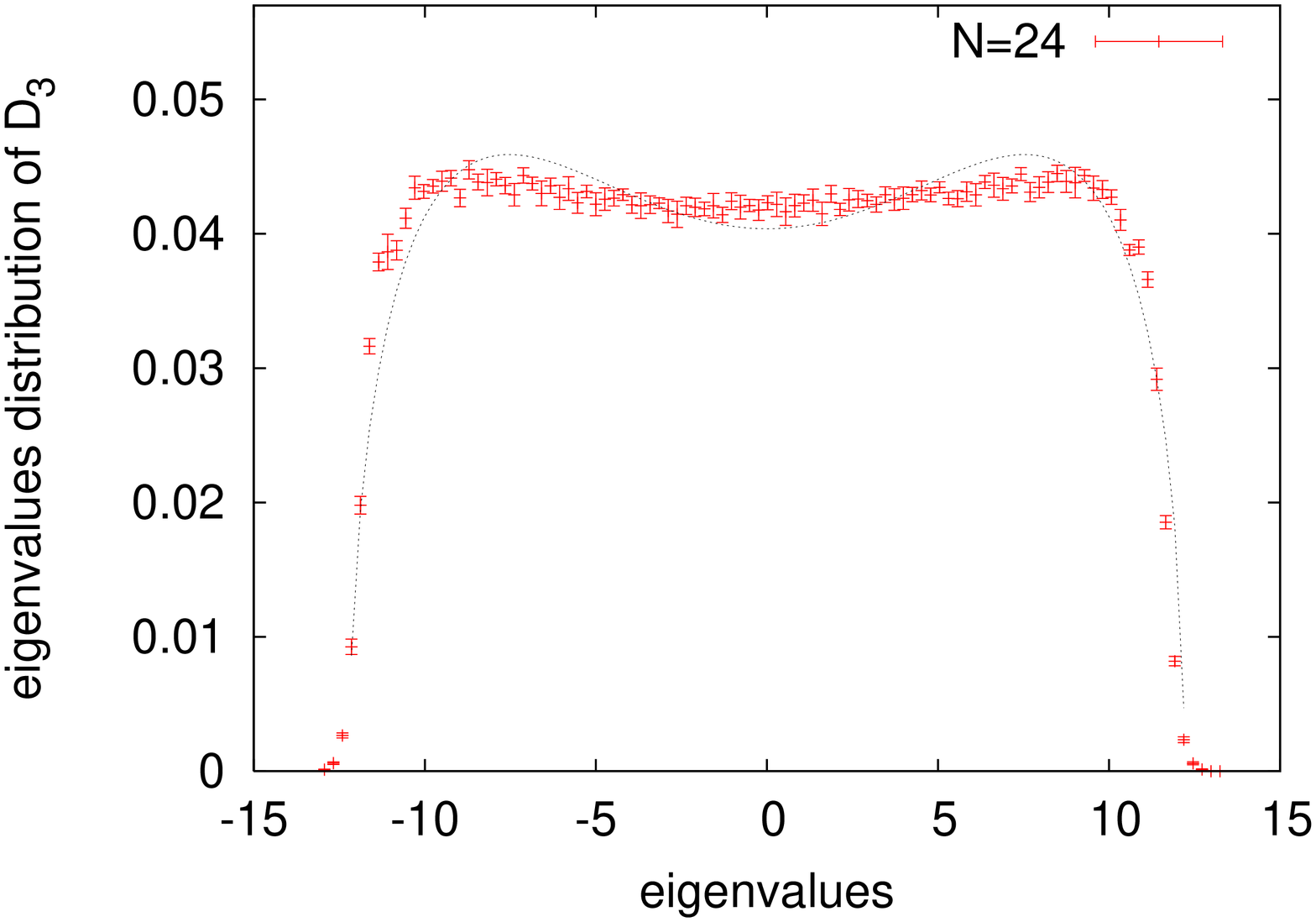}
\caption{The eigenvalue distributions for $D^2_a$, $D_3$ and $-i[D_1,D_2]$ in the matrix phase for
  $\tilde{\alpha}=0.60$ and $m^2=200$.}\label{disttransi}
\end{center}
\end{figure}

\begin{figure}[htbp!]
\begin{center}
\includegraphics[width=4.8cm,angle=-90]{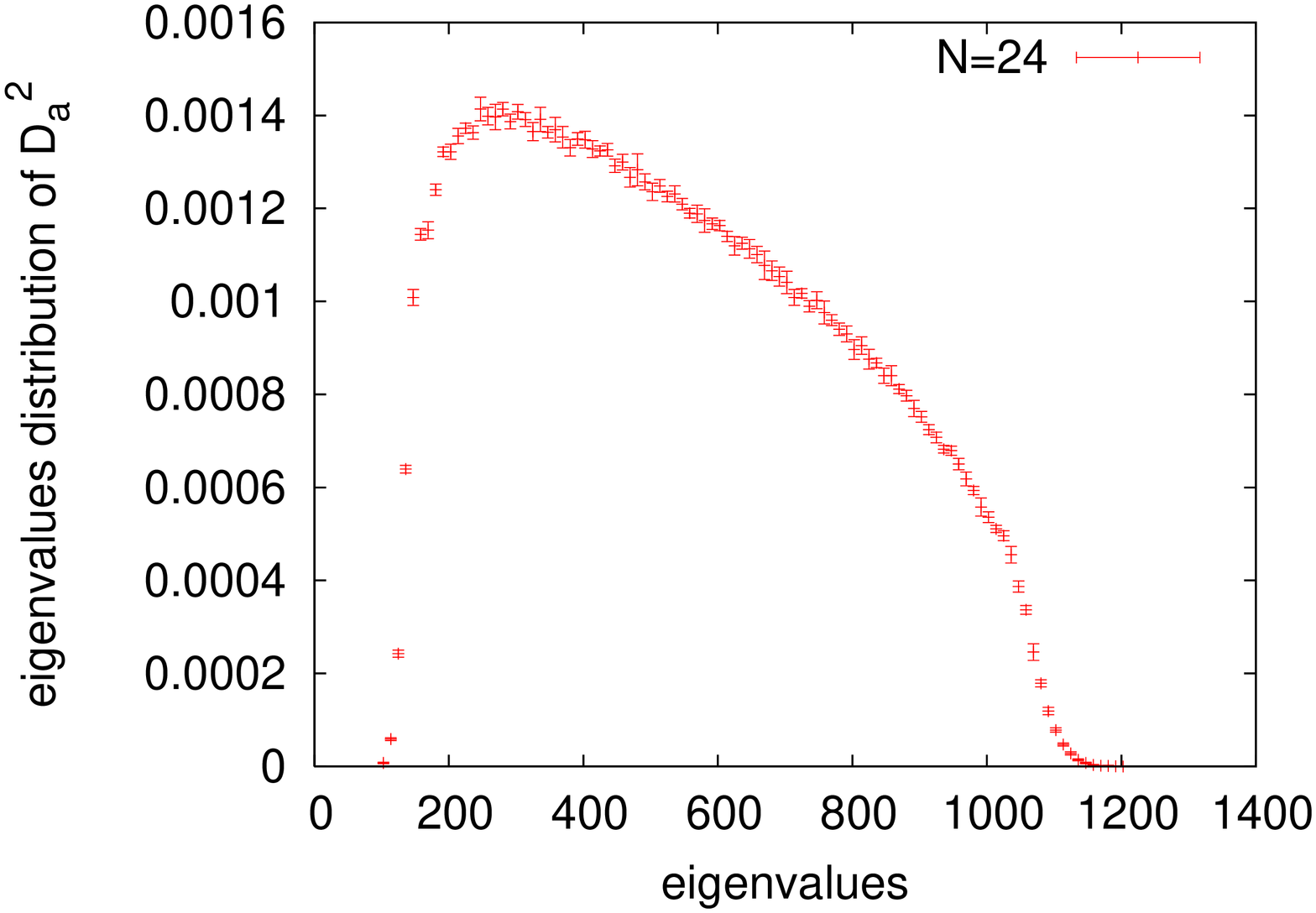}
\includegraphics[width=4.8cm,angle=-90]{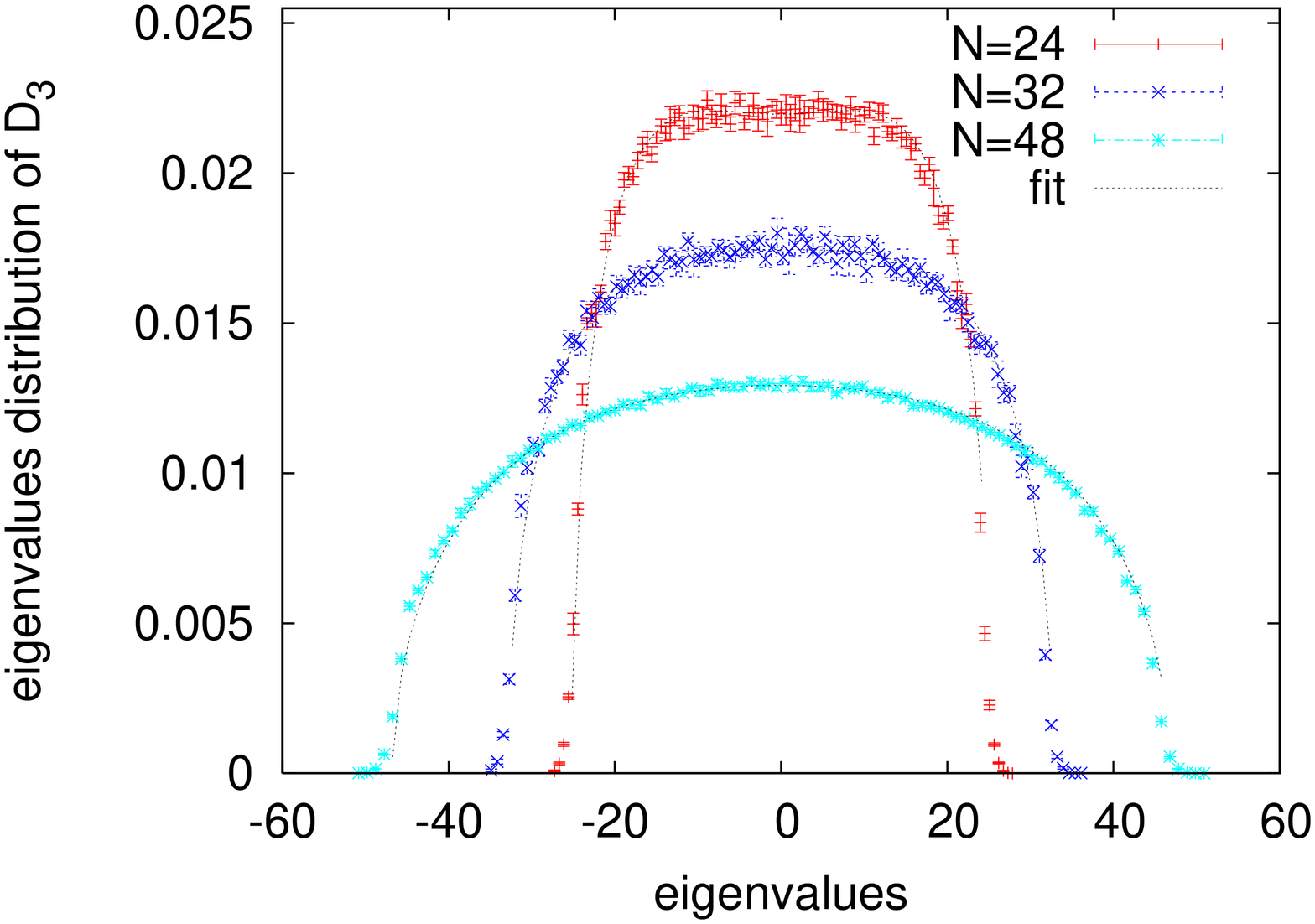}
\caption{The eigenvalue distributions for $D^2_a$ for $N=24$ and
  $D_3$ and $-i[D_1,D_2]$ for  $N=24,32,48$  in the matrix phase with $\tilde{\alpha}=0.2$
  and $m^2=200$. The dashed line corresponds with the fit
  (\ref{rhoplusgauge}) which is the the eigenvalue distribution in the
  regime of one-cut solution.}\label{distmatrix}
\end{center}
\end{figure}

The distribution of the eigenvalues of $D_a^2$ suffers a distortion as
soon as the system undergoes the matrix-to-sphere transition and
deviations from the Wigner semi-circle eigenvalue distribution
(\ref{WSClaw}) become large as we lower the coupling constant
$\tilde{\alpha}$. See figures \ref{disttransi} and \ref{distmatrix}.

In this phase the distribution for $D_3$ is symmetric around
zero and the fit is given by the one-cut solution
\begin{eqnarray}\label{rhoplusgauge}
\rho(x)=\frac{16}{c(c+4b)}\frac{1}{2\pi}\big(x^2+b\big)\sqrt{c-x^2}.\label{4.5}
\end{eqnarray}
By rotational invariance the eigenvalues of the other two matrices
$D_1$ and $D_2$ must be similarly distributed. This means in
particular that the model as a whole behaves in the ``matrix phase''
as a system of $3$ decoupled $1$-matrix models given by the effective
potentials ($i$ fixed)
\begin{eqnarray}\label{Veffmatrix}
V_{\rm eff}^{\rm matr}=\bigg(\frac{2N}{c}-\frac{3c}{a^2_{\rm eff}}\bigg)Tr D_i^2+\frac{2}{a^2_{\rm eff}}Tr D_i^4~,~a^2_{\rm eff}\equiv \frac{c^2}{2N}+\frac{2cb}{N} = \frac{4c_2N}{m^2
  \tilde{\alpha}_{\rm eff}^4}.\label{11}
\end{eqnarray}
We find numerically that

\begin{center}
\begin{tabular}{|c|c|c|c|}
\hline
$N$& $b$&$c$\\
\hline
$24$ & $1047.23 \pm 15.97$& $633.655 \pm 0.7516$\\
\hline
$32$  & $2392.16 \pm  65.09$& $1074.07 \pm 1.8680$  \\
\hline
$48$&$10350.9 \pm 236.70$ & $2185.85 \pm 1.7130$  \\
\hline
\end{tabular}
\end{center}
Again the theoretical prediction for $a^2$ goes like $N^3$ whereas the effective parameter $a_{\rm eff}^2$ is found to behave as
\begin{eqnarray}
a_{\rm eff}^2=(0.6147\pm 0.3059)\times N^{3.6580\pm 0.1553}.
\end{eqnarray} 
By going through the same argument which lead to equation (\ref{arg}) we can show that $V_{\rm eff}^{\rm matr}$ is equivalent to the addition of an extra linear correction    in $\Phi$ (which depends on $c$ and $b$ or equivalently $c$ and $a^2_{\rm eff}$)  to the classical potential
\begin{eqnarray}
-\frac{4c_2}{a^2}Tr D_i^2+\frac{2}{a^2}Tr D_i^4.
\end{eqnarray}
The result (\ref{11}) accounts for the value $C_v=0.75$ of the specific heat. Indeed  for effective potentials of the form (\ref{11}) each matrix $D_i$ contributes the amount $0.25$ .

A final remark is to  note that the eigenvalue
distributions for $D^2_a-c_2$ in figure \ref{disfuzzy} and $D_3$ in figure \ref{distmatrix}
clearly depend on $N$. It would be desirable to find
the proper scaling of the parameters for 
which the distributions are $N$-independent. This is also related 
with the predictions of the effective parameters and the corresponding 
effective potentials written in (\ref{gaussian}) and
(\ref{Veffmatrix}).

\subsection{Emergent geometry}
\begin{figure}[htbp!]
\begin{center}
\includegraphics[width=5.8cm,angle=-90]{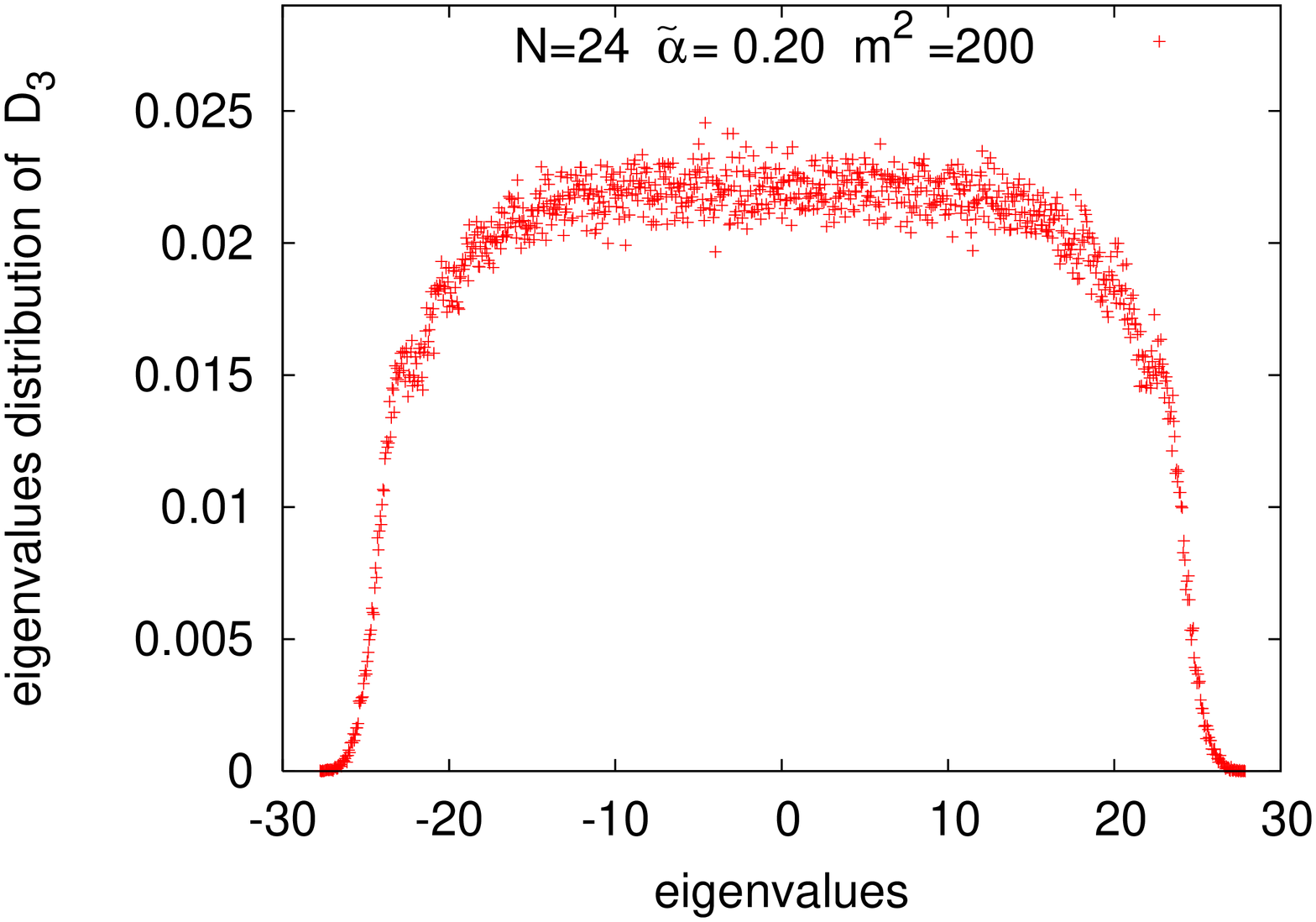}
\includegraphics[width=5.8cm,angle=-90]{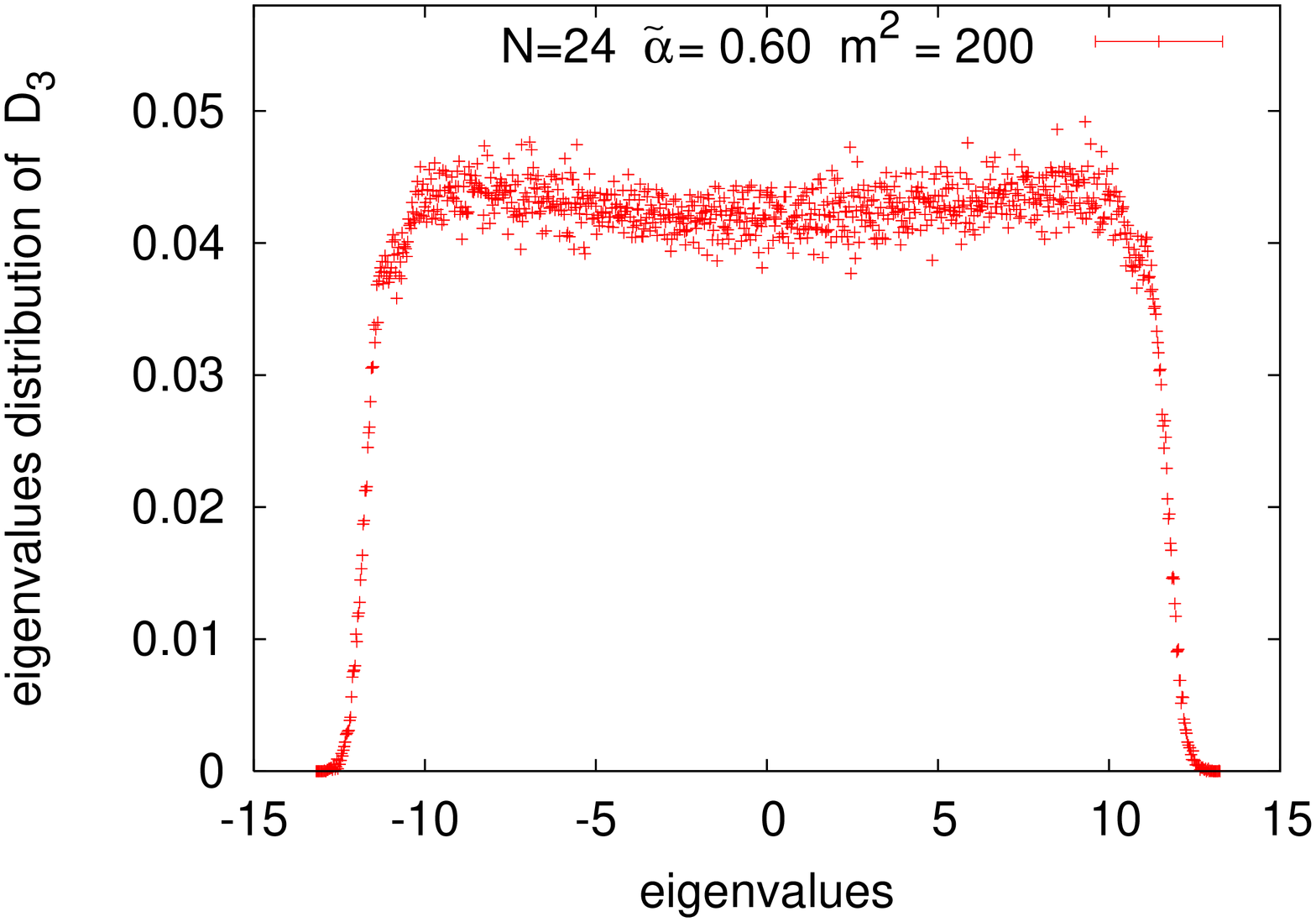}
\includegraphics[width=5.8cm,angle=-90]{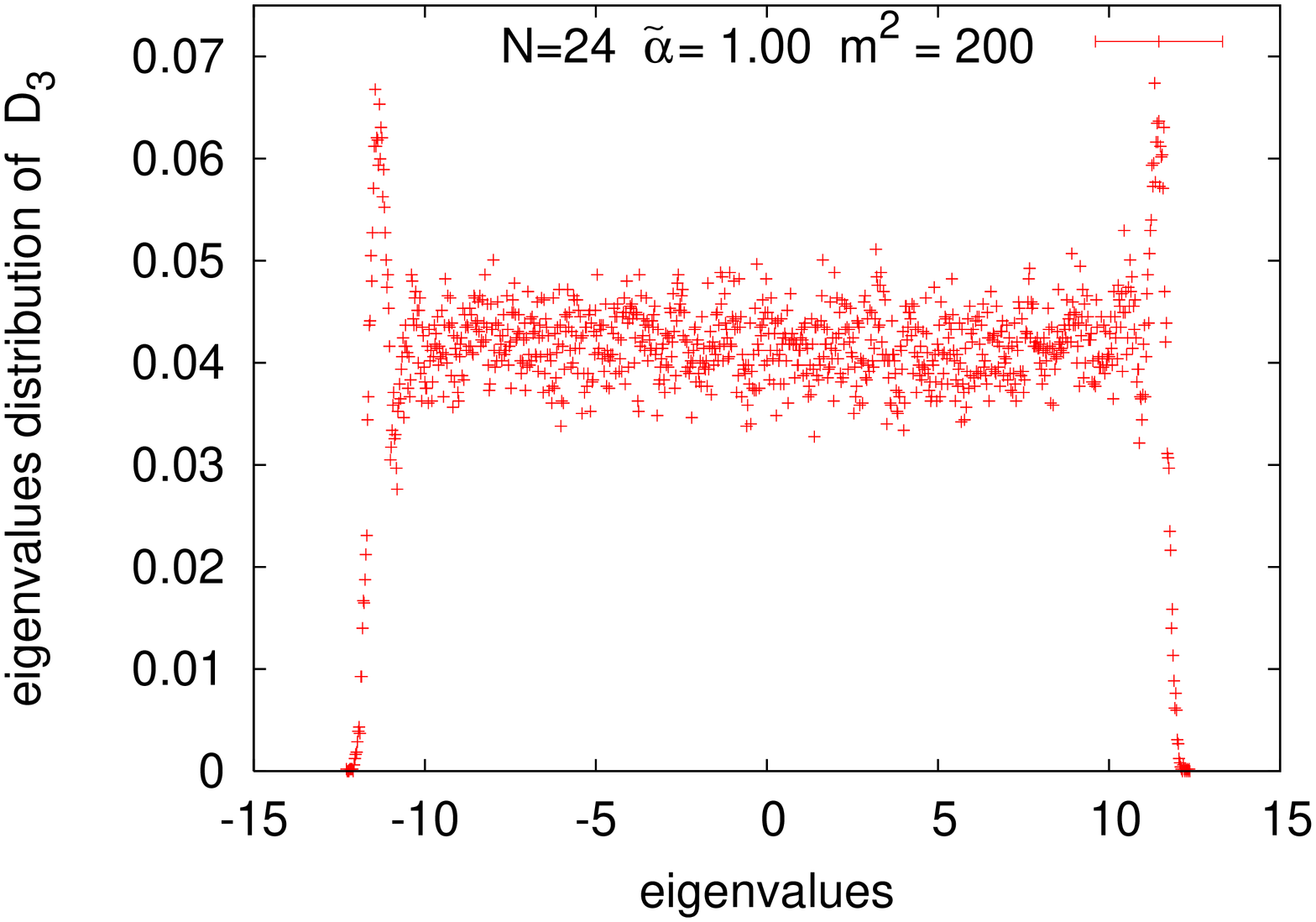}
\includegraphics[width=5.8cm,angle=-90]{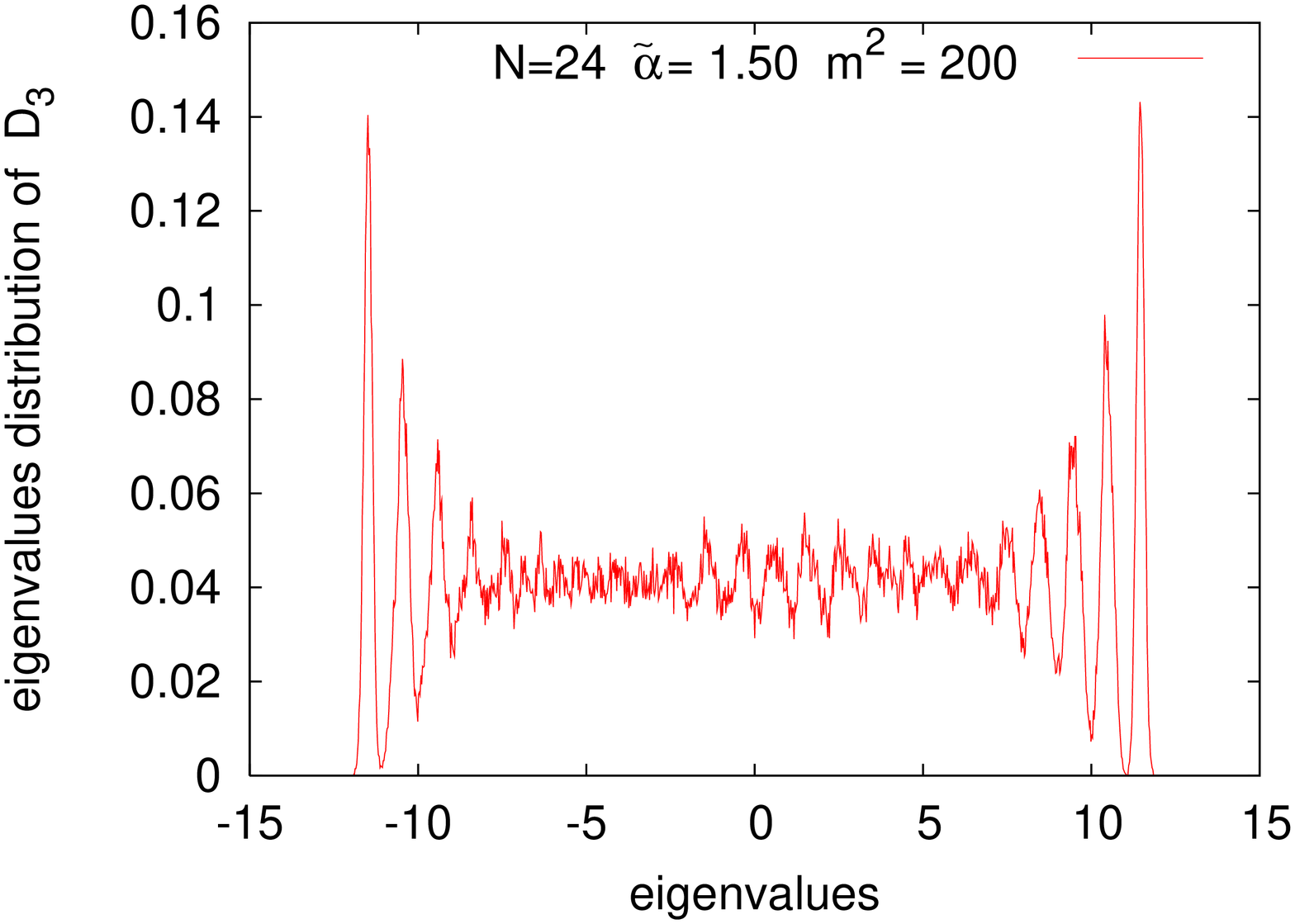}
\includegraphics[width=5.8cm,angle=-90]{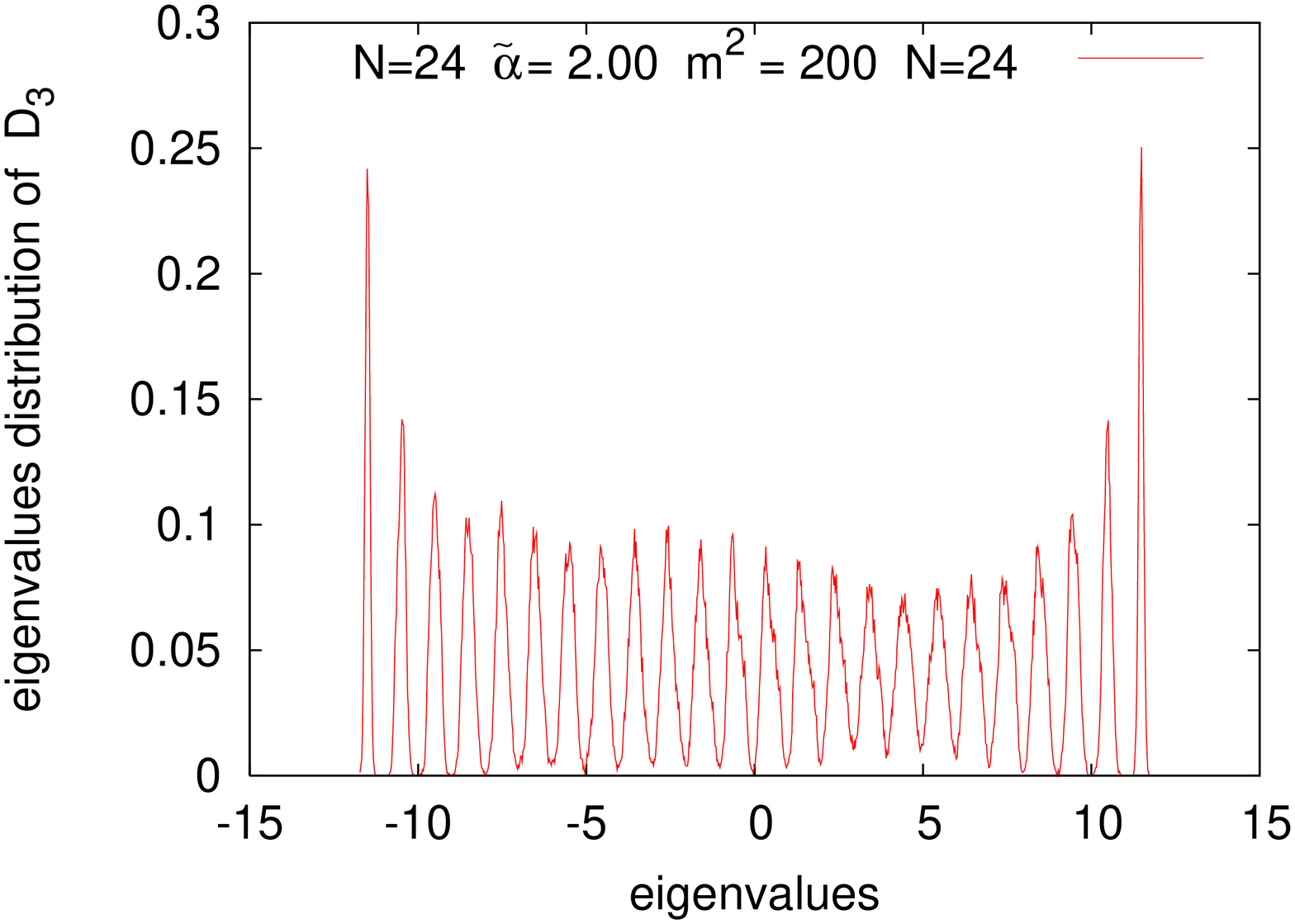}
\includegraphics[width=5.8cm,angle=-90]{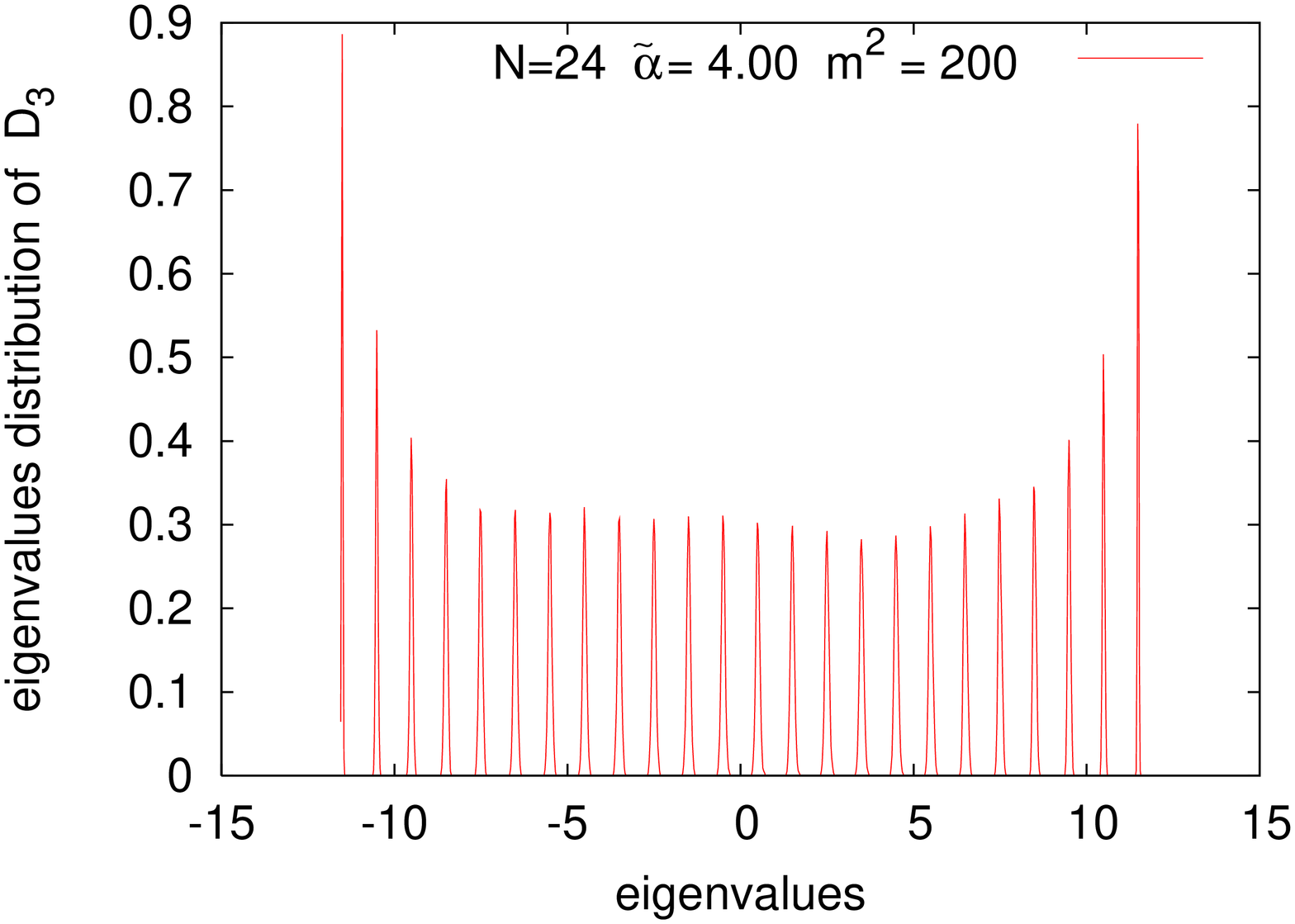}
\caption{The model undergoes a transition from the matrix phase with
  no underlying geometry to a geometrical phase as the coupling
  constant ${\tilde{\alpha}^4}=\frac{1}{T}$ is increased (or the
  temperature lowered). The geometry which emerges in this model is
  the geometry of the sphere. }\label{emerge}
\end{center}
\end{figure}

The sequence of eigenvalue distributions for
$\tilde{\alpha}=0.2,0.6,1,1.5,2$ and $4$ for $m^2=200$ and $N=24$
(figure \ref{emerge}) show clearly that a geometrical phase is
emerging as the coupling is increased or equivalently as the
temperature is reduced. The geometry that emerges here is that of the
fuzzy sphere. This geometry becomes the classical sphere as $N$ is
sent to infinity.  This is the geometry of the sphere emerging from a
pure matrix model and is in our opinion a very simple demonstration
of a novel concept and opens up the possibility of discussing emergent
geometry in a dynamical and statistical mechanical setting. We see
clearly that as $\tilde{\alpha}$ is increased from a value deep in the
matrix phase to a value well into the fuzzy sphere phase that $D_3$
goes, from a random matrix with a continuous eigenvalue distribution
centered around $0$, to a matrix whose eigenvalues are sharply
concentrated on the eigenvalues of the rotation generator $L_3$.  This
is also true for the matrices $D_1$ and $D_2$ which go over to $L_1$
and $L_2$ respectively in the fuzzy sphere phase. This can be seen
explicitly in figure \ref{comD} since the commutator $[D_1,D_2]$ is
found to be well approximated by the matrix $iL_3$.

\subsection{The model with only pure potential term}
\begin{figure}[h!]
 \begin{center}
\includegraphics[width=5.6cm,angle=-90]{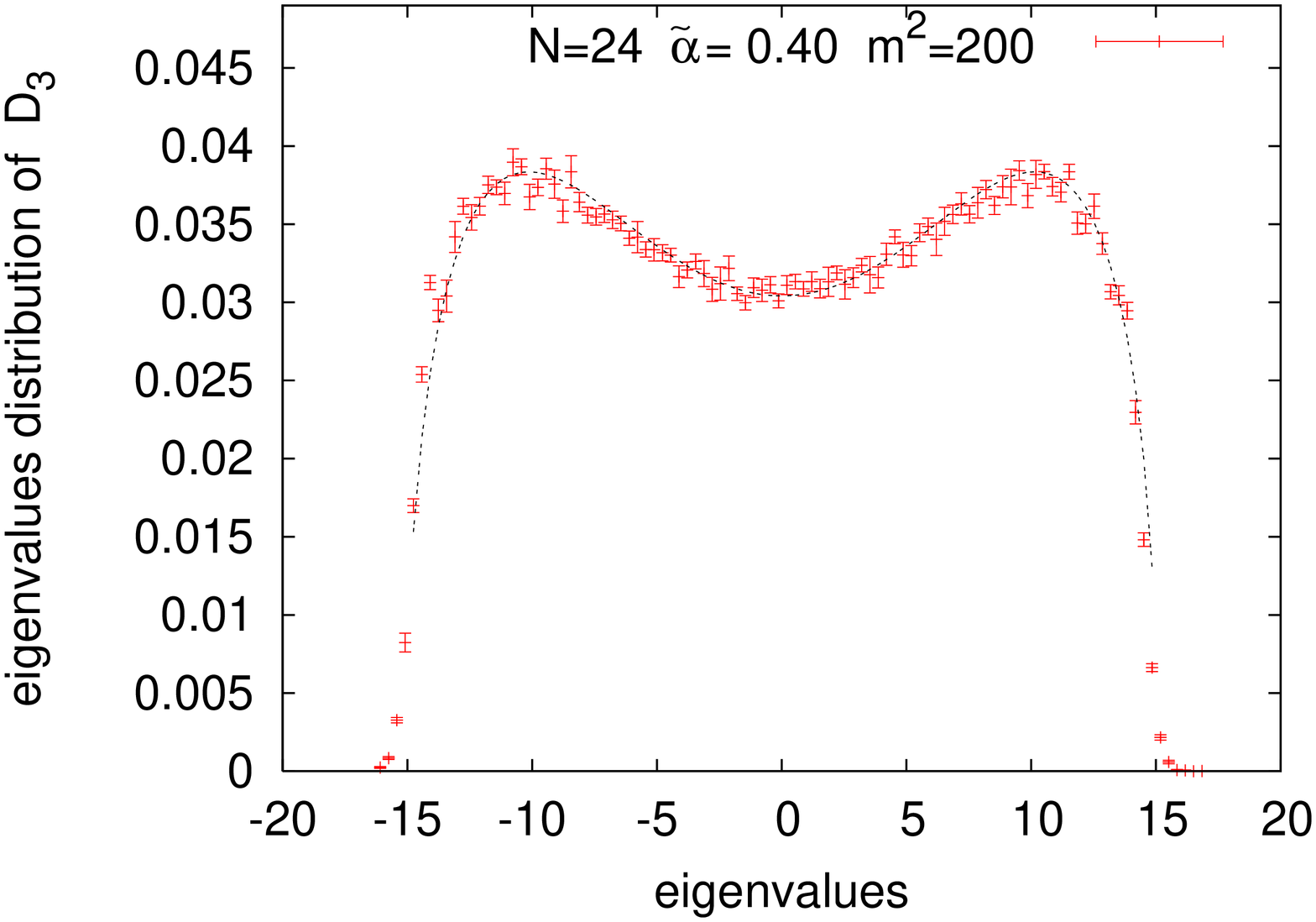}
\includegraphics[width=5.6cm,angle=-90]{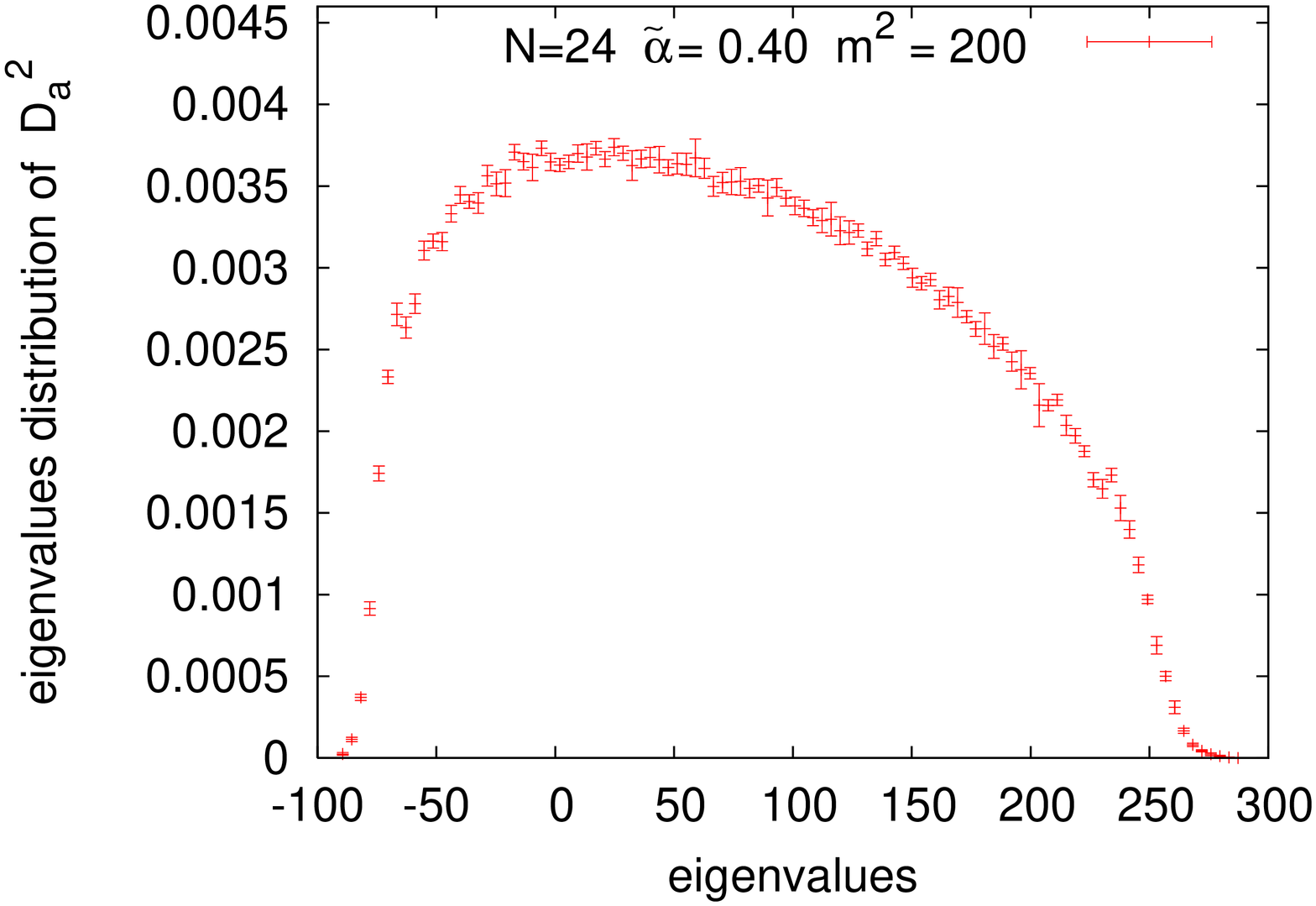}
\includegraphics[width=5.6cm,angle=-90]{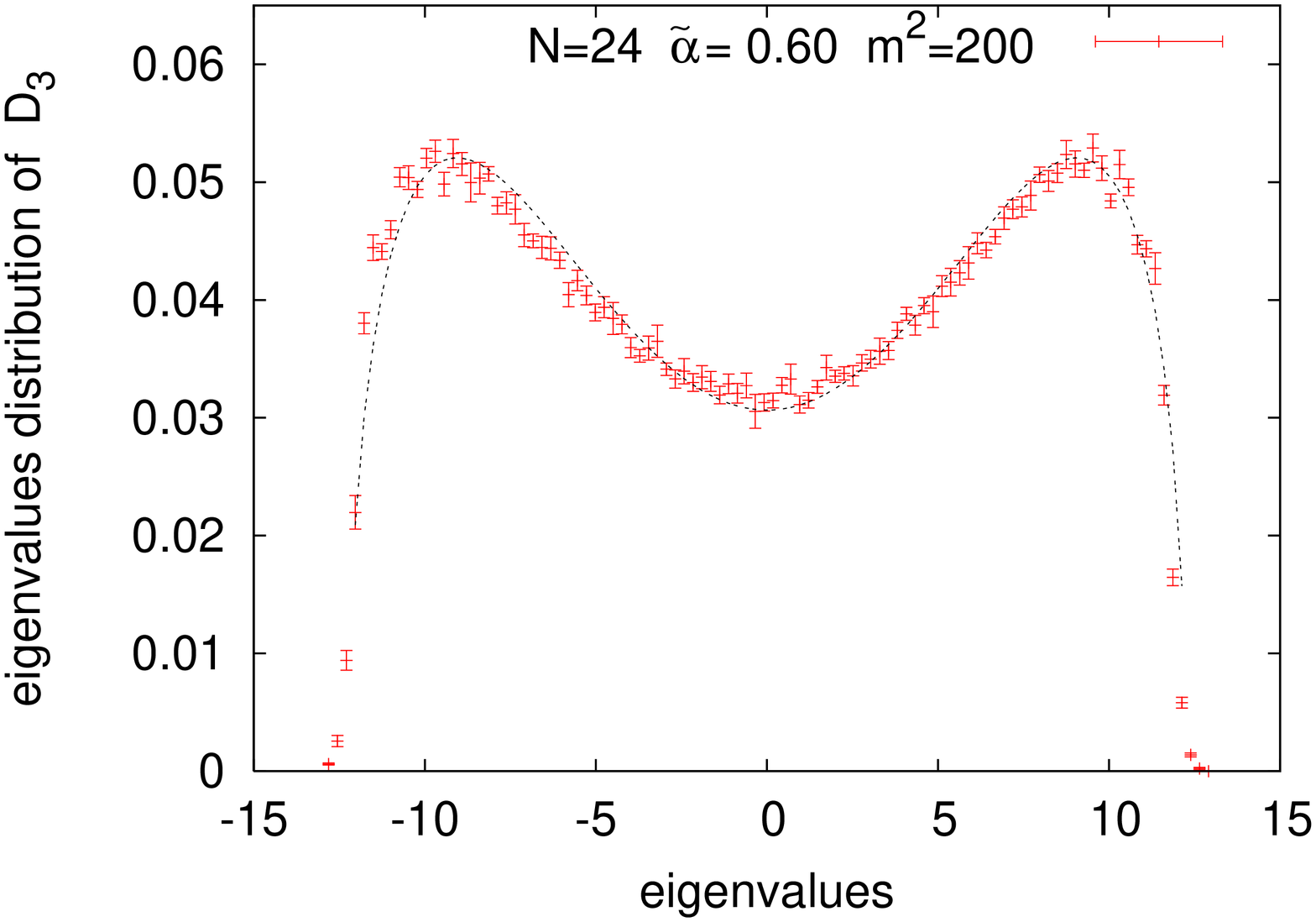}
\includegraphics[width=5.6cm,angle=-90]{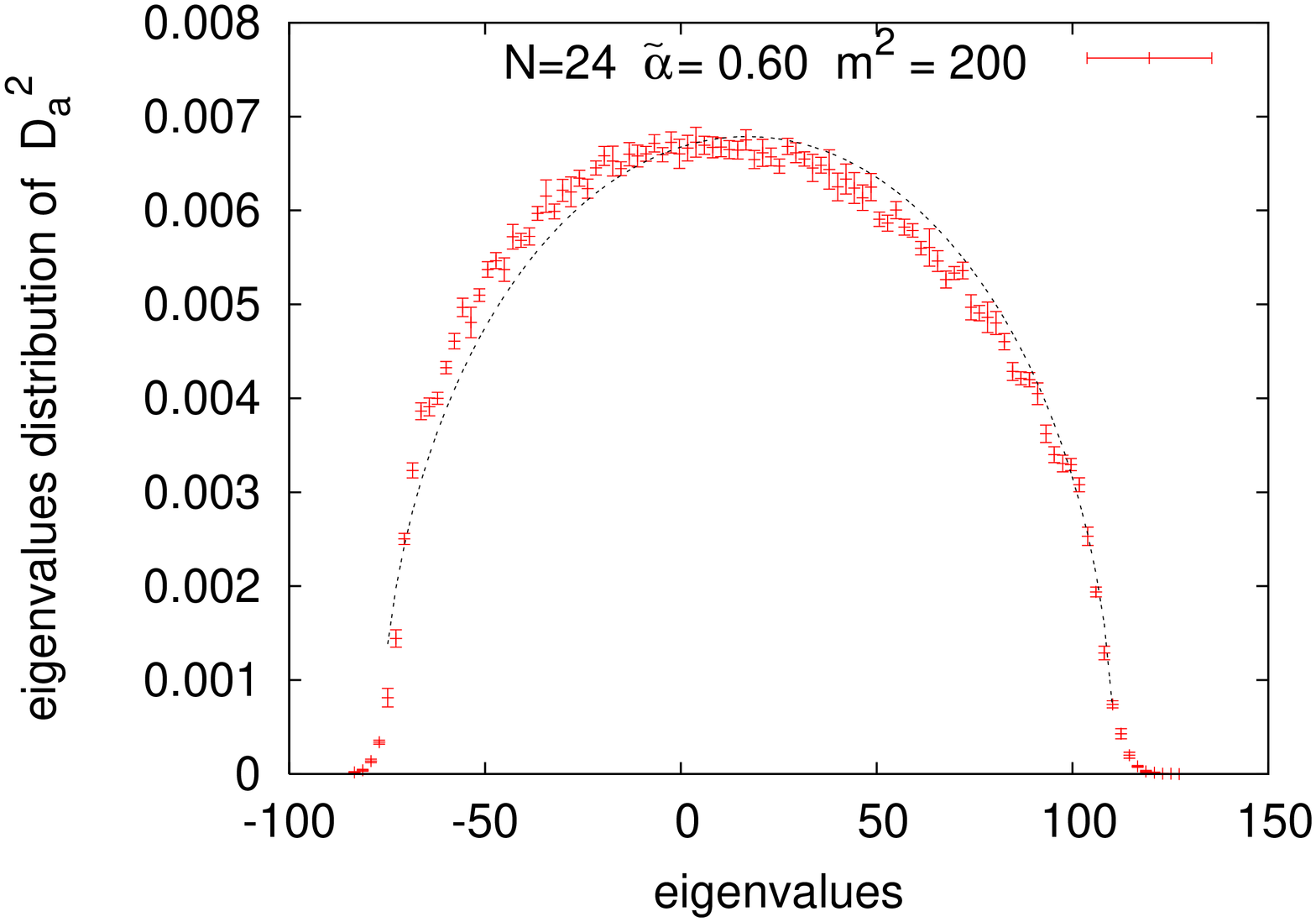}
\includegraphics[width=5.6cm,angle=-90]{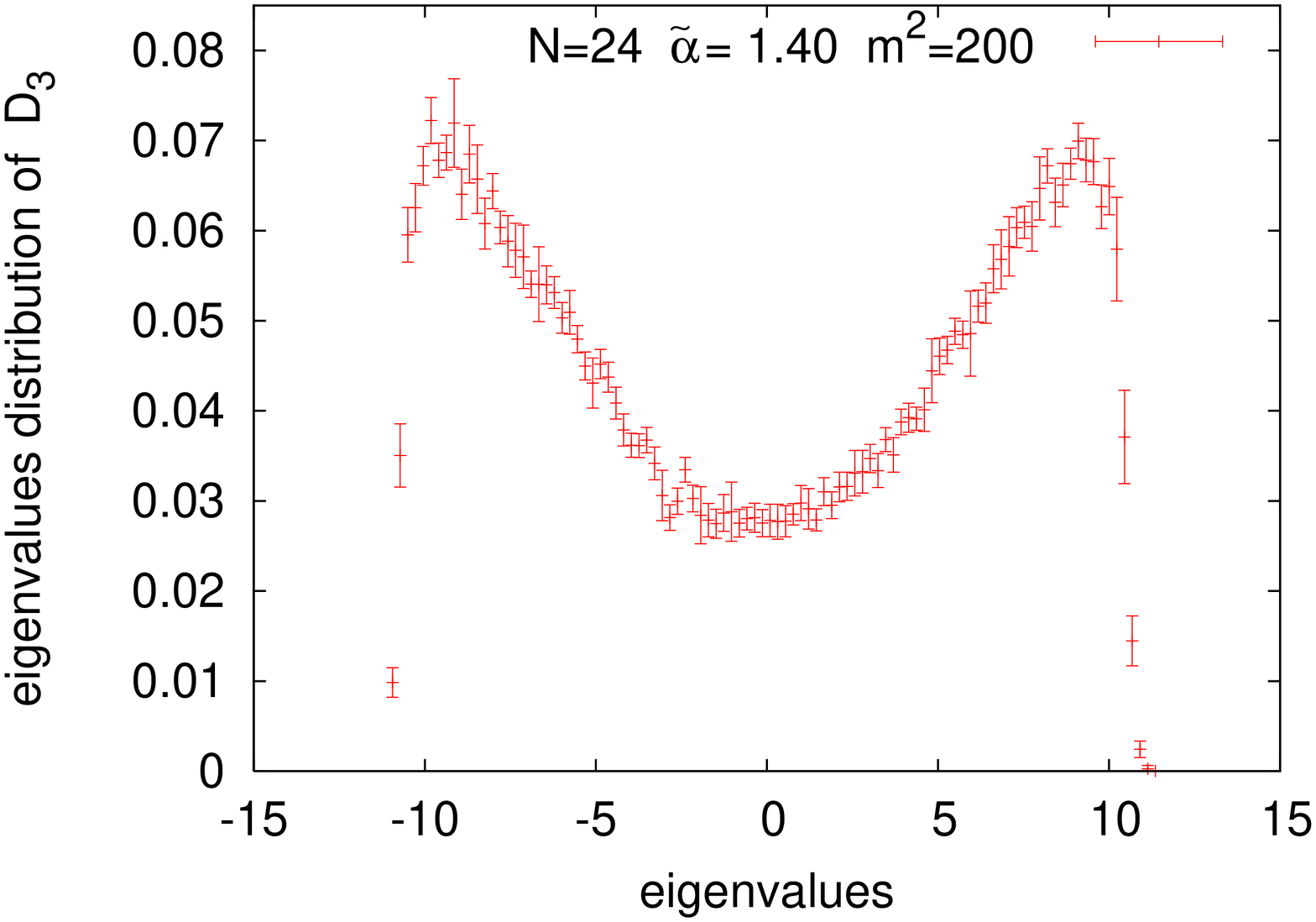}
\includegraphics[width=5.6cm,angle=-90]{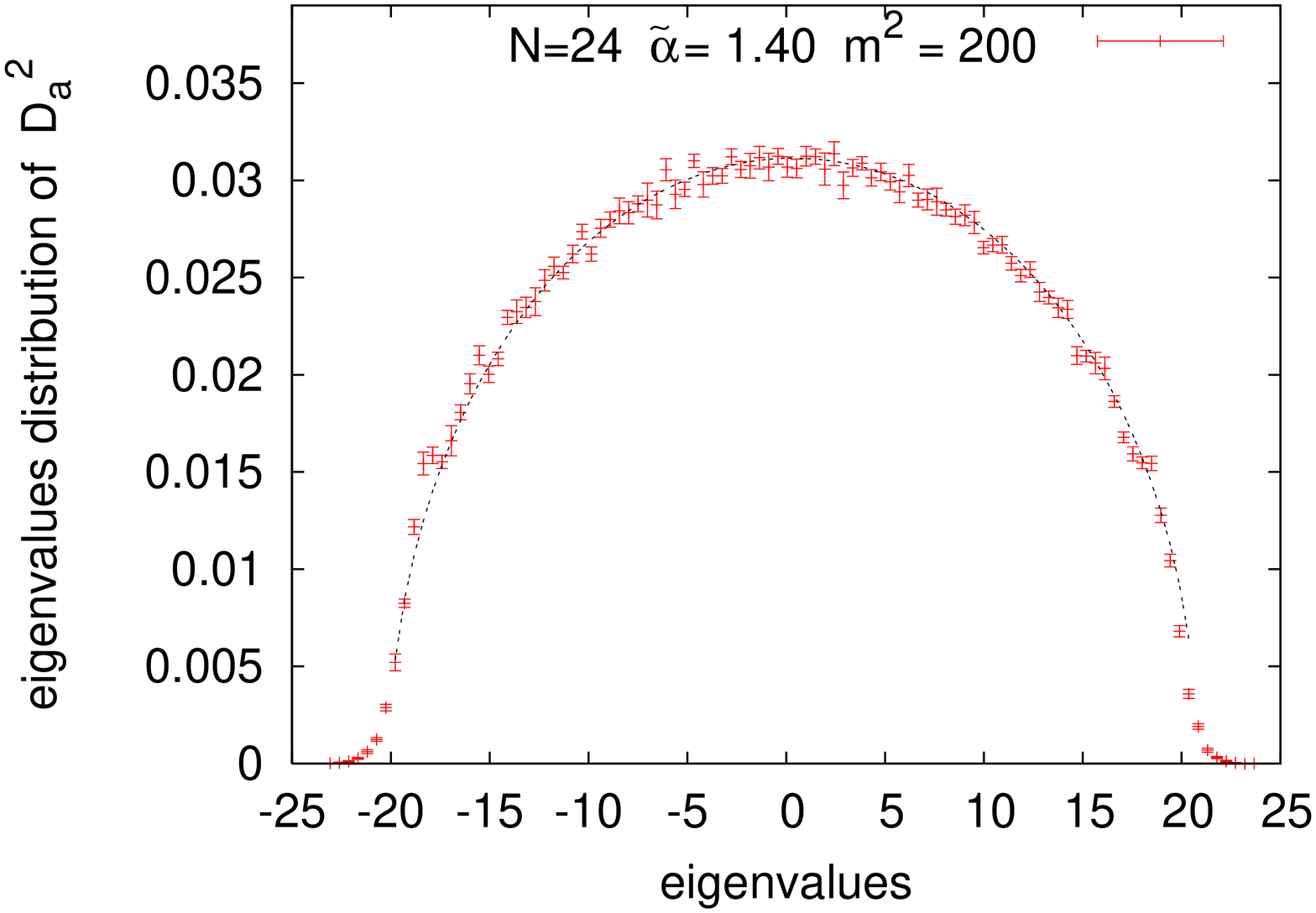}
\caption{The eigenvalue distributions of $D_a^2$ and $D_3$ for the
  pure potential term. Deep inside the nonperturbative regime (small
  $\tilde{\alpha}$ or large temperature) the distribution
  for $D_3$ is in the one-cut solution 
  (\ref{rhoplusgauge}) whereas for large $\tilde{\alpha}$ (or low
  temperature) the eigenvalue distribution for $D^2_a$ is given by the
  Wigner semi-circle law.}\label{potevo}
\end{center}
\end{figure}

The model is given by the potential
\begin{eqnarray}
V[D]=\frac{\tilde{\alpha}^4m^2}{N} Tr \big\{-D^2_a+\frac{1}{2c_2}(D^2_a)^2\big\}
\end{eqnarray}
The configurations which minimize this potential are 
matrices which satisfy $D_a^2=c_2$. We plot (figure \ref{potevo}) the
eigenvalue distributions of the matrices ${D_3}$ and $D^2_a$ for
$N=24$, $m^2=200$ and $\tilde{\alpha}=0.4,0.6,1.4$. The distributions
of $D_a^2$ seem to behave in the same way as the distributions of
$D_a^2$ computed in the full model for all values of $\tilde{\alpha}$.
The distributions of $D_3$ are similar to the corresponding
distributions of $D_3$ computed in the full model only in the region
of the matrix phase where they are found to fit to the one-cut
solution (\ref{4.5}). In the region of the fuzzy sphere phase the
distributions of $D_3$ tend to split into two cuts.  However they
never achieve this splitting completely due to the mixing terms
$TrD_1^2D_2^2$, $TrD_1^2D_3^2$ and $TrD_2^2D_3^2$. These terms 
cannot be neglected for values of the coupling $\tilde{\alpha}$
where the model is in the fuzzy sphere phase.

\subsection{The Chern-Simons+potential model}
In this final section we present the eigenvalue distributions for the
model in which we set the Yang-Mills term to zero.  The action reduces
to
\begin{eqnarray}
S[D]=\frac{1}{g^2N}Tr\bigg[\frac{i}{3}\epsilon_{abc}[D_a,D_b]D_c+\frac{m^2}{2c_2}(D^2_a-c_2)^2\bigg].
\end{eqnarray}
The equations of motion are 
\begin{eqnarray}
i\epsilon_{abc}[D_b,D_c]-2m^2D_a+\frac{m^2}{c_2}\{D^2_c,D_a\}=0.
\end{eqnarray}
The possible solutions are the commuting matrices and reducible
representations of $SU(2)$.

\begin{figure}[h!]
 \begin{center}
\includegraphics[width=70mm,angle=-90]{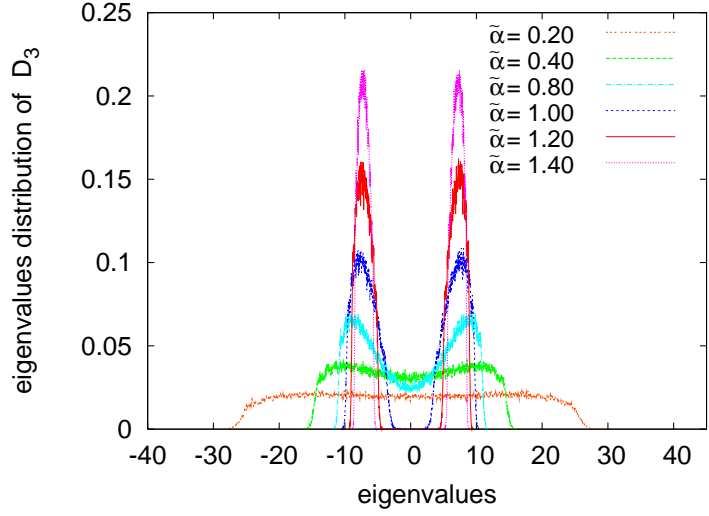}
\caption{The eigenvalue distributions of $D_3$ for the Chern-Simons +
  Potential model for different values of the coupling
  $\tilde{\alpha}$ for $N=24$ and $m^2=200$. In this case, the system
  undergoes a transition from the one-cut regime to the two-cut regime
  as the coupling (or temperature) is increased
  (lowered).}\label{disYM0}
\end{center}
\end{figure}

\begin{figure}[h!]
 \begin{center}
\includegraphics[width=5.8cm,angle=-90]{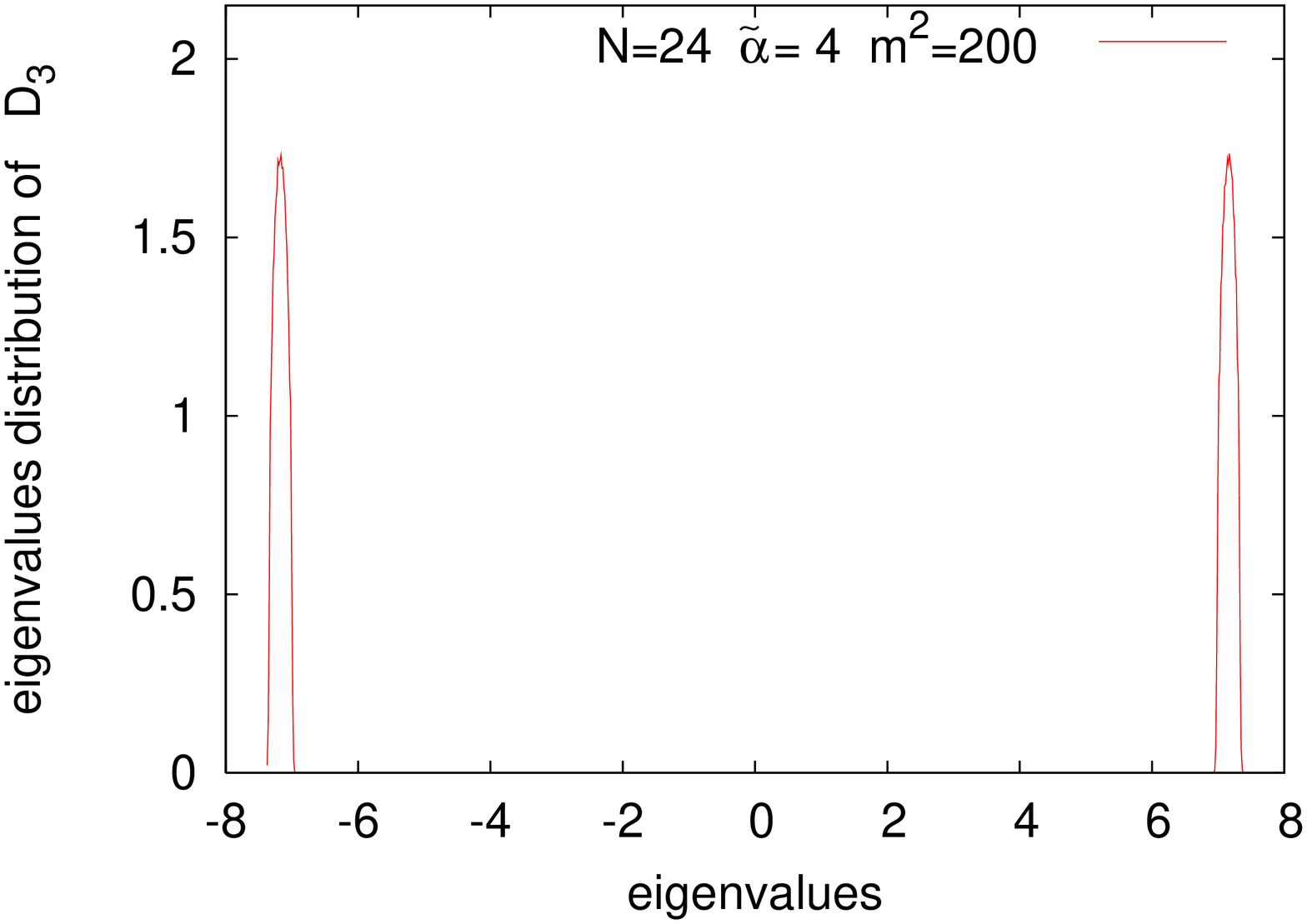}
\includegraphics[width=5.8cm,angle=-90]{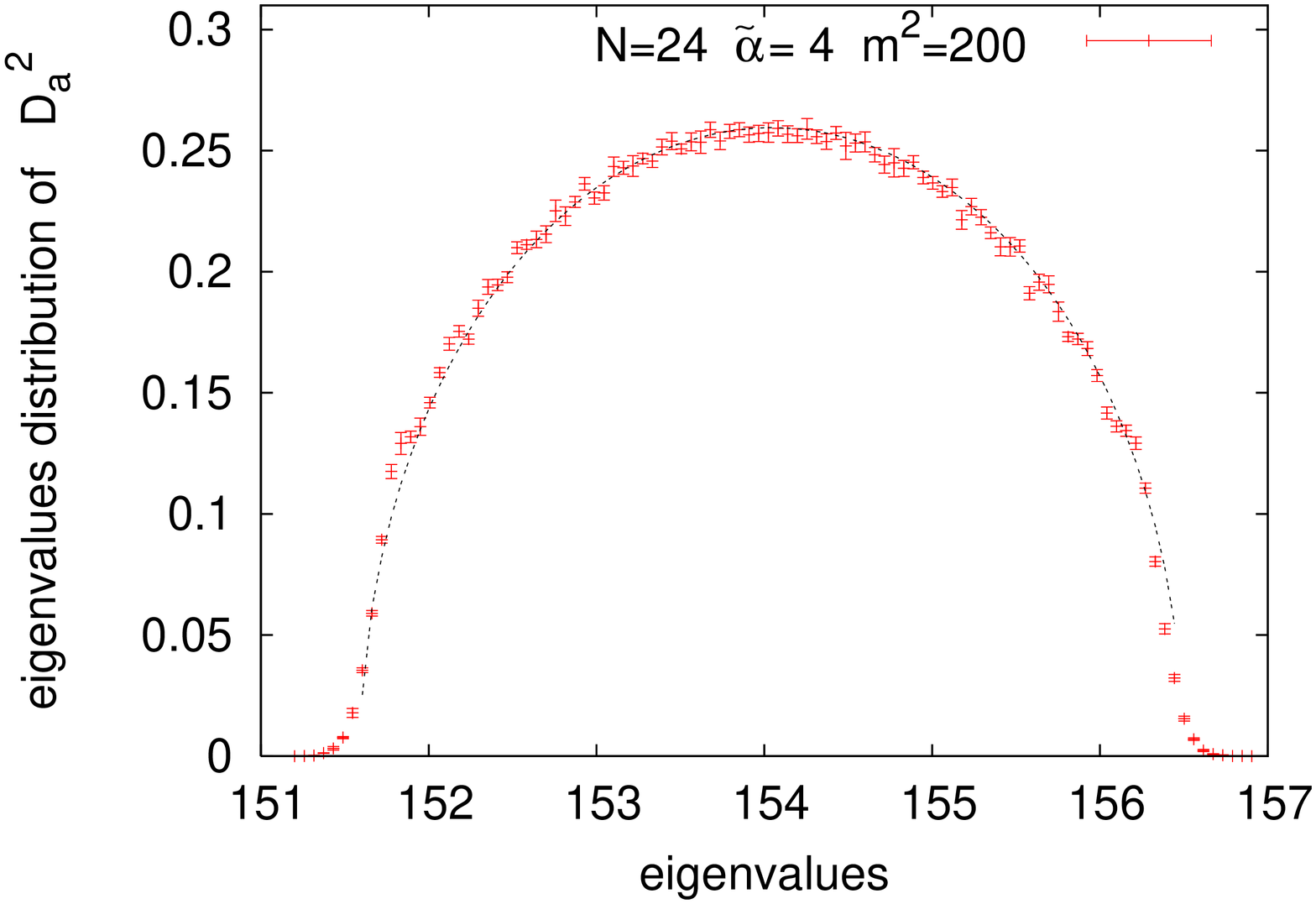}
\includegraphics[width=5.8cm,angle=-90]{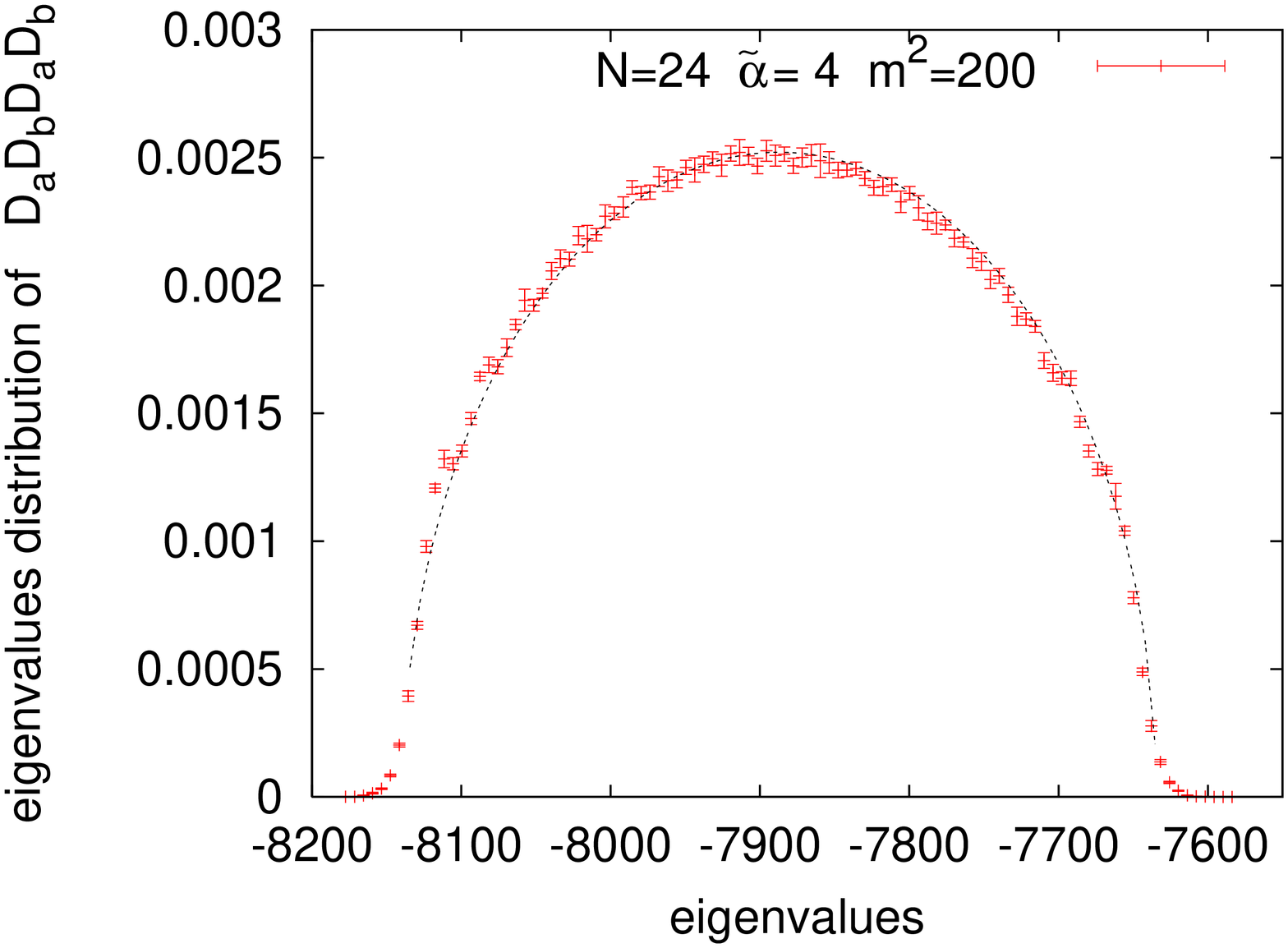}
\includegraphics[width=5.8cm,angle=-90]{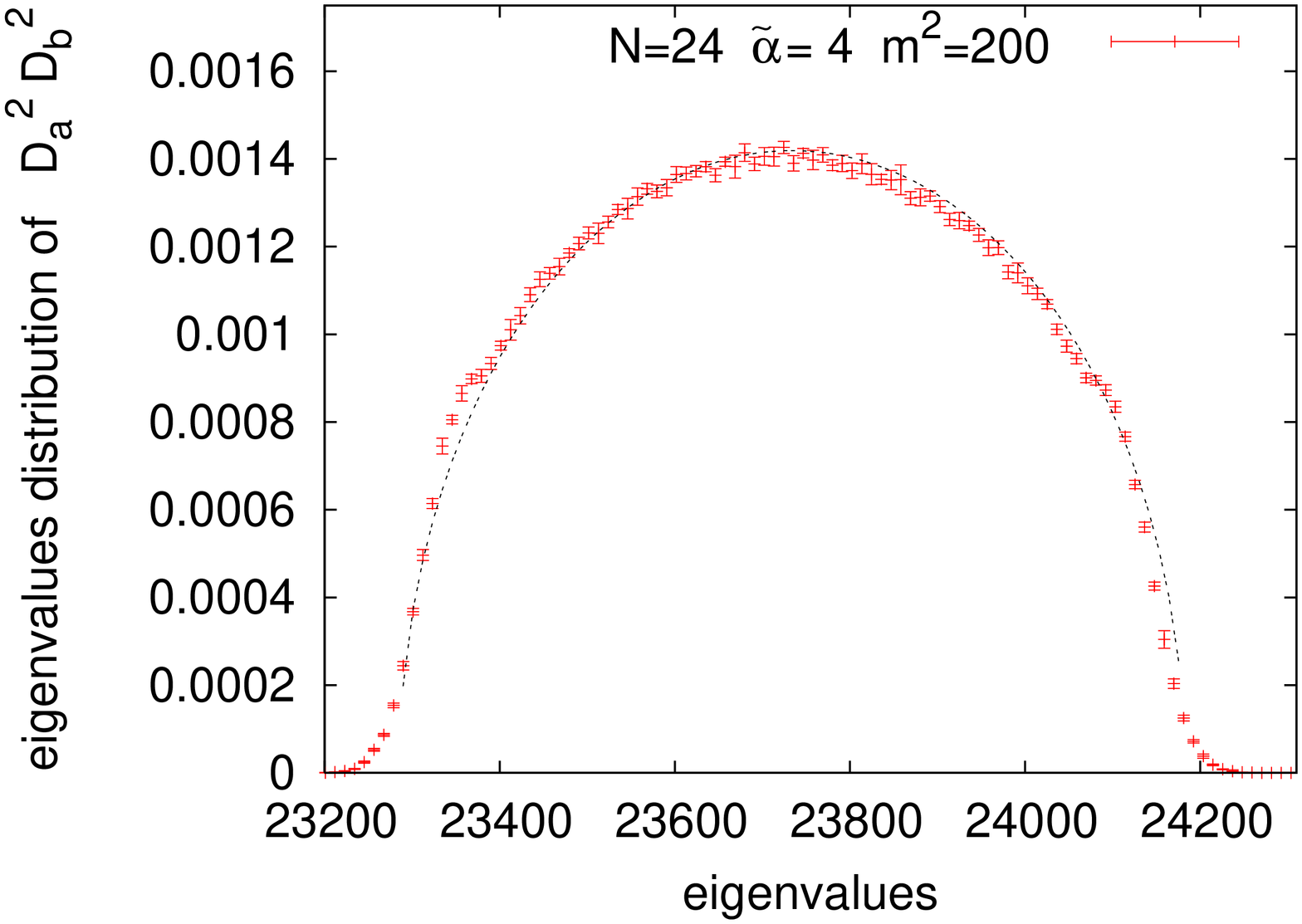}
\includegraphics[width=5.8cm,angle=-90]{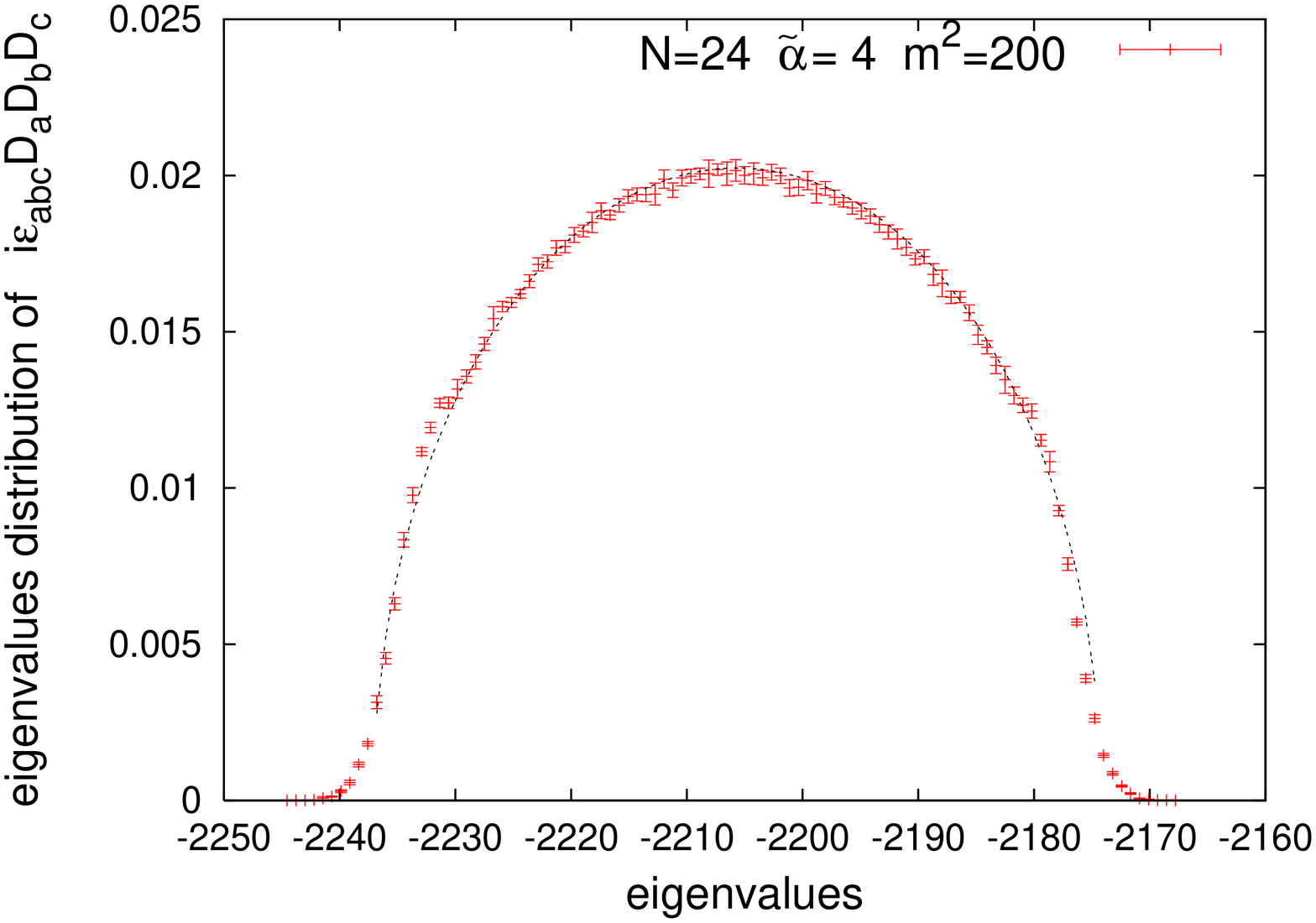}
\caption{The eigenvalue distributions of  $D_3$, ${D^2_a}$,
  $D_aD_bD_aD_b$, $D_aD_bD_bD_a$, ${i\epsilon_{abc}D_aD_bD_c}$ for the
  Chern-Simons+potential model. The solid lines correspond to the
  Wigner semicircle laws with centers given by eqs.(\ref{centers}).}\label{obsYM0}
\end{center}
\end{figure}

We measure the eigenvalue distributions of the matrix $D_3$ for
different values of $\tilde{\alpha}$. The distributions of $D_3$ in
the region of the matrix phase can be fit to the one-cut solution
(\ref{4.5}). In the region of the fuzzy sphere phase the distributions
of $D_3$ split into two well separated cuts. See figure
\ref{disYM0}. This model exhibits therefore the behaviour of a typical
quartic one-matrix model. The value of the coupling $\tilde{\alpha}$
where the transition from the one-cut phase to the two-cut phase
happens coincides with the maximum of the specific heat. From the
numerical results we can also see that in the regime of large
$\tilde{\alpha}$ the preferred configurations are given by
\begin{eqnarray}\label{confYM0}
D_a=2\lambda \frac{\sigma_a}{2}\otimes {\bf 1}_{\frac{N}{2}}.
\end{eqnarray}
$\lambda$ is fixed by the requirement $D^2_a=c_2$, i.e 
$3{\lambda}^2=c_2$ and $\sigma_a$ are the Pauli matrices. 
Indeed the eigenvalue distributions of the matrices ${D^2_a}$,
$D_aD_bD_aD_b$, $D_aD_bD_bD_a$ and ${i\epsilon_{abc}D_aD_bD_c}$
for large $\tilde{\alpha}$ are found (figure \ref{obsYM0}) to be
given by the Wigner semi circle laws
\begin{eqnarray}
\rho(x)=\frac{2}{a^2\pi}\sqrt{a^2-(x-x_0)^2}.
\end{eqnarray} 
The centers $x_0$ are given by the theoretical values
\begin{eqnarray}\label{centers}
D^2_a & = & 3\lambda^2{\bf 1}_N \nonumber \\
D_aD_bD_aD_b & = & -3\lambda^4{\bf 1}_N \nonumber \\
D_aD_bD_bD_a & = & 9\lambda^4 {\bf 1}_N\nonumber \\
i\epsilon_{abc}D_aD_bD_c & = & -6\lambda^3{\bf 1}_N.
\end{eqnarray}
For instance in figure \ref{obsYM0} we can see from the eigenvalue distribution of $D_3$ with 
$N=24$, $\tilde{\alpha}=4$ and $m^2=200$ that the preferred
configurations are given by (\ref{confYM0}) with $\lambda=\pm(7.16\pm 0.25)$ whereas  the
predicted value is $\lambda=\pm 6.9567$. We also remark that the eigenvalue distributions  of $D^2_a$ and the other operators have the same  structure as in the full
model in its matrix phase. 
\section{Conclusion and outlook}

We have studied the simple three matrix model with 
Euclidean action functional (\ref{Action-simpleform})
for general values of its parameters $\beta=\tilde{\alpha}^4$ and $m$ 
but focused on a small range of the possible values of the parameter $\mu$.

We find the model to have two clearly distinct phases. In the high
temperature regime (i.e. small $\tilde{\alpha}$) the model has a disordered 
phase. In this phase the eigenvalues distribution of an individual matrix is well approximated by the one-cut distribution of a hermitian matrix model with quartic potential. We call this phase the {\it matrix phase} of the model. 



At low temperature the model has an ordered phase. The order is unusual 
in that it describes the condensation of a background geometry as a collective
order of the matrix degrees of freedom. We call this the 
{\it geometrical} or {\it fuzzy sphere phase}, since for finite
matrix size the ground state is described by a fuzzy sphere which is 
a quantized version of the classical sphere. In the large matrix size limit 
the sphere becomes classical but at a microscopic level
the geometry always has a noncommutative character as can be seen from
the spectrum of the ``coordinate functions'' which are proportional
to the $su(2)$ generator $L_3$ in the irreducible representation of 
dimension given by the matrix size.

In the geometrical phase small fluctuations are those of a $U(1)$
gauge field and a neutral scalar field fluctuating on a round two
sphere. The two fields have non-trivial mixing at the quadratic level
but are otherwise not interacting. In this phase, with 
$\mu=m^2$, the parameter $m^2$ parameterizes the mass of the 
scalar fluctuation, otherwise for $\mu=m^2-\rho/N$, the parameter $\rho$
provides a constant external current for the scalar field.  

For $m=\mu=0$ the transition between the two phases is found to be
discontinuous. There is a jump in the internal energy (expectation of
the Euclidean action) and from a theoretical analysis (valid in the
fuzzy sphere phase) we find the entropy drops by $1/9$ per degree of
freedom \cite{emergent-geometry-prl} as one crosses from the high
temperature matrix phase to the geometrical one. Our theoretical
results also predict that: For all $m$ and $\mu>-\frac{2}{9}$, the
model has divergent critical fluctuations in the specific heat
characterised by the critical exponent $\alpha=1/2$.  The critical
regime narrows as the critical temperature decreases and the
transition temperature is sent to zero at $\mu=-\frac{1}{4}$ and so
there is no geometrical phase beyond this point. 

Our numerical simulations are in excellent agreement with these
theoretical predictions and we find the critical fluctuations are only
present in the fuzzy sphere phase so that the transition has an
asymmetrical character. We know of no other physical setting that
exhibits transitions of the type presented here and further numerical
and theoretical study are needed. However, the thermodynamic properties of the transition are similar to those found in the $6-$vertex model and the dimer model \cite{Nash:2008dk, Lieb}. 

We focus simulations on $\mu=m^2$ and observe that the discontinuity
in the entropy (or the latent heat) decreases as $m^2$ is increased
with the jump vanishing for sufficiently large $m$. We have not been
able to determine with precision where the transition becomes
continuous, however the jump in the entropy becomes too small to
measure beyond $m^2\sim 40$.  Also, the predicted critical
fluctuations are not seen in the numerical results for large $m^2$.

The data for the critical point coupling, from our different methods
of estimating it, separate in the parameter range where the latent
heat disappears indicating a possible multi-critical point or richer
structure. For larger values of $m^2=\mu$ we have not been able to
resolve the nature of the transition. The numerical evidence shows the
structure of a 3rd order transition, a behaviour typical of many
matrix models, however, the fact that $\tilde{\alpha}_s$, the crossing
point of the average action curves for different $N$, still reliably
predicts the transition line, suggests the transition is continuous
with asymmetric critical fluctuations, consistent with the theoretical
analysis.

Our conclusion is that the full model has the qualitative features of
the model with $m=\mu=0$ (model $S_0$ of eq. (\ref{Action-S0})) and
that the effect of the potential is to shift the transition
temperature and provide a non-trivial background specific heat. The
divergence of the specific heat arises from the interplay of the
Chern-Simons and Yang-Mills terms.  Given that this competition leads
to a divergent specific heat, we expect, sufficiently close to the
transition, to see the effect of this competition emerge and the
specific heat to eventually rise above the background provided by the
potential and diverge as the critical point.

The model of emergent geometry described here, though reminiscent of
the random matrix approach to two dimensional gravity
\cite{Ambjorn:2006hu} is in fact very different.  The manner in which
spacetime emerges is also different from that envisaged in string
pictures where continuous eigenvalue distributions
\cite{Seiberg:2006wf} or a Liouville mode \cite{Knizhnik:1988ak} give
rise to extra dimensions.  It is closely connected to the $D0$ brane
scenario described in \cite{Myers:1999ps} and the $m=0$ version is a
dimensionally reduced version of a boundary WZNW models in the large
$k$ limit \cite{Alekseev:2000fd}.  A two matrix model where the large $N$ limit describes a hemispherical geometry was studied in  \cite{Berenstein:2008eg}. It is not difficult to invent higher
dimensional models with essentially similar phenomenology to that
presented here (see \cite{Dou:2007in}, \cite{Steinacker:2007dq} and
\cite{Kawahara:2007nw}). For example any complex projective space
${\bf CP}^N$ can emerge from pure matrix dynamics by choosing similar
matrix models with appropriate potentials \cite{Dou:2007in}.

{\it In summary}, we have found an exotic transition in a simple three
matrix model. The nature of the transition is very different if
approached from high or low temperatures. The high temperature phase is
described by three decoupled random matrices with self interaction so
there is no background spacetime geometry. As the system cools a
geometrical phase condenses and at sufficiently low temperatures the
system is described by small fluctuations of a $U(1)$ gauge field
coupled to a massive scalar field.  The critical temperature is pushed
upwards as the scalar field mass is increased.  Once the geometrical
phase is well established the specific heat takes the value $1$ with
the gauge and scalar fields each contributing $1/2$. 

We believe that this scenario gives an appealing picture of how a
geometrical phase might emerge as the system cools and suggests a very
novel scenario for the emergence of geometry in the early universe. In
such a scenario the temperature can be viewed as an effect of other
degrees of freedom present in a more realistic model but not directly
participating in the transition we describe.  In the model described
in detail above, both the geometry and the fields are emergent
dynamical concepts as the system cools. Once the geometry is well
established the background scalar decouples from the rest of the
physics and is always massive. If a realistic cosmological model the
can be found, such decoupled matter should provide a natural candidate
for dark matter.

\paragraph{Acknowledgements}
The work of B.Y is supported by a Marie Curie Fellowship from The
Commission of the European Communities under contract number
MIF1-CT-2006-021797. The work of R.D.B. is supported by CONACYT
M\'exico. R.D.B would also like to thank the Institute f\"ur Physik,
Humboldt-Universit\"at zu Berlin for their hospitality and support
while this work was in progress. In particular he would like to thank
Prof. Michael M\"uller-Preussker and Mrs. Sylvia Richter.


\begin{thebibliography}{99}

\bibitem{connes}
A. Connes, {\it Noncommutative Geometry}, Academic Press, London,1994. 

\bibitem{Bombelli:1987aa}
  L.~Bombelli, J.~H.~Lee, D.~Meyer and R.~Sorkin,
  Phys.\ Rev.\ Lett.\  {\bf 59} (1987) 521.

\bibitem{Seiberg:2006wf}
  N.~Seiberg, ``Emergent spacetime,''
 {\it  23rd Solvay Conference In Physics: The Quantum Structure Of Space And Time}
Edited by D. Gross, M. Henneaux, A. Sevrin. Hackensack, World Scientific, 2007.
  arXiv:hep-th/0601234.

\bibitem{Ambjorn:2006hu}
  J.~Ambj\o rn, R.~Janik, W.~Westra and S.~Zohren,
  Phys.\ Lett.\  B {\bf 641} (2006) 94
  [arXiv:gr-qc/0607013].

\bibitem{Azuma:2004zq}
  T.~Azuma, S.~Bal, K.~Nagao and J.~Nishimura,
  JHEP {\bf 0405} (2004) 005
  [arXiv:hep-th/0401038].

\bibitem{CastroVillarreal:2004vh}
  P.~Castro-Villarreal, R.~Delgadillo-Blando and B.~Ydri,
  Nucl.\ Phys.\  B {\bf 704} (2005) 111
  [arXiv:hep-th/0405201].

\bibitem{O'Connor:2006wv}
  D.~O'Connor and B.~Ydri,
  JHEP {\bf 0611} (2006) 016
  [arXiv:hep-lat/0606013].


\bibitem{Ydri:2001pv}
  B.~Ydri,
  arXiv:hep-th/0110006.

\bibitem{O'Connor:2003aj}
  D.~O'Connor,
  Mod.\ Phys.\ Lett.\  A {\bf 18} (2003) 2423.

\bibitem{Balachandran:2005ew}
  A.~P.~Balachandran, S.~K\"{u}rk\c{c}\"{u}o\v{g}lu and S.~Vaidya,
  arXiv:hep-th/0511114.

\bibitem{Grosse:1996mz}
  H.~Grosse, C.~Klim\v{c}\'{\i}k and P.~Pre\v{s}najder,
  Commun.\ Math.\ Phys.\  {\bf 180} (1996) 429
  [arXiv:hep-th/9602115].

\bibitem{Myers:1999ps}
  R.~C.~Myers,
  JHEP {\bf 9912} (1999) 022
  [arXiv:hep-th/9910053].

\bibitem{Alekseev:2000fd}
  A.~Y.~Alekseev, A.~Recknagel and V.~Schomerus,
  JHEP {\bf 0005} (2000) 010
  [arXiv:hep-th/0003187].

\bibitem{HoppeMadore} J. Hoppe, MIT Ph.D. Thesis, (1982). 
J. Madore, {\it Class. Quantum. Grav.} { 9} (1992) 69.

\bibitem{emergent-geometry-prl}
  Rodrigo~Delgadillo-Blando, Denjoe~O'Connor and Badis~Ydri,
  ``Geometry in transition: A model of emergent geometry,''
  Phys.\ Rev.\ Lett.\  {\bf 100} (1987) 201601
  arXiv:0712.3011 [hep-th].

\bibitem{Steinacker:2003sd}
  H.~Steinacker,
  Nucl.\ Phys.\  B {\bf 679} (2004) 66
  [arXiv:hep-th/0307075].

\bibitem{Azuma:2004ie}
  T.~Azuma, K.~Nagao and J.~Nishimura,
  JHEP {\bf 0506} (2005) 081
  [arXiv:hep-th/0410263].

\bibitem{Steinacker:2007iq}
  H.~Steinacker and R.~J.~Szabo,
  arXiv:hep-th/0701041.

\bibitem{Dou:2007in}
  D.~Dou and B.~Ydri,
  Nucl.\ Phys.\  B {\bf 771} (2007) 167
  [arXiv:hep-th/0701160].

\bibitem{Kawahara:2007nw}
  N.~Kawahara, J.~Nishimura and S.~Takeuchi,
  JHEP {\bf 0705} (2007) 091
  [arXiv:0704.3183 [hep-th]].

\bibitem{Steinacker:2007dq}
  H.~Steinacker,
  arXiv:0708.2426 [hep-th].

\bibitem{Di Francesco:1993nw}
  P.~Di Francesco, P.~H.~Ginsparg and J.~Zinn-Justin,
  Phys.\ Rept.\  {\bf 254} (1995) 1
  [arXiv:hep-th/9306153].

\bibitem{Knizhnik:1988ak}
  V.~G.~Knizhnik, A.~M.~Polyakov and A.~B.~Zamolodchikov,
  Mod.\ Phys.\ Lett.\  A {\bf 3} (1988) 819.

\bibitem{Minwalla:1999px}
  S.~Minwalla, M.~Van Raamsdonk and N.~Seiberg,
  JHEP {\bf 0002} (2000) 020
  [arXiv:hep-th/9912072].

\bibitem{FroehlichGawedzki}
J.~Fr\"{o}hlich and  K.~Gaw\c{e}dzki,
{\sl Conformal Field Theory and Geometry of Strings},
Lectures given at Mathematical Quantum Theory Conference, 
Vancouver, Canada, 4-8 Aug 1993. 
Published in Vancouver 1993, 
Proceedings, Mathematical quantum theory, {\bf Vol. 1} 57-97,
[arXiv:hep-th/9310187].


\bibitem{Grosse:1992bm}
  H.~Grosse and J.~Madore,
  ``A Noncommutative version of the Schwinger model,''
  Phys.\ Lett.\  B {\bf 283} (1992) 218.

\bibitem{CarowWatamura:1997qw}
  U.~Carow-Watamura and S.~Watamura,
  ``Differential calculus on fuzzy sphere and scalar field,''
  Int.\ J.\ Mod.\ Phys.\  A {\bf 13} (1998) 3235
  [arXiv:q-alg/9710034].

\bibitem{CarowWatamura:1998jn}
  U.~Carow-Watamura and S.~Watamura,
  ``Noncommutative geometry and gauge theory on fuzzy sphere,''
  Commun.\ Math.\ Phys.\  {\bf 212} (2000) 395
  [arXiv:hep-th/9801195].

\bibitem{Grosse:2000gd}
  H.~Grosse, J.~Madore and H.~Steinacker,
  ``Field theory on the q-deformed fuzzy sphere. I,''
  J.\ Geom.\ Phys.\  {\bf 38} (2001) 308
  [arXiv:hep-th/0005273].

\bibitem{Azuma:2005bj}
  T.~Azuma, S.~Bal and J.~Nishimura,
  Phys.\ Rev.\  D {\bf 72} (2005) 066005
  [arXiv:hep-th/0504217].

\bibitem{faddeev}
  L.D.~Faddeev and V.N.~Popov,
  %
  Phys.Lett.{\bf 25B},29 (1967).


\bibitem{Berenstein:2008eg}
  D.~E.~Berenstein, M.~Hanada and S.~A.~Hartnoll,
  JHEP {\bf 0902} (2009) 010
  [arXiv:0805.4658 [hep-th]].


\bibitem{Nash:2008dk}
  C.~Nash and D.~O'Connor,
  J.\ Phys.\ A  {\bf 42} (2009) 012002
  [arXiv:0809.2960 [hep-th]].

\bibitem{Lieb}
  E.H.Lieb,
  Phys.Rev.Lett.{\bf 73},2158 (1994).

\end{thebibliography}
\end{document}